\documentclass[10pt,a4paper]{article}
\usepackage{inputenc}
\usepackage{setspace}
\usepackage{amsmath}
\usepackage{amsfonts}
\usepackage{amssymb}
\usepackage{amsthm}
\usepackage{bm}
\usepackage{graphicx}
\usepackage{enumitem}
\graphicspath{ {./figures/} }

\usepackage[margin=1in]{geometry}
\usepackage[title]{appendix}
\usepackage{authblk}
\usepackage{xcolor}
\usepackage{times}
\usepackage{bm}
\usepackage{natbib}
\usepackage{graphicx}
\usepackage{amsmath}
\usepackage{float}
\usepackage{xcolor}
\usepackage{amsfonts}
\usepackage{bbm}
\usepackage{booktabs} 
\usepackage{caption}
\usepackage{subcaption}
\usepackage{enumitem}
\usepackage{mathrsfs} 
\usepackage{accents}
\usepackage{xr}
\usepackage{longtable}
\usepackage{adjustbox}
\usepackage{hyperref}
\usepackage{optidef}
\usepackage[mathscr]{euscript}
\usepackage{bbm}
\newcommand{\argmin}{\arg\!\min}
\newcommand{\argmax}{\arg\!\max}

\theoremstyle{thmstyleone}
\newtheorem{thm}{Theorem} 
\newtheorem{corollary}{Corollary} 
\newtheorem{lem}[thm]{Lemma}
\newtheorem{cond}{Condition}
\newtheorem{theorem}{Theorem S\ignorespaces}
\newtheorem{lemma}[theorem]{Lemma S\ignorespaces}
\theoremstyle{thmstyletwo}
\newtheorem{exmp}{Example}[section]

\theoremstyle{thmstylethree}

\usepackage{caption}
\usepackage[noend]{algpseudocode}

\usepackage{tikz}
\usetikzlibrary{positioning}

\tikzset{
    every neuron/.style={
        circle,
        draw,
        minimum size=1cm
    },
    neuron missing/.style={
        draw=none, 
        scale=4,
        text height=0.333cm,
        execute at begin node=\color{black}$\vdots$
    },
}

\usepackage{natbib}
\title{Data fusion using weakly aligned sources}
\author[*]{Sijia Li}
\author[*$\dagger$]{Peter B. Gilbert}
\author[$\star$]{Rui Duan}
\author[$\ddagger$]{Alex Luedtke}
\affil[*]{Department of Biostatistics, University of Washington}
\affil[$\dagger$]{Vaccine and Infectious Disease Division, Fred Hutchinson Cancer Center}
\affil[$\ddagger$]{Department of Statistics, University of Washington}
\affil[$\star$]{Department of Biostatistics, Harvard T.H. Chan School of Public Health}
\date{}

\begin{document}
\maketitle

\begin{abstract}
We introduce a new data fusion method that utilizes multiple data sources to estimate a smooth, finite-dimensional parameter. Most existing methods only make use of fully aligned data sources that share common conditional distributions of one or more variables of interest. However, in many settings, the scarcity of fully aligned sources can make existing methods require unduly large sample sizes to be useful. Our approach enables the incorporation of weakly aligned data sources that are not perfectly aligned, provided their degree of misalignment is known up to finite-dimensional parameters. We quantify the additional efficiency gains achieved through the integration of these weakly aligned sources. We characterize the semiparametric efficiency bound and provide a general means to construct estimators achieving these efficiency gains. We illustrate our results by fusing data from two harmonized HIV monoclonal antibody prevention efficacy trials to study how a neutralizing antibody biomarker associates with HIV genotype.
\end{abstract}

\section{Introduction}
\label{s:intro}
The increasing availability of data has led to a growing interest in data fusion, which involves combing data from multiple sources to obtain a global summary of a target population of interest. Existing data fusion approaches require the data sources to be fused to share a common conditional distribution of one or more variables of interest with the target distribution \citep{pearl2011transportability,hernan2011compound,bareinboim2014transportability,stuart2015assessing,dahabreh2019efficient,kallus2020optimal,li2021efficient}. This is known as the exchangeability over data sources condition, or the common distributions condition, which enables the transfer of conclusions across data sources and thus facilitates data fusion. In some special cases, this condition can be depicted in a causal diagram as a lack of directional edges connecting data source nodes to those of other variables \citep{pearl2011transportability}. With the use of such aligned data sources, data fusion often result in efficiency gains compared to traditional analyses that only leverage one data source.

Several variants of the common distributions condition have been proposed in the literature. For instance, this condition can be relaxed to only require a common outcome model \citep{rudolph2017robust,dahabreh2019efficient} or other common statistical summaries \citep{taylor2023data}. These weaker conditions generally only enable valid inferences for particular summaries of the target distribution, such as an average treatment effect. To infer about a generic summary, the common distributions condition is still generally required.

An aligned conditional distribution condition is often most likely to hold if the data source contains a rich set of variables that fully explains the variability between the observed and target populations. This condition often fails when these populations have different sets of covariates measured, or in cases when the conditional distributions across data sources are fundamentally different. For example, evaluating the efficacy of the same HIV vaccine regimen to prevent HIV-1 acquisition diagnosis using pooled data from trials with different prevailing HIV strains may not be possible due to the variability in the conditional distributions of new HIV-1 diagnosis across trials, even with the inclusion of baseline covariates.  As a result, the lack of exact alignment among data sources often leaves little room for existing data fusion techniques to be applied. This raises an important question: can researchers unlock efficiency gains by making use of slightly misaligned data sources? In this work, we answer that question affirmatively by introducing a novel data fusion method that can take advantage of such data sources provided appropriate conditions are satisfied.

We examine the general case where different data sources weakly align with different parts of the distribution of the target population, in the sense that the ratio of certain conditional densities between these sources and the target distribution can be characterized by selection bias models. Selection bias models have been discussed by \cite{vardi1982nonparametric, vardi1985empirical}, \cite{gill1988large}, and \cite{bickel1993efficient} in the case of completely known density ratios, and by \cite{qin1998inferences} and \cite{gilbert2000large} in the case of density ratios that are known up to finite dimensional parameters. In the aforementioned HIV vaccine efficacy example, a selection bias model can be used to describe a covariate-dependent shift in the log odds of new HIV-1 diagnosis across trials. We show that using such a model makes it possible to calibrate the discrepancy in the conditional distributions of the outcome between data sources before data fusion. {Our main contributions consist in introducing this weakly aligned data fusion framework and developing the corresponding efficiency theory for the estimation of any any smooth parameter relative to the implied statistical model.}

{Depending on how our proposed framework is used, it can improve on standard approaches --- such as those detailed in \cite{li2021efficient} --- in one of two ways. On the one hand, it makes it possible to incorporate information from additional data sources, namely weakly aligned ones. We show that incorporating this information will typically improve statistical efficiency and will never worsen it. On the other hand, it makes it possible to robustify existing data fusion methods by weakening their required conditions. In particular, rather than assuming all usable sources are aligned with the target population, we show in this work that it suffices to know one that is aligned with the target and then assume that the others are only weakly aligned. Making these assumptions results in a larger statistical model, thereby yielding more robust inferences. Moreover, by using the efficient estimator we provide, the loss in efficiency incurred for this improved robustness can be minimized.}

\section{Notations and Assumptions}
\label{s:notations}

We will use the same notation as in \cite{li2021efficient}. We use $[d]$ to denote $\{1,\ldots,d\}$ for a natural number $d$ and let $E_{\nu}$ denote the expectation operator under a distribution $\nu$. We use $Z=(Z_1,\ldots,Z_d)$ to denote a random variable and we let $\bar{Z}_j=(Z_1,\ldots,Z_j)$ for $j\in [d]$, where we let $\bar{Z}_0$ denote an empty set. Random variables will be denoted by capital letters whereas their realizations will be denoted by corresponding lowercase letters. To indicate conditioning on a random variable taking a specific value, we will condition on lowercase letters in expectations. For instance, $E_\nu(Z_2|z_1)$ will denote $E_\nu(Z_2|Z_1=z_1)$. For a distribution $Q$ of $Z$ and $j\in [d]$, $Q_j(\,\cdot\mid \bar{z}_{j-1})$ represents the conditional distribution of $Z_j\mid \bar{Z}_{j-1}=\bar{z}_{j-1}$. Similarly, for a distribution $P$ of $(Z,S)$, $P_j(\,\cdot\mid \bar{z}_{j-1},s)$ denotes the conditional distribution of $Z_j\mid \bar{Z}_{j-1}=\bar{z}_{j-1},S=s$. We will assume sufficient regularity conditions so that all such conditional distributions are well-defined, and that all discussed distributions of $Z_j\mid \bar{Z}_{j-1}=\bar{z}_{j-1}$ and $Z_j\mid \bar{Z}_{j-1}=\bar{z}_{j-1},S=s$ are defined on some common measurable space. We will denote the conditional densities of {$P_j$} and {$Q_j$} by {$p_j$} and {$q_j$}, respectively, and if there is no ambiguity we will drop the $j$ subscript, such as by writing {$p(z_j\mid \bar{z}_{j-1},s)$} rather than {$p_j(z_j\mid \bar{z}_{j-1},s)$}. For a set $\mathcal{D}$, we use $|\mathcal{D}|$ to denote its cardinality. For a collection of vectors $v_\ell\in\mathbb{R}^{d_\ell}$, $\ell\in L$, we write $(v_\ell)_{\ell\in L}\in\mathbb{R}^{\sum_{\ell=1}^L d_\ell}$ to denote the concatenation of these vectors.

Suppose we have a collection of $k$ data sources and want to estimate the target parameter $\psi:\mathcal{Q} \rightarrow \mathbb{R}$ of a target distribution $Q^0$. This distribution is known to belong to a collection $\mathcal{Q}$ of distributions of a random variable $Z=(Z_1, \ldots, Z_d)$, where $Z$ takes values in $\mathcal{Z}=\prod_{j=1}^d \mathcal{Z}_j$. Since the summary $\psi$ may only depend on certain components of $Q^0$, we let $\mathcal{I}\subset [d]$ denote a set of irrelevant indices $j$ such that $\psi$ is not a function of the distribution of $Z_j\mid \bar{Z}_{j-1}$ --- for example, if $d=3$ and $\psi$ is the G-computation parameter $\psi(Q):=\iint Q_3(dz_3\mid Z_2=1,z_1)Q_1(dz_1)$ \citep{robins1986new}, then $\mathcal{I}=\{2\}$. We use $\mathcal{J}=[d]\backslash \mathcal{I}$ to denote the set of indices that may be relevant to the evaluation of $\psi$, which we call the set of relevant indices.

Instead of directly obtaining samples from the target distribution $Q^0$, we observe $n$ independent copies of $X = (Z, S)$ drawn from some common distribution $P^0$, where $Z$ takes values in $\mathcal{Z}$ and $S$ is a categorical random variable denoting the data source that has support $[k]$. The distribution $P^0$ is known to weakly align with $Q^0$ such that, for each $j \in \mathcal{J}$ and some $s$, the conditional density of the observed $P^0$ is such that
\begin{align}
    p^0(z_j \mid \bar{z}_{j-1},s) &= \frac{w_{j,s}(\bar{z}_j;\beta^0_{j,s})}{W_{j,s}(Q^0_j;\beta^0_{j,s})(\bar{z}_{j-1})}q^0(z_j \mid \bar{z}_{j-1}) \label{eq: select_bias}, 
\end{align}
where the form of weight function $w_{j,s}$ is known, $\beta^0_{j,s} \in \mathbbm{R}^{c_{j,s}}$ is a vector of length $c_{j,s}$, and $W_{j,s}( Q^0_j;\beta^0_{j,s})(\bar{z}_{j-1})$ is the normalizing function given by 
\begin{equation}
    W_{j,s}( Q^0_j;\beta^0_{j,s})(\bar{z}_{j-1}) : = \int_{\mathcal{Z}_j} w_{j,s}( \bar{z}_{j};\beta^0_{j,s}) \,Q^0_j(dz_j\mid \bar{z}_{j-1}) \nonumber.
\end{equation}
\begin{sloppypar}
\noindent The quantity $W_{j,s}( Q^0_j,\beta^0_{j,s})(\bar{z}_{j-1})$ is assumed to be strictly positive and finite for all possible values of $\beta^0_{j,s}$. 
The values of $\beta^0_{j,s}$ can vary in magnitude and in lengths across different conditional distribution index $j$ and data source $s$.  We let $w^*_{j,s}(\bar{z}_{j};\beta_{j,s})  := w_{j,s}(\bar{z}_{j};\beta_{j,s})/W_{j,s}(Q^0_j;\beta_{j,s})(\bar{z}_{j-1})$ hereafter, and we refer to this function as a density ratio.    
\end{sloppypar}

The selection bias model specified in \eqref{eq: select_bias} generalizes a variety of models, including univariate and multivariate logistic models \citep{gilbert2000large}, density ratios of distributions from an exponential family \citep{bickel1993efficient}, and truncated regression \citep{bickel1993efficient}. We give two examples below to illustrate the wide applicability of selection bias models and refer readers to Chapter 4.4 in \citep{bickel1993efficient} for additional examples.

\begin{exmp}
\textbf{Multivariate logistic regression model.} Suppose $Z = (Z_1,Z_2,Z_3) = (X,A,Y)$, where $X$ is a covariate, $A$ is an indicator of treatment, and $Y$ is binary outcome. When $w_{3,s}(z;\beta_{3,s}) = \exp({\beta_{3,s} y })$ for data source $s$, \eqref{eq: select_bias} rewrites as follows when $j=3$:
    \begin{align*}
    &\log \left\{\frac{p^0(Y=1 \mid A=a, X=x, S=s)}{p^0(Y=0\mid A=a, X=x, S=s)} \right\} = \log \left\{\frac{q^0(Y=1 \mid A=a, X=x)}{q^0(Y=0\mid A=a, X=x)} \right\} +\beta_{3,s}.
    \end{align*}
    The above corresponds to a constant shift in log odds between the observed population and the target population across all strata defined by treatment $A$ and covariate $X$. With a more complex choice of $w_{3,s}(z;\beta_{3,s}) = \exp({\beta^1_{3,s} y + \beta^2_{3,s} a y + \beta^3_{3,s} x y  + \beta^4_{3,s} a x  y})$, we arrive at a more flexible shift in log odds that differs across strata defined by $A$ and $X$. This biased sampling model can be generalized naturally for multi-level categorical outcomes $Y$ and can be useful in interpreting differential vaccine efficacy against different strains \citep{gilbert1999maximum}. Even more generally, with a choice of $w_{j,s}(\bar{z}_j;\beta_{j,s}) = \exp\{f(\bar{z}_j;\beta_{j,s}) \}$ for some function $f$, the selection bias model \eqref{eq: select_bias} can accurately model shifts between any two distributions that are from the same exponential family \citep{patil1978weighted}. 
\end{exmp}

\begin{exmp}
\textbf{Truncated regression model \citep{bickel1993efficient}.} Truncated regression models arise naturally in a variety of fields, including astronomy and biostatistics \citep{bhattacharya1983nonparametric,jewell1985least}. Suppose $Z = (Z_1,Z_2)$ and that data source $1$ consists of observations that are directly sampled from $Q^0$. In data source $2$, the outcome $Z_2$ is only observed if $Z_2 \geq \beta_{2,2}$ for some threshold $\beta_{2,2} \in \mathbbm{R}$. For example, $Z_2$ may denote the luminosity of a distant celestial object, and data source $2$ may come from a low-sensitivity telescope that only observes objects whose luminosity exceeds some threshold $\beta_{2,2}$. To use both data sources to study the association between the size of a celestial object, $Z_1$, and its luminosity, $Z_2$, it is helpful to note that the observed conditional density in data source $s=2$ is given by
    \begin{align*}
    p^0(z_2 \mid z_1,s ) & = \frac{\mathbbm{1}(z_2 \geq \beta_{2,2})q^0(z_2 \mid z_1 )}{E_{Q^0}[\mathbbm{1}(Z_2 \geq \beta_{2,2}) \mid z_1]}.
    \end{align*}
    Hence, \eqref{eq: select_bias} is satisfied when $j=2$ with $w_{2,s}(\bar{z}_2;\beta_{2,2})=\mathbbm{1}(s=1)+\mathbbm{1}(s=2,z_2 \geq \beta_{2,2})$.
\end{exmp}

Other forms of the weight function can also be appropriate to model density ratios \citep{cook1974model,rao1965discrete,patil1978weighted,bickel1993efficient}.  Depending on the extent of shifts between the observed and the target distribution, the form of the weight function will vary.  In what follows, we formally outline an alignment condition that makes it possible to relate the conditional distributions $P^0_j(\cdot \mid \bar{z}_{j-1},s)$ and $Q^0_j(\cdot \mid \bar{z}_{j-1})$.

\begin{cond} \label{cond:sufficient_alignment}
\textit{(Sufficient alignment)}\; For each relevant index $j\in \mathcal{J}$, there exist known, disjoint subsets $\mathcal{A}_j$ and $\mathcal{W}_j$ of $[k]$ such that all of the following hold:
\begin{enumerate}[label=\textbf{1\alph*},ref=1\alph*,leftmargin=*]
     \item\label{suff_overlap} \textit{(Sufficient overlap)}\; for all $s\in \mathcal{S}_j:=\mathcal{A}_j\cup\mathcal{W}_j$, the marginal distribution of $\bar{Z}_{j-1}$ under sampling from $Q^0$ is absolutely continuous with respect to the conditional distribution of $\bar{Z}_{j-1}\mid S=s$ under sampling from $P^0$;
    \item \label{identB} \textit{(At least one aligned data source)} $\mathcal{A}_j\not=\emptyset$ and, for all $s\in \mathcal{A}_j$,  $p^0(z_j \mid \bar{z}_{j-1},s) = q^0(z_j \mid \bar{z}_{j-1})$;
    \item \label{weak_alignment} \textit{(Weakly aligned data sources)} for all $s\in \mathcal{W}_j$, $p^0(z_j \mid \bar{z}_{j-1},s) = w^*_{j,s}(\bar{z}_j;\beta^0_{j,s})q^0(z_j \mid \bar{z}_{j-1})$.
\end{enumerate}
\end{cond}
For a given $j$, we refer to $\mathcal{A}_j$ and $\mathcal{W}_j$ as the aligned fusion set and weakly aligned fusion set, respectively. We denote the concatenation of vectors $\beta_{j,s}$ over all $s \in \mathcal{W}_j$ as $\beta_j  = (\beta_{j,s})_{s \in \mathcal{W}_j}$, and $\beta_{j,s}$ over all $j \in \mathcal{J}$ and $s\in \mathcal{W}_j$ as $\beta = (\beta_{j,s})_{j\in \mathcal{J},s\in \mathcal{W}_j}$. The true parameter $\beta^0$ is known to belong to a collection $\mathcal{B}:= \mathbb{R}^t$ of vectors of length $t :=\sum_{j \in \mathcal{J}}\sum_{s \in \mathcal{W}_j}c_{j,s}$. For ease of notation, we define a mapping $\mathrm{B}: \mathcal{P}_{\mathcal{Q},\mathcal{B}} \rightarrow \mathcal{B}$ such that $\beta^0:= \mathrm{B}(P^0)$. Similarly, the true parameter $\beta^0_{j,s}$ for each $j\in\mathcal{J}$ and $s \in \mathcal{W}_j$ is known to belong to a collection $\mathcal{B}_{j,s}:=\mathbbm{R}^{c_{j,s}}$ of vectors of length $c_{j,s}$. Condition~\ref{cond:sufficient_alignment} implies that $P^0$ is known to belong to the collection $\mathcal{P}_{\mathcal{Q},\mathcal{B}}$ of distributions $P_{Q,\beta}$ with support on $\mathcal{Z} \times [k]$ for which there exists a $Q \in \mathcal{Q}$ and $\beta_{j,s} \in \mathcal{B}_{j,s}$ such that, for all $j \in \mathcal{J}$ and $s \in \mathcal{S}_j$: (a) the marginal distribution of $\bar{Z}_{j-1}$ under sampling from $Q$ is absolutely continuous with respect to the conditional distribution $\bar{Z}_{j-1}\mid S=s$ under sampling from $P$, (b) $P_j(\cdot \mid \bar{z}_{j-1},s) =  Q_j(\cdot \mid \bar{z}_{j-1})$ $Q$-almost everywhere for all $s \in \mathcal{A}_j$, and (c) $p(z_j \mid \bar{z}_{j-1},s) = w^*_{j,s}(\bar{z}_j;\beta_{j,s})q(z_j \mid \bar{z}_{j-1})$ for all $s \in \mathcal{W}_j$.  In addition, we define $\mathcal{P}_{Q^0,\mathcal{B}}=\{P_{Q^0,\beta}:\beta \in \mathcal{B}\}$ and $\mathcal{P}_{\mathcal{Q},\beta^0}=\{P_{Q,\beta^0}:Q \in \mathcal{Q}\}$. We will refer to $\mathcal{Q}$, $\mathcal{P}_{Q^0,\mathcal{B}}$, $\mathcal{P}_{\mathcal{Q},\beta^0}$, and $\mathcal{P}_{\mathcal{Q},\mathcal{B}}$ as models.

\begin{figure}[htb]
    \centering
    \includegraphics[width=0.8\textwidth]{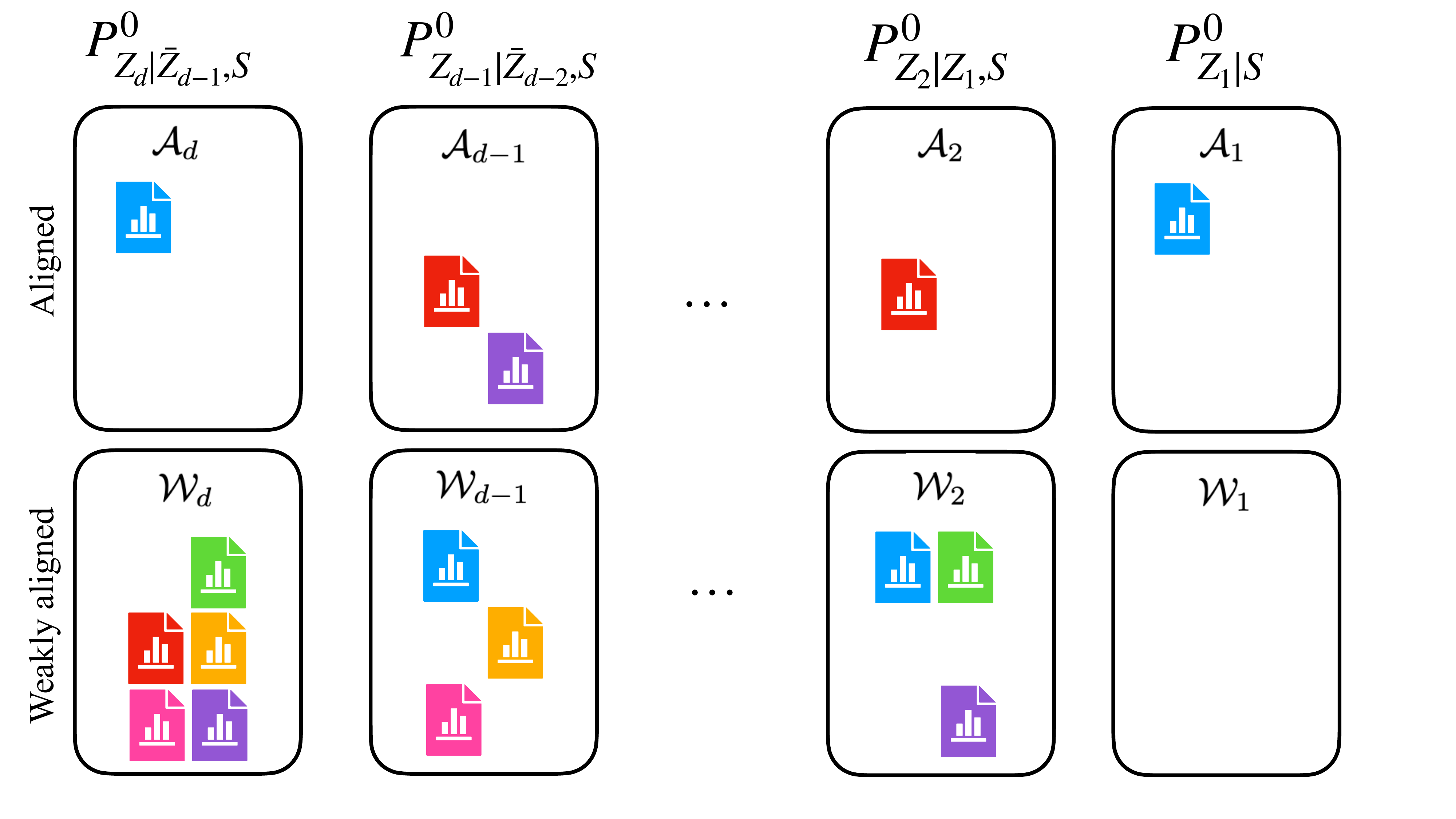}
    \caption{Previous data fusion methods propose to use aligned fusion sets only (top row). We advocate fusing weakly aligned data sources (bottom row) in addition to aligned data sources. While aligned data sources might be scarce (small $ |\mathcal{A}_{d}| $), there can be more weakly aligned sources (larger $|\mathcal{W}_{d}|$). If $\mathcal{W}_j$ is empty for some $j \in [d]$, as illustrated in $\mathcal{W}_1$ in above, we can still identify the target estimand as long as $\mathcal{A}_j$ is nonempty. }
    \label{fig:visualization}
\end{figure}

Condition \ref{suff_overlap} ensures the conditional distributions appearing in Conditions \ref{identB} and \ref{weak_alignment} are uniquely defined up to $Q^0$-null sets. Condition \ref{weak_alignment} imposes that data sources in $\mathcal{W}_j$ are weakly aligned with a known  $w_{j,s}$ but unknown $\beta^0_{j,s}$. Each function $w_{j,s}(\,\cdot\,;\beta^0_{j,s})$ calibrates the conditional distribution of $Z_j$ given $\bar{Z}_{j-1}$ in data source $s$ to that of the target population. As the name suggests, weak alignment is typically a less restrictive property than alignment. It is important to note that if there are not any data sources known to be aligned with the target distribution at an index $j$, then the parameters $\beta^0_{j,s}$ describing the shift of the weakly aligned data sources $s\in\mathcal{W}_j$ will not generally be identifiable. Therefore, we impose Condition \ref{identB}, which requires having and correctly identifying nonempty sets $\mathcal{A}_j$ of data sources that align with the target. Condition \ref{identB} is the same as the alignment condition imposed in \cite{li2021efficient}. Since we can always take the weakly aligned fusion sets $\mathcal{W}_j$ to be empty for all $j$, Condition~\ref{cond:sufficient_alignment} in the current work is no more restrictive than Condition~1 from \cite{li2021efficient}. However, we will show that, when there is at least one $j$ such that $\mathcal{W}_j$ is nonempty, it will generally be possible to gain statistical efficiency when estimating summaries of the target distribution $Q^0$. See Figure~\ref{fig:visualization} for an illustration comparing the alignment condition from the current work to that in \cite{li2021efficient}. {In addition, we also provide one toy example and two real-word examples of aligned, weakly aligned and unaligned data sources in Supplementary Appendix~\ref{sec:app:example}}.

Together, Conditions~\ref{suff_overlap} and \ref{identB} enable the identification of the summary $\psi(Q^0)$ of the target population using the observed data sources in $\mathcal{A}_j, j\in \mathcal{J}$ only, thus making it feasible to learn about $\psi(Q^0)$ solely from the observed data distribution $P^0$. To express this identifiability result, we define a mapping $\theta : \mathcal{P}_{\mathcal{Q}, \mathcal{B}}\rightarrow\mathcal{Q}$. In particular, for any $P\in\mathcal{P}_{\mathcal{Q}, \mathcal{B}}$, we let $\theta(P)$ denote an arbitrarily selected distribution from the set $\mathcal{Q}(P)$ of distributions $Q\in\mathcal{Q}$ that are such that, for each $j\in \mathcal{J}$, $Z_j\mid \bar{Z}_{j-1}$ under sampling from $Q$ has the same distribution as $Z_j\mid \bar{Z}_{j-1},S\in\mathcal{A}_j$ under sampling from $P$. We let $\phi = \psi \circ \theta$. Then it can be shown that under Condition~\ref{cond:sufficient_alignment}, $\psi(Q^0) = \phi(P^0)$. The proof of this result can be found in Theorem 1 of \cite{li2021efficient}.

\section{Methods}
\label{s:methods}

We now outline strategies to construct regular and asymptotically linear estimators of $\psi(Q^0)$ with the observed data when $\mathcal{Q}$ is a collection of nonparametric distributions $Q$. Such estimators are appealing since they are consistent and asymptotically normal under mild conditions, thus facilitating the construction of confidence intervals and hypothesis tests. The construction of such estimators requires a key object, namely, gradients of statistical parameters \citep{bickel1993efficient}. It is desirable to seek for the canonical gradient as it has the least variance among all gradients. We begin by deriving the form of the canonical gradient $D^\mathrm{eff}_{P^0}$ in the model $\mathcal{P}_{\mathcal{Q},\mathcal{B}}$ implied by $\mathcal{Q}$ and the data fusion conditions. We then use knowledge of this gradient to construct a one-step estimator of $\psi(Q^0)$ \citep{bickel1982adaptive}. Given an estimate $\widehat{P}$ of $P^0$, the one-step estimator takes the form of $\hat{\phi} \equiv \phi(\widehat{P}) + \sum_{i=1}^n D^\mathrm{eff}_{\widehat{P}}(X_i)/n$. When the remainder term $R(\widehat{P},P^0)\equiv \phi(\widehat{P}) - \phi(P^0) + E_{P^0}[D^\mathrm{eff}_{\widehat{P}}(X)]$ is $o_p(n^{-1/2})$ and the empirical mean of $D^\mathrm{eff}_{\widehat{P}}(X)-D^\mathrm{eff}_{P^0}(X)$ is within $o_p(n^{-1/2})$ of the mean of this term when $X\sim P^0$, this estimator will be asymptotically linear with the efficient influence function $D^\mathrm{eff}_{P^0}$, that is, 
$$\sqrt{n} [\hat{\phi} -  \phi(P^0)] \xrightarrow{d} N(0, \sigma^2_{P^0}),$$
where $\sigma^2_{P^0}$ is the variance of $D^\mathrm{eff}_{P^0}(X)$ when $X \sim P^0$. We refer the readers to \cite{bickel1993efficient}, \cite{ibragimov1981statistical}, and \cite{bickel1982adaptive} for overviews of influence functions, gradients of statistical parameters, semiparametric efficiency theory, and one-step estimators. 

Our proposed data fusion procedure proceeds as follows. The first step is to obtain a gradient of $\psi$ at $Q^0$ in a model where direct sampling from the target distribution $Q^0$ is possible. For many estimands of interest, this step can be accomplished with relative ease as these gradients are readily accessible in existing literature. As the second step,  we leverage the findings from this section to establish a correspondence between the aforementioned gradient and the canonical gradient of $\phi$ at $P^0$ within our data fusion framework. Lastly, we use the derived canonical gradient to construct a one-step estimator. {We refer readers to Supplementary Appendix~\ref{sec:app:regularity} for a detailed outline of how to construct such an estimator.} There are also alternative approaches for using knowledge of a gradient to construct asymptotically linear estimators, such as those based on estimating equations \citep{van2003unified,tsiatis2006semiparametric} and targeted minimum loss-based estimation \citep{van2006targeted}.

We start by deriving the canonical gradients of (i) $\phi$ in a model where $\beta^0$ is known and (ii) $\beta^0$ in a model where it is not known. We will subsequently use these functions to obtain the canonical gradient of $\phi$ under $\mathcal{P}_{\mathcal{Q}, \mathcal{B}}$. To derive the form of a gradient of the target estimand when $\beta^0$ is known, we require one additional regularity condition, which we now introduce. To express this condition, we let $\lambda_{j-1}$ denote the Radon-Nikodym derivative of the marginal distribution of $\bar{Z}_{j-1}$ under sampling from $Q^0$ relative to the conditional distribution of $\bar{Z}_{j-1} \mid S \in \mathcal{S}_j$ under sampling from $P^0$ and   $\lambda^{\dagger}_{j}(\bar{z}_j,s): = \lambda_{j-1}(\bar{z}_{j-1}) q^0(z_j\mid \bar{z}_{j-1})/p^0(z_j \mid \bar{z}_{j-1}, s)$.

\begin{cond}\textit{(Strong overlap)}
\label{cond:overlaps}
For each fixed $j\in\mathcal{J}$ and $s \in \mathcal{W}_j$, there exists a $u_{j,s}\in (0,\infty)$ such that $Q^0\{u_{j,s}^{-1}\le \lambda^\dagger_{j}(\bar{Z}_{j},s)\le u_{j,s}\}=1$.
\end{cond}

Condition~\ref{cond:overlaps} strengthens the overlap in $\bar{Z}_{j-1}$ between data sources specified by Condition~\ref{suff_overlap}. Under this strengthening, for $Q^0$-almost all $\bar{z}_{j-1}$, an observation with $\bar{Z}_{j-1} = \bar{z}_{j-1}$ has a positive probability of being observed in any data source $s \in \mathcal{S}_j$.  In the meantime, it also requires the Radon-Nikodym derivative of $Q^0_j(\cdot \mid \bar{z}_{j-1})$ relative to $P^0_j(\cdot \mid \bar{z}_{j-1},s)$ for each $s \in \mathcal{W}_j$ and $j\in \mathcal{J}$ to be bounded for each $\bar{z}_{j-1}$ in the support of $Q^0$. 

Suppose that Conditions~\ref{cond:sufficient_alignment}  and \ref{cond:overlaps} hold, $\beta^0$ is known and $\psi$ is pathwise differentiable at $Q^0$ relative to $\mathcal{Q}$ with gradient $D_{Q^0}$. Then a gradient of $\phi$ relative to $\mathcal{P}_{\mathcal{Q}, \beta^0}$ is given by
\begin{align}
    D_{P^0}(z,s;\beta^0) =  \sum_{j\in\mathcal{J}} \mathbbm{1}(\bar{z}_{j-1}\in\bar{\mathcal{Z}}_{j-1}^\dagger) \frac{\mathbbm{1}(s\in\mathcal{S}_j)}{P^0(S\in\mathcal{S}_j)} \lambda^\dagger_{j}(\bar{z}_j,s;\beta_{j,s}^0) D_{Q^0,j} (\bar{z}_j),\label{eq:gradient_known}
\end{align}
\begin{sloppypar}
\noindent where $\bar{\mathcal{Z}}_{j-1}^\dagger$ denotes the support of $\bar{Z}_{j-1}$ under sampling from $Q^0$ 
and $D_{Q^0,j}(\bar{z}_j)= E_{Q^0}[D_{Q^0}(Z)\mid \bar{Z}_j = \bar{z}_j] -E_{Q^0}[D_{Q^0}(Z)\mid \bar{Z}_{j-1} = \bar{z}_{j-1}] $. {We provide the proof in Supplementary Appendix~\ref{sec:app:lem_known}.} This result resembles Corollary 1 in \cite{li2021efficient} which proposes to use only aligned data sources by constructing an estimator using the following gradient:
\begin{align}
    D^{\mathcal{A}}_{P^0}(z,s)=\sum_{j\in\mathcal{J}} \mathbbm{1}(\bar{z}_{j-1}\in\bar{\mathcal{Z}}_{j-1}^\dagger)\frac{\mathbbm{1}(s\in\mathcal{A}_j)}{P(S\in\mathcal{A}_j)} \lambda^\mathsection_{j-1}(\bar{z}_{j-1})D_{Q^0,j}(\bar{z}_j), \label{eq:gradient_alignedonly}
\end{align}
where $\lambda^\mathsection_{j-1}$ denote the Radon-Nikodym derivative of the marginal distribution of $\bar{Z}_{j-1}$ under sampling from $Q^0$ relative to the conditional distribution of $\bar{Z}_{j-1}\mid S\in \mathcal{A}_j$ under $P^0$. The gradients in \eqref{eq:gradient_known} and \eqref{eq:gradient_alignedonly} differ in some key aspects. The gradient in \eqref{eq:gradient_known} uses not only aligned data sources in $\mathcal{A}_j$ but also weakly aligned data sources in $\mathcal{W}_j$ as reflected by the indicator term $\mathbbm{1}(s\in\mathcal{S}_j)$. Because of the inclusion of these slightly misaligned sources, we need to additionally correct for the potential shifts in the conditional distribution of $Z_j \mid \bar{Z}_{j-1}, S$ for data sources in $\mathcal{W}_j$, besides correcting the shifts in the joint distribution of $\bar{Z}_{j-1} \mid S$. As a result, we have $\lambda^\dagger_j$ compared to $\lambda^\mathsection_{j-1}$ in \eqref{eq:gradient_alignedonly}. When $\mathcal{W}_j = \emptyset$ for all $j$, all the data sources in $\mathcal{S}_j$ are aligned in the sense that $p^0(z_j \mid \bar{z}_{j-1}, s) = q^0(z_j \mid \bar{z}_{j-1})$, which in turn, gives  $\lambda^\dagger_j(\bar{z}_{j},s)=\lambda^\mathsection_{j-1}(\bar{z}_{j-1})$ $Q^0$-almost surely. In this special case, \eqref{eq:gradient_known} recovers the results from Corollary 1 from \cite{li2021efficient}. 

{The canonical gradient of $\phi$ under $\mathcal{P}_{\mathcal{Q},\beta^0}$ can be derived by projecting any valid gradients of $\phi$, such as $D_{P^0}$ or $D^\mathcal{A}_{P^0}$, onto the tangent space $\mathcal{T}(P^0,\mathcal{P}_{\mathcal{Q},\beta^0})$. We let $\delta_{s}:= P^0(S=s)$,  $r_j(\bar{z}_{j};\beta_j) := \left\{\sum_{m \in \mathcal{S}_j} \delta_{m} w^*_{j,m}(\bar{z}_{j};\beta_{j,m})p^0(\bar{z}_{j-1}\mid m)/q^0(\bar{z}_{j-1})\right\}^{-1}$, and $r_{j,s}(\bar{z}_{j};\beta_{j,s}) :=  \delta_{s}w^*_{j,s}(\bar{z}_{j};\beta_{j,s})r_j(\bar{z}_{j};\beta_j)p^0(\bar{z}_{j-1}\mid s)/q^0(\bar{z}_{j-1})$. In addition, we let $\bar{w}^*_j:= (w^*_{j,m})_{m \in \mathcal{S}_j}$, $\bar{r}_j:= (r_{j,m})_{m\in \mathcal{S}_j}$, and $\Delta$ be the diagonal matrix with diagonal $((\delta_{m})_{m \in \mathcal{S}_j})^{\top}$. We define an $|\mathcal{S}_j| \times |\mathcal{S}_j| $ matrix $M_j(\bar{z}_{j-1};\beta_j) = \Delta^{-1} - \int r_j(\bar{z}_j;\beta_j)\bar{w}^*_j(\bar{z}_j;\beta_j){\bar{w}_j}^{*\top}(\bar{z}_j;\beta_j)\,Q^0_j(dz_j\mid\bar{z}_{j-1}) $ and let $M_j^{-}$ be the generalized inverse of $M_j$. We denote $d_j(\bar{z}_j;\beta^0_j) := \sum_{m \in \mathcal{S}_j} r_{j,m}(\bar{z}_j;\beta^0_{j,m})D_{P^0,j}(\bar{z}_{j},m;\beta^0_{j,m})$, where $D_{P^0,j}(\bar{z}_{j},s;\beta^0_{j,s}) = E_{P^0}[D_{P^0}(Z,S;\beta^0)\mid \bar{Z}_{j} = \bar{z}_{j},S=s] -E_{P^0}[D_{P^0}(Z,S;\beta^0)\mid \bar{Z}_{j-1} = \bar{z}_{j-1},S=s]$. We denote $\tilde{d}_j(\bar{z}_j;\beta^0_j):=d_j(\bar{z}_j;\beta^0_j)- E_{P^0}\left[d_j(\bar{z}_j;\beta^0_j)\mid  \bar{z}_{j-1}, S \in \mathcal{A}_j\right] + E_{P^0}\big[d_j(\bar{z}_j;\beta^0_j) {\bar{w}_j}^{*\top}(\bar{Z}_{j};\beta^0_j)\mid  \bar{z}_{j-1}, S \in \mathcal{A}_j\big]  M^{-}_j(\bar{z}_{j-1};\beta^0_j)^{\top}  \bigg\{{\bar{w}_j}^{*\top}(\bar{z}_{j};\beta^0_j)r_j(\bar{z}_j;\beta^0_j)-E_{P^0}\left[{\bar{w}_j}^{*\top}(\bar{Z}_{j};\beta^0_j)r_j(\bar{Z}_{j};\beta^0_j)\mid  \bar{z}_{j-1}, S \in \mathcal{A}_j\right] \bigg\}$.}
\end{sloppypar}

{
\begin{lem}\textit{(The canonical gradient of $\phi$ under $\mathcal{P}_{\mathcal{Q},\beta^0})$}.\label{lem:canonical_beta_known}
        Suppose that Conditions 1 and 2 hold, $\beta^0$ is known and $\psi$ is pathwise differentiable at $Q^0$ relative to $\mathcal{Q}$ with gradient $D_{Q^0}$. The canonical gradient of $\phi$ relative to $\mathcal{P}_{\mathcal{Q},\beta^0}$ is given by
        \begin{align}
        \tilde{D}_{P^0}(z,s;\beta^0) & = \sum_{j\in\mathcal{J}}\mathbbm{1}(s\in\mathcal{S}_j) \left\{\tilde{d}_j(\bar{z}_j;\beta^0_j) - E_{P^0}[\tilde{d}_j(\bar{Z}_j;\beta^0_j)\mid \bar{Z}_{j-1} = \bar{z}_{j-1},s]\right\}. \label{beta_known_canonical}
        \end{align}
\end{lem}}

\begin{sloppypar}
While Lemma~\ref{lem:canonical_beta_known} provides a preliminary solution for estimating $\psi(Q^0)$, it relies on the assumption that $\beta^0$ is known. However, in practice, the value of $\beta^0$ needs to be estimated. To this end, we propose to efficiently estimate $\beta^0$ based on its corresponding canonical gradient. 
 We denote $\dot{w}_{j,s}(\bar{z}_m,s'; \beta^0_{m,s'})$ be the derivative of $w_{m,s'}(\bar{z}_m; \beta_{m,s'})$ with respect to $\beta_{j,s}$ evaluated at $\beta^0_{j,s}$. For a fixed $s\in \mathcal{W}_j$ and $j \in \mathcal{J}$, the score function of $\beta^0_{j,s}$ when $Q^0$ is known is $\dot{\ell}_{\beta_{j,s}}(\bar{z}_{j},s';\beta^0_{j,s}):= \frac{\dot{w}_{j,s}(\bar{z}_{j},s';\beta^0_{j,s})}{w_{j,s}(\bar{z}_{j},s';\beta^0_{j,s})} -E_{P^0}\left[ \frac{\dot{w}_{j,s}(\bar{Z}_{j},S;\beta^0_{j,s})}{w_{j,s}(\bar{Z}_{j},S;\beta^0_{j,s})}\mid   \bar{z}_{j-1}, S=s'\right]$. We denote $a^*_j(\bar{z}_j;\beta^0_j):=a_j(\bar{z}_j;\beta^0_j)- E_{P^0}\left[a_j(\bar{z}_j;\beta^0_j)\mid  \bar{z}_{j-1}, S \in \mathcal{A}_j\right] + E_{P^0}\big[a_j(\bar{z}_j;\beta^0_j) {\bar{w}_j}^{*\top}(\bar{Z}_{j};\beta^0_j)\mid  \bar{z}_{j-1}, S \in \mathcal{A}_j\big] M^{-}_j(\bar{z}_{j-1};\beta^0_j)^{\top}  \bigg\{{\bar{w}_j}^{*\top}(\bar{z}_{j};\beta^0_j)r_j(\bar{z}_j;\beta^0_j)-E_{P^0}\left[{\bar{w}_j}^{*\top}(\bar{Z}_{j};\beta^0_j)r_j(\bar{Z}_{j};\beta^0_j)\mid  \bar{z}_{j-1}, S \in \mathcal{A}_j\right] \bigg\}$ with $a_j(\bar{z}_j;\beta^0_j):= \sum_{m \in \mathcal{S}_j} r_{j,m}(\bar{z}_j;\beta^0_{j,m})\dot{\ell}_{\beta_{j,s}}(\bar{z}_{j},m;\beta^0_{j,m})$.
\end{sloppypar}

\begin{lem}[\textit{The canonical gradient of $\beta^0$}]\label{lem: gradient_beta}
The efficient score function of $\beta^0_{j,s}$ under $\mathcal{P}_{\mathcal{Q},\mathcal{B}}$ for each $s \in \mathcal{W}_j$ and $j \in \mathcal{J}$ is , 
\begin{align}
    & \dot{\ell}^*_{\beta_{j,s}}(\bar{z}_{j},s';\beta^0_{j,s}) \nonumber\\
    & = \dot{\ell}_{\beta_{j,s}}(\bar{z}_{j},s';\beta^0_{j,s}) - \mathbbm{1}(s' \in \mathcal{S}_j)\left\{a^*_{j}(\bar{z}_j;\beta^0_{j})- E_{P^0}\left[ a^*_j(\bar{Z}_{j};\beta^0_{j})\mid  \bar{Z}_{j-1} = \bar{z}_{j-1}, S=s'\right]\right\}.
\end{align}
\sloppy Furthermore, the canonical gradient of $\beta^0$ is $D^{\beta}_{P^0}(z,s;\beta^0) = I_{\beta^0}^{-1} \dot{\ell}^*_{\beta}(z,s;\beta^0)$, where $I_{\beta^0}:=  E_{P^0}[\dot{\ell}_{\beta}^*(Z,S;\beta^0){\dot{\ell}^{*\top}_{\beta}}(Z,S;\beta^0)]$.
\end{lem}
The above results can be derived via a score function argument for finite-dimensional parameters under semiparametric models --- see Supplementary Appendix~\ref{sec:app:lem_beta}
for details.  The results of \cite{gilbert2000large} emerge as a special case of Lemma~\ref{lem: gradient_beta} when $Z$ is one-dimensional and $\mathcal{S}_1 = [k]$.

Now we make use of the preceding lemmas and derive the canonical gradient via a variant of an efficient score function argument for semiparametric models, where some adaptation is needed to account for the fact that the unknown parameter $Q^0$ is infinite-dimensional. To begin with, we define a new mapping $\gamma:\mathcal{B}\rightarrow \mathbbm{R}$ so that $\gamma(\beta) := E_{P_{\underline{Q}^0, \beta}}[ \tilde{D}_{P^0}(Z,S;\beta^0)]$, where $\underline{Q}^0:=\theta(P^0)$. Note that $\gamma$ depends on $P^0$ through both $D_{P^0}$ and the $\underline{Q}^0$ indexing $P_{\underline{Q}^0,\beta}$ in the expectation.

{
\begin{thm}[\textit{The canonical gradient of $\phi$}]\label{thm: canonical}
Suppose that
$\gamma$ is differentiable in $\beta$, and Conditions~\ref{cond:sufficient_alignment} and \ref{cond:overlaps} hold. The canonical gradient of $\phi$ relative to $\mathcal{P}_{\mathcal{Q},\mathcal{B}}$ is given by
\begin{equation}
    D^\mathrm{eff}_{P^0}(z,s;\beta^0) = \tilde{D}_{P^0}(z,s;\beta^0) - \{\nabla_{\beta}\gamma(\beta)\mid_{\beta = \beta^0} \}^{\top}D^{\beta}_{P^0}(z,s;\beta^0) \label{eq:them3},
\end{equation}
where $\nabla_{\beta}\gamma(\beta)\mid_{\beta = \beta^0}$ denotes the partial derivative of $\gamma(\beta)$ with respect to $\beta$ evaluated at $\beta = \beta^0$.
\end{thm}}
The first term in \eqref{eq:them3} is  the canonical gradient of $\phi$ under $\mathcal{P}_{\mathcal{Q},\beta^0}$. The second term is the transformation of the canonical gradient of $\beta^0$ under $\mathcal{P}_{\mathcal{Q}, \mathcal{B}}$, whose inclusion accounts for the fact that $\beta^0$ needs to be estimated. It can be shown that the coefficient $\nabla_{\beta}\gamma(\beta)\mid_{\beta = \beta^0}$ equals $E_{P^0}[\tilde{D}_{P^0}(Z,S;\beta^0) \dot{\ell}_{\beta}(Z,S;\beta^0) ]$, and hence this result is consistent with Theorem 2.1 and part C of Proposition 4.1 in \cite{klaassen2005efficient}.

{

    Figure~\ref{fig:two settings} illustrates tradeoffs between efficiency and robustness that arise when constructing an estimator with $D^{\mathrm{eff}}_{P^0}$. 
    As shown in the top panel, working with more flexible density ratio models results in greater robustness, since the implied statistical model is larger. Estimation in a larger model is no easier than in a smaller model, 
    and so the efficiency will also generally worsen. The bottom panel displays a different version of the tradeoff, fixing the density ratio model and instead adding more data sources. Adding these sources results in a smaller statistical model, worsening robustness and generally improving efficiency. The following corollary makes this statement precise.
\begin{figure}[htb]
    \centering
    \includegraphics[scale=0.2]{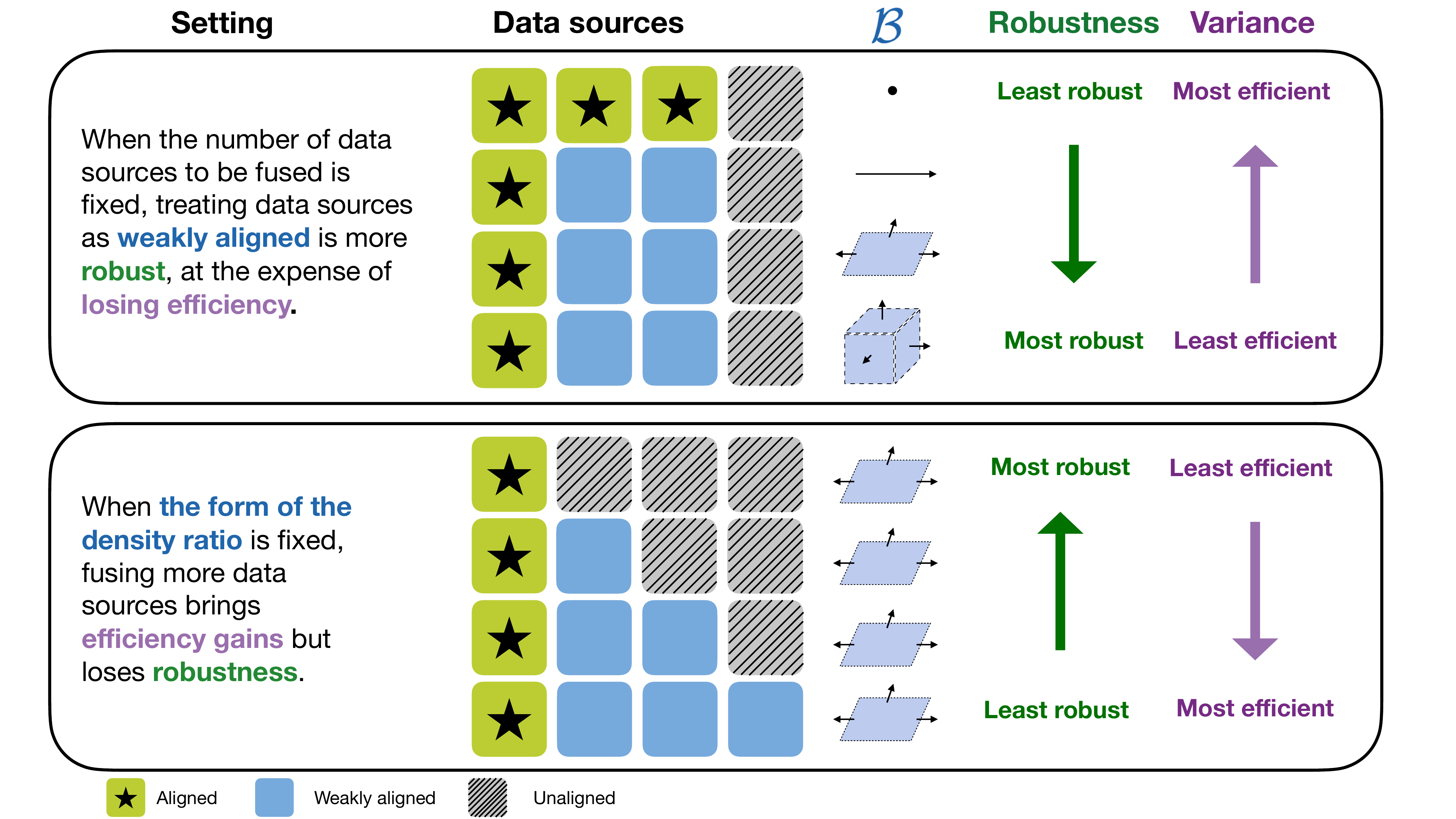}
    \caption{Efficiency-robustness tradeoff when performing data fusion. In general, using simpler weight functions and fusing more data sources offer greater potential efficiency gains at the expense of poorer robustness.}
    \label{fig:two settings}
\end{figure}}

\setcounter{corollary}{3}
{ \begin{corollary}\textit{(Leveraging weakly aligned data sources improves efficiency).}\label{corollary:rule}
            Let $\mathcal{T}^*_{\mathcal{B}}(P^0)$ denote the closed linear space spanned by $\dot{\ell}^*_{\beta}$. Denote $\mathcal{R}^\mathcal{A}_{P^0}(z,s):= D^\mathcal{A}_{P^0}(z,s) - \Tilde{D}_{P^0}(z,s)$. If  $\mathcal{R}^\mathcal{A}_{P^0} \notin \mathcal{T}^*_{\mathcal{B}}(P^0)$,  
            then
            $$\mathrm{var}(D^\mathrm{eff}_{P^0}) < \mathrm{var}(D^\mathcal{A}).$$
    \end{corollary}
Corollary~\ref{corollary:rule} states that, unless $D^\mathcal{A}_{P^0}$ can be decomposed as $\Tilde{D}_{P^0}(z,s) = \Pi_{P^0,\beta^0}\{ D^\mathcal{A}_{P^0}\mid \mathcal{T}(P^0,\mathcal{P}_{\mathcal{Q},\beta^0})\}(z,s)$ and a component that is in the space spanned by $\dot{\ell}^*_{\beta}$, incorporating weakly aligned sources is strictly better than using aligned sources only regardless of the magnitude of $\beta^0$. To illustrate the idea, we provide a visualization in Figure~\ref{fig:tangent_space} and three specific examples in Supplementary Appendix~\ref{sec:app:examples of gains}. One important implication of Corollary~\ref{corollary:rule} is that practitioners would benefit from selecting the smallest density ratio model that they know to contain the true weight function. This is important because, as $\mathcal{T}^*_{\mathcal{B}}(P^0)$ grows, it is more likely to contain $\mathcal{R}^\mathcal{A}_{P^0}$.  As a result, the cost of estimating unnecessary $\beta^0$ will cancel out the benefits as shown in Figure~\ref{fig:tangent_space}. }
    \begin{figure}[H]
        \centering
        \includegraphics[scale = 0.22]{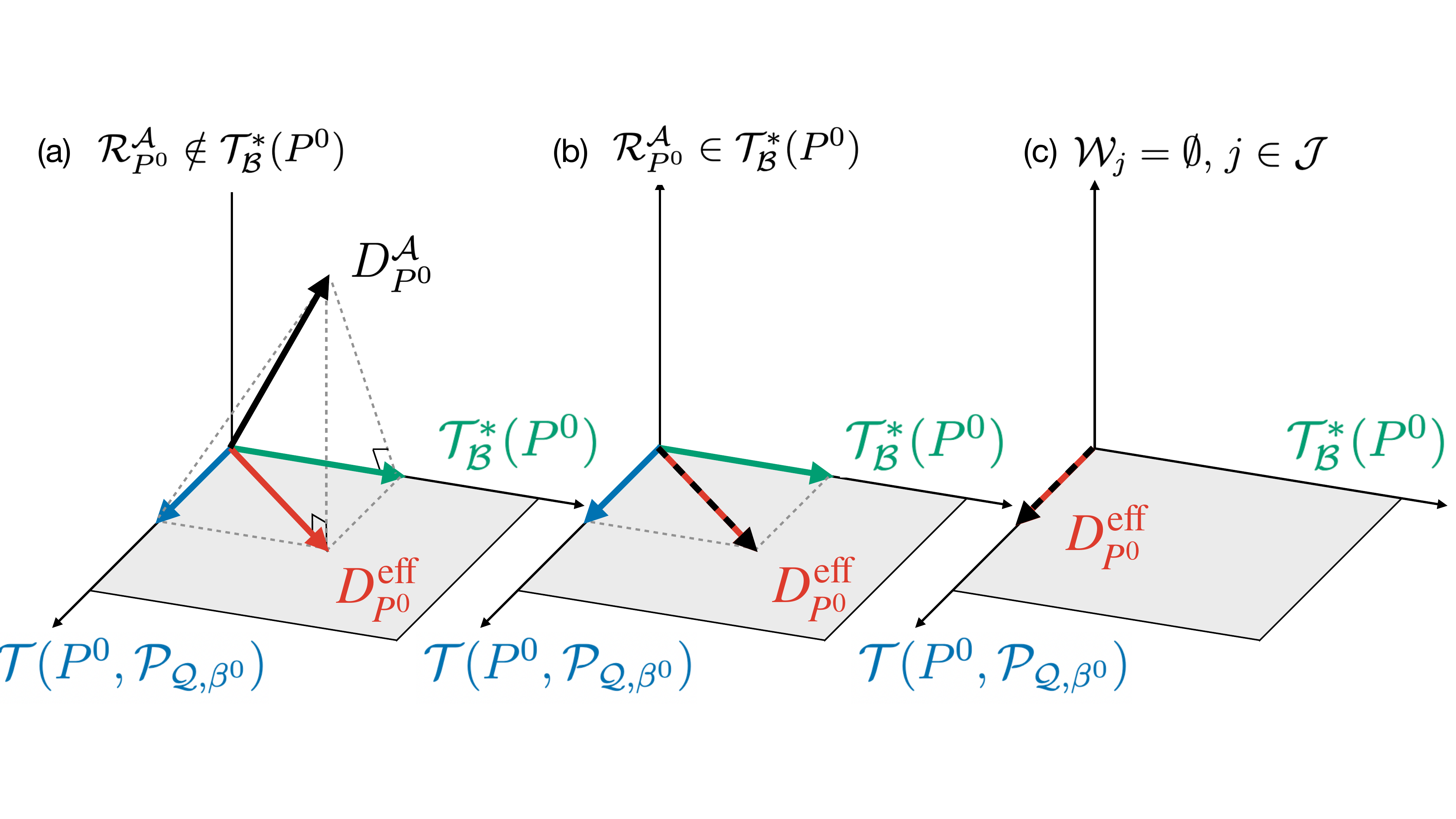}
        \caption{Visualization of efficiency gain. In general, incorporating weakly aligned sources renders additional efficiency gains compared to using aligned sources only (a). When the projection residue $\mathcal{R}^\mathcal{A}_{P^0}$ resides in $\mathcal{T}^*_{\mathcal{B}}(P^0)$, there is no gain from using weakly aligned sources (b). As a special case where there are only aligned sources available (c), $D^\mathrm{eff}_{P^0} = D^\mathcal{A}_{P^0}$.}
        \label{fig:tangent_space}
    \end{figure}

{In the extreme case where the weight function is nonparametric, including weakly aligned data sources will no yield any efficiency gain. This is not surprising given that, under an absolute continuity condition, there is always a density ratio between $Q^0$ and $P^0$. Since absolute continuity conditions do not restrict the tangent space, knowing that this density ratio exists -- without knowing anything about its form -- does not restrict the tangent space of the statistical model. Consequently, the semiparametric efficiency bound will remain unchanged. This highlights that the benefits of leveraging weakly aligned data sources are contingent on the additional structure imposed by the parametric assumptions in Condition~\ref{weak_alignment}. }
{When the weight function is misspecified, the resulting estimand $\phi$ becomes a projection-based parameter. We defer reader's attention to Supplementary Appendix~\ref{sec:app:sens} for detailed discussions on the potential sensitivity analysis.}

{Lastly, we note that even if the lower bound in Condition~\ref{cond:overlaps} fails, $D^{\mathrm{eff}}_{P^0}$ remains a valid gradient of $\phi(P^0)$ relative to $\mathcal{P}$, albeit potentially an inefficient one. This is due to the fact that when $\lambda^\dagger_{j}$ is upper bounded, $D^{\mathrm{eff}}_{P^0}$ is a valid gradient in a superset of the tangent space $\mathcal{T}(P^0,\mathcal{P})$.  }

\section{Simulation}
\label{s:simulation}

{We simulate $Z = (Z_1, Z_2, Z_3)$ from $k=4$ data sources with fixed sample sizes of $2000$ observations per data source. Conditional on $S$ and the covariate $Z_1$, the data are distributed as follows:  $Z_2 \mid Z_1, S \sim \textnormal{Bernoulli}(1/2)$, $Z_3 \mid Z_2, Z_1, S  \sim \textnormal{Beta}\{(2 - \epsilon \mathbbm{1}(S=2))(Z_1+Z_1Z_2) - \epsilon \mathbbm{1}(S=4) Z_1Z_2 , (2 - \epsilon \mathbbm{1}(S=3))Z_1 \}$ with $\epsilon = (0, 0.2,0.5,0.7)$, where $\epsilon$ measures the magnitude of misalignment between data sources and the target population. We term these four settings with increasing $\epsilon$ as: fully aligned , strongly aligned, moderately aligned, and poorly aligned, respectively. We consider two scenarios where (1) covariate $Z_1$ is identically distributed across data sources such that $Z_1\mid S \sim \textnormal{Uniform}(1,2)$, and (2) covariate shift is present such that $Z_1\mid S \sim \textnormal{Beta}(0.5(S-1)+4,5) + 1$. }

{Data source 1 perfectly aligns with the target distribution. All other sources are aligned in the distributions of $Z_2\mid Z_1$ and $Z_1$, and are weakly aligned in the distribution of $Z_3\mid Z_2, Z_1$ for appropriately parameterized weight functions. However, Condition~\ref{cond:overlaps} is violated to different degrees under the current setup. Specifically, $\lambda^\dagger_3(\bar{z}_3,s)$ is only upper bounded for all weakly aligned sources in the absence of covariate shift, and becomes unbounded in the presence of covariate shift. See Table~\ref{tab:quantiles}, Figures~\ref{fig:z_1 density} and \ref{fig:shifts_visual} in Supplementary Section~\ref{sec:app:simulation_continued} for a summary of the corresponding values of $\lambda^\dagger_3(\bar{z}_3,s)$ and visualizations of the data generating mechanisms.  Under parsimonious correctly specified weight functions,  each data source has a distinct form of $w^*_{3,s}$ with distinct values of $\beta^0_{3,s} \in \mathbbm{R}$. Specifically, $w_{3,2}(\bar{z}_3;\beta^0_{3,2}) = \exp ( (z_1 \log z_3,z_1 z_2\log z_3)^\top \beta^0_{3,2})$, $w_{3,3}(\bar{z}_3;\beta^0_{3,3}) = \exp [\beta^0_{3,3} z_1 \log (1 - z_3)]$ and $w_{3,4}(\bar{z}_3;\beta^0_{3,4}) = \exp [\beta^0_{3,4} z_1z_2 \log (z_3)]$. In addition, we also consider a ``complex weight parameterization", $w_{3,s} \in \mathcal{G}$ for all $s\in\{2,3,4\}$, where 
\begin{align*}
    \mathcal{G}&: = \Big\{(\bar{z}_3,s) \mapsto \exp \{(\log z_3, z_1 \log z_3, z_2 \log z_3 ,z_1z_2 \log  z_3 ,\log (1-z_3),\\
    & \hspace{8em} z_1 \log (1-z_3),z_2 \log (1-z_3), z_1z_2 \log  (1-z_3) )^\top \beta^0_{3,s}\}: \beta^0_{3,s} \in \mathbbm{R}^8\Big\}.
\end{align*}
Even though only one or two out of the eight coefficients is non-zero for each data source, we estimate a selected subset of $\beta^0_{3,s}$ that is necessarily bigger than the correct parimonious model for each source $s$ across a range of subset size. This mimics a scenario where there is limited knowledge on the possible form of the density ratio model, so an overparametrized one is used.}

{We aim to estimate the average treatment effect defined as $\psi(Q^0) =E_{Q^0}[ E_{Q^0}(Z_3 \mid Z_2 = 1, Z_1)] - E_{Q^0}[ E_{Q^0}(Z_3 \mid Z_2 = 0, Z_1)]$ under positivity, consistency, and no unmeasured confounding assumptions \citep{rubin1980randomization,rosenbaum1983central}.} We construct the following one-step estimators of $\psi(Q^0)$: (1) a target-only estimator with $\mathcal{S}_1 = \mathcal{S}_2 = \mathcal{S}_3 = \{1\}$ and $\mathcal{W}_3=\emptyset$, (2) a na\"ive fusion estimator that assumes  $\mathcal{A}_1 = \mathcal{A}_2 = \mathcal{A}_3 = \{1,2,3,4\}$ and $\mathcal{W}_3=\emptyset$, 
(3) the efficient fusion estimator constructed via $D^\mathrm{eff}_{P^0}$ with $\mathcal{S}_1 = \mathcal{S}_2 = \mathcal{S}_3 = \{1,2,3,4\}$ with $\mathcal{W}_3 = \{2,3,4\}$, and (4) three estimators constructed via $D^\mathrm{eff}_{P^0}$ using the same fusion sets as in (3) but using overparametrized weight functions. We also examined the aligned fusion scenario using methods proposed in \cite{li2021efficient}, where  $\mathcal{S}_1=\mathcal{S}_2=\{1,2,3,4\}$ with $\mathcal{W}_1=\mathcal{W}_2=\emptyset$ and $\mathcal{S}_3=\{1\}$ with $\mathcal{W}_3=\{2,3,4\}$. Due to the data generating mechanism in our simulation setting, the resulting one-step estimators had nearly identical performance to those under (1); thus, only estimator from (1) is reported, and any observed improvements on that estimator also represent improvements over the one proposed in \cite{li2021efficient}.

Initial estimates of $\beta^0$ were obtained via matching moments. Then one-step estimators of $\beta^0$ were constructed using these initial estimates and the canonical gradient of $\beta^0$. The normalizing functions of the weight functions were estimated via kernel regression under default settings in the \texttt{np} R package \citep{np} while propensity scores were estimated via main terms linear-logistic regression. While it may seem intuitive to model the density ratios $q_j^0(\cdot \mid \bar{z}_{j-1})/p_j^0(\cdot \mid \bar{z}_{j-1},s)$ nonparametrically, it is important to note that the initial estimate $\hat{P}$ of $P^0$ must belong to $\mathcal{P}$. We ensured this by estimating this quantity by using the corresponding parametric form imposed by $w_{j,s}$ and estimates of $\beta^0$. For each simulation study presented in this work, 1000 Monte Carlo replications were conducted.

As shown in Table~\ref{tab:main_results}, efficient and overparametrized fusion estimators are consistent and always have smaller variances than the target only estimators. By contrast, except for the fully aligned setting, the na\"ive fusion estimator are biased as it fails to calibrate the misalignment between data sources. When data sources are fully aligned, the na\"ive fusion estimator achieves the efficiency bound \citep{li2021efficient} while the efficient fusion estimator achieves the efficiency bound in a larger statistical model, thereby yielding more robust inferences. Examining across different level of alignments, fusion estimators achieve more substantial efficiency gain when there is better alignment in the conditional outcome distribution. 
Similar trends are observed when $Z_1$ is shifted among data sources. 
{Overall, the proposed estimators are fairly robust to the violation of the strong overlap condition in settings we examined, and the efficiency gain is smaller when overlap is poor. 
Across all scenarios, although estimating unnecessarily high-dimensional $\beta^0$ will reduce the overall efficiency gain compared to the one using a correct parsimonious model, they still render efficiency gains compared to the target-only estimator. We have also performed sensitivity analyses to examine the impact of model misspecification of weight functions on the resulting estimators and direct reader's attention to Supplementary Appendix~\ref{sec:app:sens} for detailed results. }

{\begin{longtable}[H]{lrrrrrr}
\caption{Bias, variance and coverage of estimators of the average treatment effect under various levels of alignment in $Z_3$. Bias$^2$ and variance are scaled up by $10^5$ for clarity. For overparametrized fusion estimators, $^{+}$ indicates the dimension of redundant parameters in $\beta^0$.}
\label{tab:main_results} \\
\toprule
       & \multicolumn{3}{c}{No shifts in $Z_1$} & \multicolumn{3}{c}{Shifts in $Z_1$}  \\
    \cmidrule(l){2-4} \cmidrule(l){5-7}
    &  Bias$^2$ & Var & Coverage &  Bias$^2$ & Var & Coverage \\
    \midrule
    \endfirsthead
\caption[]{Bias, variance and coverage of estimators of the average treatment effect under various levels of alignment in $Z_3$ (continued).} \\
\toprule
       & \multicolumn{3}{c}{No shifts in $Z_1$} & \multicolumn{3}{c}{Shifts in $Z_1$}  \\
    \cmidrule(l){2-4} \cmidrule(l){5-7}
    &  Bias$^2$ & Var & Coverage &  Bias$^2$ & Var & Coverage \\
    \midrule
    \endhead
\bottomrule
\endfoot
    \textbf{Fully aligned }\\
    \hspace{1em} Target only &0.02 & 5.76 & 0.97 & 0.01 & 5.53 & 0.97\\
    \hspace{1em} Na\"ive fusion  & 0.00 & 1.50 & 0.96 & 0.01 & 1.56 & 0.96\\
    \hspace{1em} Efficient fusion  &0.00 & 2.26 & 0.95 & 0.00 & 2.44 & 0.96\\
    \hspace{1em} Overparametrized$^{+1}$ fusion  &0.00 & 2.53 & 0.95 & 0.00 & 2.74 & 0.96\\
    \hspace{1em} Overparametrized$^{+2}$ fusion  &0.01 & 2.74 & 0.95 & 0.01 & 3.06 & 0.96\\
    \hspace{1em} Overparametrized$^{+5}$ fusion  &0.00 & 3.59 & 0.94 & 0.02 & 3.71 & 0.95\\
    \textbf{Strongly Aligned }\\
    \hspace{1em} Target only &0.02 & 5.76 & 0.97 & 0.01 & 5.53 & 0.97\\
    \hspace{1em} Na\"ive fusion  & 0.97 & 1.65 & 0.88 & 0.92 & 1.61 & 0.89\\
    \hspace{1em} Efficient fusion  &0.00 & 2.53 & 0.94 & 0.00 & 2.47 & 0.95\\
    \hspace{1em} Overparametrized$^{+1}$ fusion  &0.00 & 2.75 & 0.94 & 0.00 & 2.77 & 0.96\\
    \hspace{1em} Overparametrized$^{+2}$ fusion  &0.01 & 3.09 & 0.94 & 0.01 & 3.09 & 0.95\\
    \hspace{1em} Overparametrized$^{+5}$ fusion  &0.00 & 3.86 & 0.93 & 0.02 & 3.62 & 0.95\\
    \textbf{Moderately Aligned }\\
    \hspace{1em} Target only &0.02 & 5.76 & 0.97 & 0.01 & 5.53 & 0.97\\
    \hspace{1em} Na\"ive fusion  & 8.29 & 1.73 & 0.41 & 7.86 & 1.81 & 0.48\\
    \hspace{1em} Efficient fusion  &0.00 & 2.55 & 0.94 & 0.04 & 2.53 & 0.95\\
    \hspace{1em} Overparametrized$^{+1}$ fusion  &0.00 & 2.78 & 0.95 & 0.05 & 2.81 & 0.95\\
    \hspace{1em} Overparametrized$^{+2}$ fusion  &0.00 & 3.14 & 0.95 & 0.09 & 3.14 & 0.96\\
    \hspace{1em} Overparametrized$^{+5}$ fusion  &0.07 & 4.31 & 0.93 & 0.12 & 3.71 & 0.96\\
    \textbf{Poorly Aligned }\\
    \hspace{1em} Target only &0.02 & 5.76 & 0.97 & 0.01 & 5.53 & 0.97\\
    \hspace{1em} Na\"ive fusion & 20.6 & 1.88 & 0.09 & 18.9 & 1.97 & 0.16\\
    \hspace{1em} Efficient fusion  &0.00 & 2.66 & 0.94 & 0.03 & 2.74 & 0.95\\
    \hspace{1em} Overparametrized$^{+1}$ fusion  &0.06 & 2.97 & 0.93 & 0.08 & 3.17 & 0.95\\
    \hspace{1em} Overparametrized$^{+2}$ fusion  &0.04 & 3.59 & 0.95 & 0.16 & 3.58 & 0.94\\
    \hspace{1em} Overparametrized$^{+5}$ fusion  &0.87 & 5.21 & 0.91 & 0.19 & 4.24 & 0.95\\
\end{longtable}}

\section{Data illustration}
\label{s:data application}
We illustrate our approach using data from two harmonized phase IIb trials that evaluated the efficacy of the broadly neutralizing antibody (bnAb) VRC01 to prevent new HIV-1 diagnosis in different populations. One of these studies, HVTN 703, enrolled women in sub-Saharan Africa, while the other, HVTN 704, enrolled men and transgender people who have sex with men in North America, South America, and Switzerland. Results from both trials supported that the bnAb did not prevent overall HIV-1 acquisition, but did prevent acquisition of HIV-1 strains that were sensitive to neutralization by VRC01 \citep{corey2021two}, providing proof-of-concept evidence supporting development of bnAb cocktails that provide neutralization coverage of more HIV-1 strains. Using data from both trials, \cite{gilbert2022neutralization} showed that a biomarker that quantifies the neutralization potency of antibodies in an individual's serum against a given HIV-1 isolate, the predicted 80\% neutralizing antibody titer (PT80), is a promising potential surrogate endpoint for new HIV-1 diagnosis with application to future HIV-1 bnAb studies. This putative surrogate endpoint can be used for ranking candidate bnAb cocktails by their predicted HIV-1 prevention efficacy, aiding down-selection of the most promising cocktails for study in future prevention efficacy trials. In this article, for VRC01 recipients (pooling over the low-dose and high-dose arms) who acquired the new HIV-1 diagnosis primary endpoint, we studied the association of the PT80 biomarker with amino acid sequence features of the HIV-1 virus.  The viral sequence was measured from a blood sample drawn at the time of HIV-1 diagnosis, and the PT80 was calculated as the concentration of VRC01 in this same blood sample divided by the measured level of resistance of the virus to neutralization by VRC01.
Typically about 200 HIV-1 sequences were measured in an individual's blood sample with the Pacbio sequencing technology, and the sequence that had greatest sequence-based predicted resistance \citep{williamson2021super} to neutralization by 
VRC01 was used. Studying the associations of PT80 marker with sequence features of HIV-1 can provide insights into ``viral genetic signatures" that may mark whether VRC01 can protect against a given viral strain. 

The data consist of 98 VRC01 new HIV-1 diagnosis endpoint cases with PT80 measured ($n_{703} = 43$, $n_{704}=55$). 
There are 246 residue indicator variables for the 50 amino acid positions of the HIV-1 Envelope protein that comprise CD4 binding sites and the VRC01 binding footprint that are most relevant for potential protection by VRC01. Only the subset of these residue indicators with enough variation to reasonably study an association, quantified by requiring at least 10 participants with observed residue at the position and at least 10 participants without the observed residue at the position, are analyzed. This process yielded 22 amino acid features of interest for HVTN 703 and 30 for HVTN 704. The data also includes a single genetic distance variable defined based on all 50 amino acid positions mentioned previously: the physicochemical-weighted Hamming distance between the observed sequence and the HIV-1 sequence  
in the CATNAP database \citep{yoon2015catnap} with least amount of neutralization resistance to VRC01, calculated matched to a given participant's geographic region (HVTN 703 South Africa, HVTN 703 outside South Africa, HVTN 704 U.S. or Switzerland, HVTN 704 South America) and the subtype of their acquired virus (mostly subtype C for HVTN 703 and subtype B for HVTN 704). Our objective is to study the coefficients corresponding to the selected residue indicators and the genetic distance in univariate linear working models with outcome $\log 10$ PT80. The definition of these parameters, models and the forms of the gradients can be found in Supplementary Appendix~\ref{sec:app:analysis_table_continued}.

\begin{figure}[htb]
\centering\includegraphics[scale = 0.7]{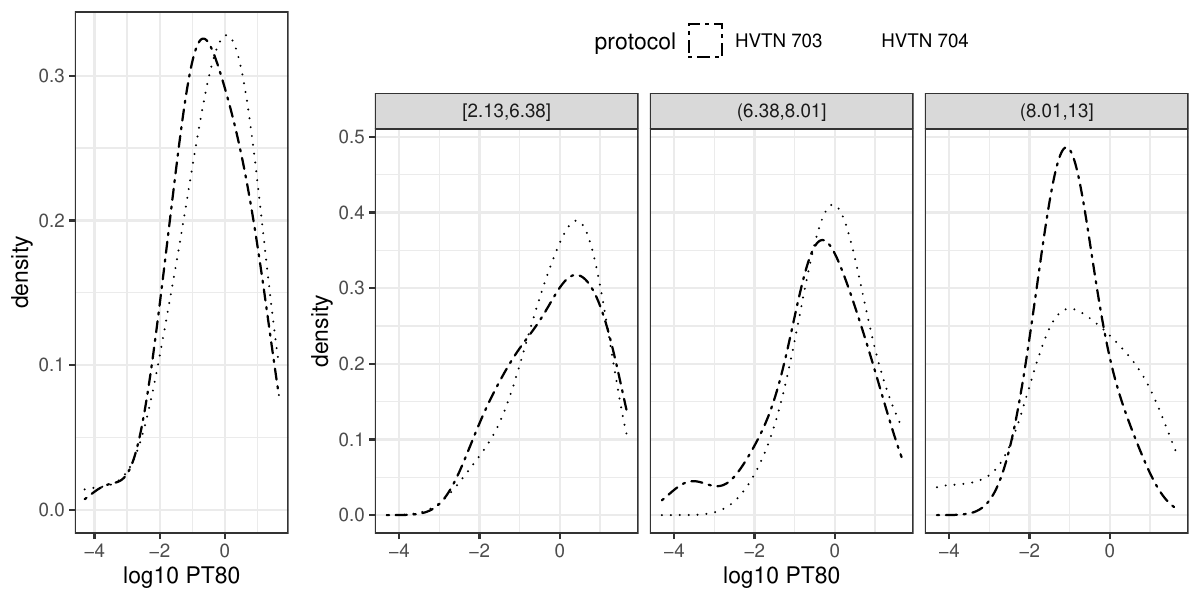}
\caption{Density plots of the $\log_{10}$ PT80 marker using a kernel density estimator with a bandwidth selected by cross-validation separately for HVTN 703 and HVTN 704. The left figure compares the density of $\log_{10}$ PT80 in all groups. The right panels compare density plots of $\log_{10}$ PT80 stratified by the genetic distance in tertiles. From left to right, we observe a monotone shift in the relative positions of the means and also in the spread of the curves. }
\label{Fig:PT80dens}
\end{figure}

Data fusion is especially appealing in this problem due to the limited HIV-1 diagnosis endpoints in each study \citep{corey2021two}. There is a need to use a fusion method that only requires weak alignment because participants in HVTN 703 and HVTN 704 had different sexes assigned at birth and HIV-1 acquisition routes, and were exposed to different HIV-1 circulating subtypes. As a result, the common distribution condition considered in previous data fusion literature likely fails \citep{li2021efficient}. The density plot of the $\log_{10}$ PT80 marker also suggests the conditional distribution indeed differs between the two studies (Figure~\ref{Fig:PT80dens}). We assume the density ratio between the conditional density of biomarkers takes the following form: 
\begin{align*}
    \frac{p_{703}(Y \mid W,X)}{p_{704}(Y \mid W,X)} = \frac{\exp\{\beta_0 Y + \beta_1 XY + \beta_2 Y^2\}}{E_{704}\left[\exp\{\beta_0 Y + \beta_1 XY +\beta_2 Y^2\} \mid  X \right]},
\end{align*}
where $Y = \log_{10} \textnormal{PT80}$, $X$ denotes genetic distance, and $W$ are the HIV-1 amino acid features. If $Y$ were to follow a normal distribution in both trials conditional on $W$ and $X$,  then the form of the above density ratio allows shifts in the mean of $Y$ stratified by genetic distance $X$, while also accommodating a different variance of $Y$ across the two studies due to the inclusion of $\beta_2$. We separately treated each of the study populations from the two studies as the target population and compared the estimation results generated by using one single dataset vs. using both datasets. 

We estimated density ratios via kernel regression using the Nadaraya-Watson estimator, adaptive nearest neighbors bandwidths, and expected Kullback-Leibler cross-validation for bandwidth selection in the \texttt{np} R package \citep{np}. Nuisance parameters such as conditional expectations were estimated using SuperLearner \citep{van2007super,polley2010super} with a library of a generalized linear model, LASSO regression, generalized additive model, random forests, and xgboost.

Results for a selected list of the residue indicators based on alphabetical order and genetic distance are presented in Table~\ref{beta_PT80}. Additional results for the remaining residue indicators can be found in Table~\ref{beta_PT80_appendix} in the supplement. Augmenting HVTN 703 efficiently leads to a {0.5\% to 44\% } reduction in confidence interval width, corresponding to a {1\% to 68\%} reduction in the required sample size to achieve the same level of precision. Similarly, augmenting HVTN 704 yields a {13\% to 52\%} reduction in confidence interval width, resulting in a {24\% to 77\%} reduction in sample size. Notably, the residue indicator Glycine at position 471 (G471) demonstrates significant associations with relatively small standard errors. The 95\% confidence interval for the G471 coefficient is {[0.16, 1.23]} for augmenting HVTN 703. This feature has also been identified as having high variable importance in predicting neutralization sensitivity \citep{hake2017prediction,magaret2019prediction}. By reducing variability in the estimators of coefficients in working models, these findings provide a clearer understanding of the relative predictive importance of HIV-1 Envelope amino acid sequence features for PT80. These results are particularly valuable for feature selection and the development of models aimed at predicting neutralization readouts, as amino-acid sequence sieve analysis often faces the challenge of analyzing a large number of sequence features \citep{magaret2019prediction}.

\begin{table}[htb]
\begin{center}
\caption{{Estimated coefficient using the HVTN 703 and HVTN 704 data. Results are presented as estimates (standard errors).}\label{beta_PT80}}%
\begin{adjustbox}{width=0.8\textwidth} 
{
\begin{tabular}{llrrrr}
\toprule
   & & \multicolumn{2}{c}{Augmenting HVTN 703} 
& \multicolumn{2}{c}{Augmenting HVTN 704} \\
\cmidrule(l){3-4} \cmidrule(l){5-6}
  & &  703 only & Efficient Fusion  &  704 only  & Efficient Fusion  \\
Residue & Site &  (N=43) &  (N=98) &  (N=55) & (N=98) \\
\midrule
A & 281 & 0.66 (0.31) & 0.76 (0.21) & 0.61 (0.31) & 0.51 (0.19)\\
D & 279 & -0.22 (0.34) & -0.36 (0.24) & -0.08 (0.31) & -0.15 (0.17)\\
D & 474 & -0.23 (0.32) & 0.21 (0.23) & 0.56 (0.30) & 0.43 (0.19)\\
E & 429 & -0.07 (0.32) & -0.10 (0.19) & 0.09 (0.32) & 0.16 (0.20)\\
G & 429 & 0.20 (0.37) & 0.27 (0.21) &  & \\
G & 471 & 0.80 (0.28) & 0.69 (0.27) & 0.39 (0.42) & 0.81 (0.23)\\
gap & 463 & -0.21 (0.31) & -0.01 (0.23) & -0.06 (0.32) & -0.28 (0.19)\\
Genetic & distance  & -0.32 (0.14) &  -0.42 (0.14)&-0.43 (0.16) & -0.47 (0.08)\\
\bottomrule
\end{tabular}}
\end{adjustbox}
\end{center}
\end{table}

\section{Discussion}
\label{s:discussion}

Our method relies on a key assumption: that the density ratio model, as defined in Condition 1c, is correctly specified. {While Condition 1c relaxes the exchangeability condition, misspecification is still possible. To this end, we propose the following strategies when specifying weight functions. To begin with, aligned and weakly aligned sets should reflect prior knowledge that practitioners can make educated guesses on based on domain knowledge and historical data. When dealing with clinical data, the parameter $\beta^0$ often corresponds to shifts in log odds, or relative risks in stratified groups that one can infer based on previous findings on prognostic and therapeutic biomarkers, as well as on domain knowledge on the underlying populations. Sometimes, data sources contain different sets of covariates measurements, or there is minimal overlap in key effect modifiers, as in our HVTN 703 and 704 data illustration. In such cases, few sources will be fully aligned with the target data source, but some may be weakly aligned. In practice, goodness-of-fit tests could guide the selection of density ratio models \citep{gilbert2004goodness}, though may raise concerns about making inference after model selection. Lastly to avoid misspecification, the weight functions can always be over-parameterized. Our findings in Section~\ref{s:methods} show over-parameterization can result in efficiency loss compared to a correctly specified parsimonious model; however, the efficiency of the proposed estimator will never be worse than the target-only estimator, providing a safeguard even when there is a lack of prior domain knowledge. Alternatively, practitioners could perform sensitivity analyses, as detailed in Supplementary Appendix~\ref{sec:app:sens}.}

{It is also important to point out that there is a trade-off between the plausibility of the overlap condition and the sufficient alignment condition. While Condition~\ref{cond:sufficient_alignment} is more likely to hold when conditioning on a sufficiently rich set of variables, Condition~\ref{cond:overlaps} becomes less plausible with more variables \citep{d2021overlap}. While the proposed estimators are fairly robust to the violation of Condition~\ref{cond:overlaps} in the simulation settings that we explored, in practice there can be other scenarios where stabilizing methods in estimating density ratios are needed. One approach is to minimize entropy while matching on moment conditions to reduce estimation variability \cite{lee2023improving}. }

Our focus has been on nonparametric $\mathcal{Q}$, where we aim to leverage data from both aligned and weakly aligned sources. We attained lower asymptotic variance than estimators from \cite{li2021efficient}, which are efficient among all estimators that rely solely on aligned data. 
In future work, it would be interesting to explore what further gains are available when the model $\mathcal{Q}$ is semiparametric, rather than nonparametric. {In a parallel direction, we are interested in incorporating adaptive methods in the proposed data fusion framework. \cite{chen2021minimax} proposed an adaptive anchored thresholding estimator for the average treatment effect that balanced the bias-variance trade-off.  \cite{yang2023elastic} incorporated a preliminary test statistic to assess the extent of exchangeability and proposed a test-based elastic integration method that learns the amount of information to borrow from other sources adaptively.  In both of these works, the adaptive use of data sources renders the proposed estimator irregular. In the light of work by \cite{van2023adaptive, van2024adaptive}, developing estimators for an oracle projection parameter that provides regular and locally uniformly valid inference in the local submodel is an interesting future direction worth exploring.}

\section*{Acknowledgements}
\label{sec:acknowledgement}
We thank the participants and investigators of HVTN 703 and HVTN 704. Research reported in this publication was supported by the NIH under award number DP2-LM013340 and R37AI054165, and the National Institute Of Allergy And Infectious Diseases of the National Institutes of Health under the U.S. Public Health Service Grant AI068635.  The content is solely the responsibility of the authors and does not necessarily represent the official views of the National Institutes of Health. We thank Xiudi Li for the helpful discussions.

\section*{Appendix}

\begin{appendices}

\section{{Examples of aligned, weakly aligned and unaligned data sources}}
\label{sec:app:example}
\textbf{Toy example}. Suppose $Z_1 \sim N(\mu_0,\sigma^2_0)$ in the target population. Suppose we observe the following data sources:
\begin{align*}
    & \textnormal{Data Source 1 :  } Z_1\mid S=1 \sim N(\mu_0,\sigma^2_0)\\
    & \textnormal{Data Source 2 :  } Z_1\mid S=2 \sim N(\mu_2,\sigma^2_0)\\
    & \textnormal{Data Source 3 :  } Z_1\mid S=3 \sim N(\mu_0,\sigma^2_3)\\
    & \textnormal{Data Source 4 :  } Z_1\mid S=4 \sim N(\mu_4,\sigma^2_4)\\
    & \textnormal{Data Source 5 :  } Z_1\mid S=5 \sim \mathrm{Cauchy}(\alpha,\gamma),
\end{align*}
where $\mu_2, \mu_4 \neq \mu_0$, $\sigma^2_3,\sigma^2_4 \neq \sigma^2_0$. Data source $S=1$ is a perfectly aligned source, and we assume that this is known to the analyst. If the same functional form $w_1(z_1;\beta^0_1) = \exp(\beta^0_1z_1)$ is assumed for all weakly aligned weight functions, then only data source $S=2$ is weakly aligned. By contrast, if the weight function takes the more flexible form of $w_1(z_1;\beta^0_1) = \exp((z_1, z_1^2)^\top \beta^0_1)$, then data sources $S= \{2,3,4\}$ are all weakly aligned, and source $S=5$ is unaligned. \\~\\
\textbf{Efficacy of COVID-19 BNT162b2 vaccine in Children.}\\ 
Large-scale phase II/III trials assessing the efficacy of the BNT162b2 vaccine against COVID-19 have only been conducted among persons 16 years of age or older \citep{polack2020safety,moreira2022safety}, leaving a critical gap in understanding its efficacy in children. Addressing this gap is particularly important given children's susceptibility to severe complications such as multisystem inflammatory syndrome (MIS-C) \citep{tian2022safety}, and the significant impact of COVID on their schooling and social interactions. To date, there is only one phase II/III trial (NCT04816643) that evaluated the efficacy of the BNT162b2 vaccine among children 5 to 11 years of age. In that trial, 2285 underwent randomization across sites in the United states, Spain, Finland and Poland. The observed vaccine efficacy was 90.7\% (95\% CI,  67.7 to 98.3) with 3 cases among BNT162b2 recipients and 16 among placebo recipients \citep{walter2022evaluation}. Due to the limited sample size, the study was not powered to detect potential rare side effects.\\~\\
There is another trial, NCT04368728, which evaluated the same vaccine regimen but among children age from 12 to 15 years in the United States \citep{frenck2021safety}. Augmenting the previous trial with this study would bring substantial efficiency gain in the final estimated vaccine efficacy among children 5 to 11 years of age, due to the similarity between the two study populations and moderate sample size (N=2260). However, there are some key differences between the two trials: (1) although participants in both studies are children,
they have non-overlapping age groups (5-11 vs. 12-15) where age has been shown
to be an important marker for immunogenicity, and (2) the two studies were conducted
in non-overlapping geographical locations and in different timelines, where the circulating
strains might be different. As a result, the adjustment set of covariates excluding age, a non-overlapping yet important prognostic marker, may not fully justify the variability in infection risk between the two populations. Although there is a pressing need to data fusion in this example, practitioners do not have enough data to fuse with due to the implausibility of the required exchangeability condition. \\~\\
Assuming weak alignment remains the only viable approach in this scenario. A reasonable weight function can take the form $$w_3(\bar{z}_3;\beta^0) = \exp((Z_1Z_3,Z_2Z_3,Z_3)^\top \beta^0_1),$$ where $Z_3$ is the indicator of COVID-19 infection, $Z_2$ is the indicator of receiving the vaccine and $Z_1$  are baseline covariates including demographic features and clinical measurements of (1) perinatal factors such as gestational age and birth weight, (2) pre-existing coronavirus antibodies and microbiotas, and (3) nutritional factors such as body mass index and enteropathy. Studies mentioned by \cite{zimmermann2019factors} have shown that these are important factors that influence the immune response to vaccination, thus becomes potential effect modifiers of vaccine efficacy between different age groups. Specifically, \cite{ng2020preexisting} have shown that preexisting antibodies to other coronaviruses were present in varying degrees across different age groups. These antibodies could cross-react with SARS-CoV-2, leading to potential different vaccine efficacy between two groups of children.  Implied by the form of the weight function, we allow a shift in log odds of COVID-19 infection between the two populations across each of the covariates $Z_1$ and treatment $Z_2$ strata, as well as an overall shift marginally as reflected by the $Z_3$ term.\\~\\
While it is tempting to use the larger trial NCT04368728 (N=43548) which evaluated the same vaccine regimen among persons from 16 years and older, that population is possibly unaligned due to the drastic distinct population characteristics and age-related variations in immune systems \citep{ranjeva2019age,wagner2018age}. To name a few, adults have more pre-existing conditions, and differ from children in basic demographic and clinical features. As a result, researchers have a minimal set of overlapping baseline covariates to adjust for and so even the weak alignment condition will likely fail. \\~\\
\textbf{Efficacy of the monoclonal broadly neutralizing antibody VRC01 for preventing HIV-1 acquisition}\\ 
The HIV Vaccine Trials Network (HVTN) has conducted extensive research on the prevention of HIV-1 acquisition through administration of monoclonal broadly neutralizing antibodies (bnAbs) that attack one of three different epitope regions of HIV’s outer Envelope protein \citep{williamson2023application}: (1) the CD4 binding site, (2) the V2 loop, and (3) the V3 loop. The CD4 binding site bnAbs, such as VRC01 and VRC07, represent one class of antibodies that interfere with the interaction between HIV-1 Envelope and the CD4 receptor, while bnAbs targeting V2 (e.g., PGDM1400 and PGDM1400LS) and V3 (e.g., PGT121) exhibit distinct binding profiles and mechanisms of neutralization. \\~\\
Table 1, Figure S1 and Table S1 in \cite{williamson2023application} summarize early-phase HVTN clinical trials that studied bnAbs in each of the three antibody classes.  If PT80 and sequence data from the small number of participants that acquired HIV-1 in these early-phase trials were available, then they could be potentially fused in our data illustration example. While weak alignment may be feasible within each class of bnAbs, such as those targeting the CD4 binding site, the structural biology is different enough across the three different classes of bnAbs that weak alignment may be considered too implausible. Specifically, protocols HVTN 104, HVTN 116, HVTN 127 and HVTN 129 are potentially weakly aligned sources for studying the efficacy of VRC01 since the bnAbs studied in these phase 1 trials all target the CD4 binding site. In contrast, HVTN 136 and HVTN 140 are examples of unaligned sources since they studied bnAbs that target V3 and V2, respectively.

\section{Proof of Lemma~\ref{lem:canonical_beta_known}}
\label{sec:app:lem_known}
{We first show that $D_{P^0}$ in \eqref{eq:gradient_known} is a valid gradient of $\phi$ in $\mathcal{P}_{\mathcal{Q},\beta^0}$. Then we prove  Lemma~\ref{lem:canonical_beta_known} by showing that $\tilde{D}$ is the projection of $D_{P^0}$ onto $\mathcal{T}(P^0,\mathcal{P}_{\mathcal{Q},\beta^0})$.} To begin with, we first characterize the tangent space. Let $\mathcal{T}(P^0,\mathcal{P}_{\mathcal{Q},\mathcal{B}})$ denote the tangent set of model $\mathcal{P}_{\mathcal{Q},\mathcal{B}}$ at $P^0$. We assume that $\mathcal{T}(Q^0,\mathcal{Q})$ is a closed linear subspace of $L_0^2(Q^0)$, and it can be verified that $\mathcal{T}(P^0,\mathcal{P}_{\mathcal{Q},\beta^0})$, which is the tangent set of a model that is nonparametric model up to the restriction imposed by the data fusion alignment condition and with known value of $\beta^0$, is itself a closed linear subspace of $L_0^2(P^0)$. The same is true for $\mathcal{T}(P^0,\mathcal{P}_{\mathcal{Q},\mathcal{B}})$, which is the tangent set of a model that is nonparametric model up to the restriction imposed by the data fusion alignment condition with $\beta^0$ unknown. Therefore, we also refer to $\mathcal{T}(P^0,\mathcal{P}_{\mathcal{Q},\beta^0})$ and $\mathcal{T}(P^0,\mathcal{P}_{\mathcal{Q},\mathcal{B}})$ as the tangent space of $\mathcal{P}_{\mathcal{Q},\beta^0}$ and $\mathcal{P}_{\mathcal{Q},\mathcal{B}}$ at $P^0$, respectively. Let $L_0^2(Q^0_j)$ denote the subspace of $L_0^2(Q^0)$ consisting of all functions $f:\prod_{j=1}^j \mathcal{Z}_i \mapsto \mathbbm{R}$ such that $E_{Q^0}[f(\bar{Z}_j) \mid \bar{Z}_{j-1}] =0$ with $Q^0$-probability one. Similarly let $L_0^2(P^0_j)$ denote the subspace of $L_0^2(P^0)$ consisting of all functions $g:(\prod_{i=1}^j \mathcal{Z}_i) \times \mathcal{S} \mapsto \mathbbm{R}$ such that $E_{P^0}[g(\bar{Z}_j,S)\mid \bar{Z}_{j-1},S] = 0$ with $P^0$-probability one. For each $j \in [d]$, let $\mathcal{T}(Q^0,\mathcal{Q}_j)$ be the subspace of $L_0^2(Q^0_j)$ that consists of all $f_j$ that arise as scores of univariate submodels $\{Q^{(\epsilon)} :\epsilon \in [0,\delta )\}$ for which $Q^{(\epsilon)} = Q^0$ when $\epsilon=0$. Similarly, we define $\mathcal{T}(P^0, \mathcal{P}_{\mathcal{Q}_j,\beta^0_j})$ be the subspace of $L_0^2(P^0_j)$ that consists of all $h_j$ that arise as scores of univariate submodels of $\{P^{(\epsilon)} :\epsilon \in [0,\delta )\}$ for which $P^{(\epsilon)} = P^0$ when $\epsilon=0$. 

Throughout this work, we assume $\mathcal{Q}$ is a collection of nonparametric distributions $Q$, i.e, $\mathcal{T}(Q^0,\mathcal{Q}) = L_0^2(Q^0)$. When $Q$ is nonparametric, there exist sets $\mathcal{Q}_j$ of conditional distributions of $Z_j\mid \bar{Z}_{j-1}$, $j\in [d]$, such that $\mathcal{Q}$ is equal to the set of all distributions $Q$ such that, for all $j\in [d]$, the conditional distribution $Q_j$ belongs to $\mathcal{Q}_j$. In other words, $Q$ is variation independent such that distributions in $\mathcal{Q}$ can be defined separately via their conditional distributions such that it is possible to modify a conditional distribution $Q^0_j$ without affecting the others, namely $Q^0_{j'}$ with $j'\not=j$. This condition enables the derivation of a gradient by summing up individual projections of a gradient onto different sub-spaces of the $L_0^2$ space defined by perturbing the conditional distributions of $Q^0_j$ or $P^0_j$ separately over $j\in [d]$. By Lemma 1.6 of \cite{van2003unified}, and the fact that the tangent set of $\mathcal{P}_{\mathcal{Q},\beta^0}$ at $P^0$ is a closed linear space, the tanget space $\mathcal{T}(P^0,\mathcal{P}_{\mathcal{Q},\beta^0})$ of $\mathcal{P}_{\mathcal{Q},\beta^0}$ at $P^0$ takes the form $\bigoplus_{j=0}^d \mathcal{T}(P^0,\mathcal{P}_{\mathcal{Q}_j,\beta^0_j}):= \{\sum_{j=0}^d h_j :h_j \in \mathcal{T}(P^0,\mathcal{P}_{\mathcal{Q}_j,\beta_j^0})\}$, and the $L_0^2(P)$ projection of a function onto $\mathcal{T}(P^0,\mathcal{P}_{\mathcal{Q},\beta^0})$ is equal to the sum of projections onto $\mathcal{T}(P^0,\mathcal{P}_{\mathcal{Q}_j,\beta_j^0})$, $j = 0,1,\ldots,d$. Since the marginal distribution of $S$ is unrestricted and therefore independent of $\beta^0$, $\mathcal{T}(P^0_0, \mathcal{P}_0) = L_0^2(P^0_0)$ and $ \mathcal{T}(P^0,\mathcal{P}_{\mathcal{Q}_0,\beta^0_0})= \mathcal{T}(P^0, \mathcal{P}_0)$. Moreover, for all $j\in \mathcal{I}$, the conditional distribution of $Z_j \mid \bar{Z}_{j-1},S$ is also unrestricted and so $\mathcal{T}(P^0,\mathcal{P}_{\mathcal{Q}_j,\beta^0_j}) =L_0^2(P^0_j)$. The following result characterizes the other tangent spaces that appear in the direct sum defining $\mathcal{T}(P^0,\mathcal{P}_{\mathcal{Q},\beta^0})$.

\begin{lemma}[\textit{Tangent Space of $\mathcal{P}_{\mathcal{Q},\beta^0}$}]\label{lem: tangent_space}  
If Conditions~\ref{cond:sufficient_alignment} and \ref{cond:overlaps} from the main text hold and $j\in\mathcal{J}$, then
\begin{align}
    &\mathcal{T}(P^0,\mathcal{P}_{\mathcal{Q}_j,\beta^0_j}) \nonumber\\
    &= \{(z,s) \mapsto g_j(\bar{z}_j,s) + \mathbbm{1}_{\mathcal{S}_j}(s)\mathbbm{1}_{\bar{\mathcal{Z}}^{\dagger}_{j-1}}(\bar{z}_{j-1}) \left[\left(f_j(\bar{z}_j) -E_{P^0}[f_j(\bar{Z}_j)\mid \bar{z}_{j-1}, s]\right)  - g_j(\bar{z}_j,s)\right]\nonumber\\
    & \hspace{5em} :   f_j \in \mathcal{T}(Q^0, \mathcal{Q}_j), g_j \in L_0^2(P^0_j)\}. \label{eq:tangent_space}
\end{align}
\end{lemma}
When $\mathcal{S}_j = \mathcal{A}_j$, then this result collapses to Lemma S1 in \cite{li2021efficient} since $E_{P^0}[f_j(\bar{Z}_j)\mid \bar{z}_{j-1}, s] =0$ for all $s \in \mathcal{A}_j$. We choose to provide a heuristic proof here and direct the reader's attention to the aforementioned paper for more details. 

\begin{proof}[\textbf{Proof of Lemma S}\ref{lem: tangent_space}]
Throughout this appendix, we assume enough regularity so that the scores considered correspond to derivatives of log-likelihoods. The following proof could be modified to apply even when this fails by adopting similar arguments to those used in the proof of Lemma S1 from \cite{li2021efficient}. Fix $j \in \mathcal{J}$ and let $\mathcal{R}_j$ denote the right-hand side of \eqref{eq:tangent_space}. We first show that $ \mathcal{R}_j \subseteq \mathcal{T}(P^0,\mathcal{P}_{\mathcal{Q}_j,\beta^0_j}) $, and then we show $\mathcal{T}(P^0,\mathcal{P}_{\mathcal{Q}_j,\beta^0_j}) \subseteq \mathcal{R}_j$.

\textbf{Part 1 of proof: $\mathcal{R}_j \subseteq \mathcal{T}(P^0,\mathcal{P}_{\mathcal{Q}_j,\beta^0_j})$. } Fix $f_j \in \mathcal{T}(Q^0, \mathcal{Q}_j)$ and $g_j \in L_0^2(P^0_j)$. Since $f_j \in \mathcal{T}(Q^0, \mathcal{Q}_j)$, there exists a univariate submodel $\{Q^{(\epsilon)}: \epsilon \in [0,\delta]\}$ with score $f_j$ at $\epsilon = 0$ for which $Q^{(\epsilon)} = Q^0$. For each $\epsilon \in [0,\delta)$, $s \in \mathcal{S}_j$, and $j\in \mathcal{J}$, we let $P_j^{(\epsilon)} \in \mathcal{P}_{\mathcal{Q}_j,\beta^0_j}$ be such that $dP_j^{(\epsilon)}(\cdot \mid \bar{z}_{j-1}, s) = w_{j,s}(\bar{z}_j;\beta^0_{j,s})/E_{Q^{(\epsilon)}}[w_{j,s}(\bar{Z}_j;\beta^0_{j,s})\mid \bar{z}_{j-1}]dQ_j^{(\epsilon)}(\cdot \mid \bar{z}_{j-1})$ for $P^0$-almost all $\bar{z}_{j-1} \in \bar{\mathcal{Z}}^\dagger_{j-1}$. By the variation independence of $P^0$, we can suppose without loss of generality that $P^{(\epsilon)}_i = P^0_i$ for all $i\neq j$ and also the marginal distribution of $S$ under $P^{(\epsilon)}$ is equal to the marginal distribution of $S$ under $P^0$. It can be shown that $P^{(\epsilon)}$ belongs to $\mathcal{P}_{\mathcal{Q}_j,\beta^0_j}$. In addition, for $s\in\mathcal{S}_j$ and all $\bar{z}_{j-1} \in \mathcal{Z}^\dagger_{j-1}$, $\{P^{(\epsilon)}: \epsilon \in [0,\delta)\}$ is quadratic mean differentiable at $\epsilon = 0$ with the following score: 
\begin{align*}
     h_j(\bar{z}_j,s) 
     & = \left.\frac{d}{d \epsilon} \log p_j^{\epsilon}(\bar{z}_j,s;\beta^0_{j,s})\right|_{\epsilon=0} \\
    & = \left.\frac{d}{d \epsilon} \log\frac{w_{j,s}(\bar{z}_{j};\beta^0_{j,s})q^{(\epsilon)}_j}{E_{Q^{(\epsilon)}}[w_{j,s}(\bar{Z}_{j};\beta^0_{j,s})\mid \bar{z}_{j-1}]} \right|_{\epsilon=0} \\
    & =  \left.\frac{d}{d \epsilon} \log q_j^{\epsilon}(\bar{z}_{j})\right|_{\epsilon=0}- \frac{1}{E_{Q^0}[w_{j,s}(\bar{Z}_{j};\beta^0_{j,s})\mid \bar{z}_{j-1}]} \left.\frac{d}{d \epsilon} E_{Q^{(\epsilon)}}\left[w_{j,s}(\bar{Z}_{j};\beta^0_{j,s}) \mid \bar{z}_{j-1}\right]\right|_{\epsilon=0} \\
    & = f_j(\bar{z}_{j}) - \frac{E_{Q^0}[w_{j,s}(\bar{Z}_{j};\beta^0_{j,s}) f_j(\bar{Z}_{j}) \mid \bar{z}_{j-1}]}{E_{Q^0}[w_{j,s}(\bar{Z}_{j};\beta^0_{j,s})\mid \bar{z}_{j-1}]} \\
    & =f_j(\bar{z}_{j}) - E_{P^0}[f_j(\bar{Z}_{j}) \mid \bar{z}_{j-1},s].
\end{align*}
As $f_j \in \mathcal{T}(Q^0,\mathcal{Q}_j)$ and $g_j \in L_0^2(P^0_j)$ were arbitrary, $\mathcal{R}_j \subseteq \mathcal{T}(P^0,\mathcal{P}_{\mathcal{Q}_j,\beta^0_j})$.

\textbf{Part 2 of proof: $\mathcal{T}(P^0,\mathcal{P}_{\mathcal{Q}_j,\beta^0_j}) \subseteq \mathcal{R}_j$.} Fix $h_j \in \mathcal{T}(P^0, \mathcal{P}_{\mathcal{Q}_j,\beta^0_j})$ and let $P^{(\epsilon)}$ be the submodel such that $P^{(\epsilon)} = P^0$ at $\epsilon=0$ with score $h_j(\bar{z}_j,s)$. We will show that there exists $f_j \in \mathcal{T}(Q^0,\mathcal{Q}_j)$ and $g_j \in L_0^2(P^0_j)$ such that any $h_j \in \mathcal{T}(P^0,\mathcal{P}_{\mathcal{Q}_j,\beta^0_j})$ takes the form that $h_j(z,s) = g_j(\bar{z}_j,s) + \mathbbm{1}_{\mathcal{S}_j}(s)\mathbbm{1}_{\bar{\mathcal{Z}}^{\dagger}_{j-1}}(\bar{z}_{j-1}) [\left(f_j(\bar{z}_j) -E_{P^0}[f_j(\bar{Z}_j)\mid \bar{z}_{j-1}, s]\right)  - g_j(\bar{z}_j,s)]$ $P^0$-almost everywhere. Since $\mathcal{T}(P^0,\mathcal{P}_{\mathcal{Q}_j,\beta^0_j})$ is a subset of the maximal tangent space $L_0^2(P^0_j)$, we can let $g_j = h_j$. Then it remains to show that there exists an $f_j \in \mathcal{T}(Q^0,\mathcal{Q}_j)$ such that $h_j(z,s) = f_j(z) - E_{P^0}[f_j(\bar{Z}_j)\mid \bar{z}_{j-1},s]$ for $P^0$-almost all $\bar{z}_{j-1} \in \bar{Z}^\dagger_{j-1}$ and all $s \in \mathcal{S}_j$. 

For each $\epsilon \in [0,\delta)$, $s \in \mathcal{S}_j$, and $j\in \mathcal{J}$, we let $Q_j^{(\epsilon)} \in \mathcal{Q}$ be such that $dQ_j^{(\epsilon)}(\cdot \mid \bar{z}_{j-1}) w_{j,s}(\bar{z}_j;\beta^0_j)/E_{Q^{(\epsilon)}}[w_{j,s}(\bar{Z}_j;\beta^0_j) \mid \bar{z}_{j-1}] = dP_j^{(\epsilon)}(\cdot\mid \bar{z}_{j-1},s)$ for $Q^0$-almost all $\bar{z}_{j-1} \in \bar{\mathcal{Z}}^\dagger_{j-1}$ and $Q^{(\epsilon)}_i = Q^{(\epsilon)}_i$ for all $i \neq j$. Then $\{Q^{(\epsilon)}:\epsilon \in [0,\delta)\}$ is a submodel of $\mathcal{Q}$ that is quadratic mean differentaible at $\epsilon=0$ with the following score $f_j$:
\begin{align*}
    f_j(\bar{z}_j)& = \left.\frac{d}{d \epsilon} \log q_j^{\epsilon}(\bar{z}_j)\right|_{\epsilon=0} \\
    & = \left.\frac{d}{d \epsilon} \log\frac{p^{(\epsilon)}_j E_{Q^{(\epsilon)}}[w_{j,s}(\bar{Z}_{j};\beta^0_{j,s})\mid \bar{z}_{j-1}]}{w_{j,s}(\bar{z}_{j};\beta^0_{j,s})} \right|_{\epsilon=0} \\
    & = h_j(\bar{z}_{j},s) + \frac{E_{Q^0}[w_{j,s}(\bar{Z}_{j};\beta^0_{j,s}) f_j(\bar{Z}_{j}) \mid \bar{z}_{j-1}]}{E_{Q^0}[w_{j,s}(\bar{Z}_{j};\beta^0_{j,s})\mid \bar{z}_{j-1}]} \\
    & = h_j(\bar{z}_j,s) + E_{P^0}\left[f_j(\bar{Z}_j)\mid \bar{z}_{j-1},s\right].
\end{align*}
\begin{sloppypar}
From above, we have $h_j(\bar{z}_j,s) = f_j(\bar{z}_j) - E_{P^0}[f_j(\bar{Z}_j)\mid \bar{z}_{j-1},s]$ for $P^0$-almost all $(\bar{z}_j,s)$ that are such that $\bar{z}_{j-1} \in \mathcal{Z}^\dagger_{j-1}$ and $s \in \mathcal{S}_j$ and we see that $h_j(z,s) =g_j(\bar{z}_j,s) + \mathbbm{1}_{\mathcal{S}_j}(s)\mathbbm{1}_{\bar{\mathcal{Z}}^{\dagger}_{j-1}}(\bar{z}_{j-1}) [(f_j(\bar{z}_j) -E_{P^0}[f_j(\bar{Z}_j)\mid \bar{z}_{j-1}, s])  - g_j(\bar{z}_j,s)] $ $P^0$-almost everywhere. Hence, $h_j \in \mathcal{R}_j$. As $h_j$ was an arbitrary element of $\mathcal{T}(P^0,\mathcal{P}_{\mathcal{Q}_j,\beta^0_j})$, we have proved that $\mathcal{T}(P^0,\mathcal{P}_{\mathcal{Q}_j,\beta^0_j}) \subseteq \mathcal{R}_j$.
\end{sloppypar}
\end{proof}

{Now we show that $D_{P^0}$ in \eqref{eq:gradient_known} is a valid gradient of $\phi$ relative to $\mathcal{P}_{\mathcal{Q},\beta^0}$.}
The proof shares great similarity to the proof of Theorem 2 in Supplementary Section A.3 in \cite{li2021efficient}. We suppose that $\psi$ is pathwise differentiable at $Q^0$ relative to $\mathcal{Q}$ and fix a gradient $D_{Q^0}$ of $\psi$. We will show that, for any submodel $\{P^{(\epsilon)} : \epsilon\in [0,\delta)\}$ with score $h\in \mathcal{T}(P^0,\mathcal{P}_{\mathcal{Q},\beta^0})$ and with $P^{(\epsilon)}=P^0$ when $\epsilon=0$, it holds that $\frac{\partial}{\partial \epsilon}\phi(P^{(\epsilon)}) \mid_{\epsilon=0} = E_{P^0}\{D_{P^0}(Z,S)h(Z,S)\}$, where $D_{P^0}$ takes the form given in \eqref{eq:gradient_known}. This will show that $D_{P^0}$ is a gradient of $\phi$, which will complete the proof of \eqref{eq:gradient_known}. We first provide a useful lemma that will be used in later proof. 
\begin{lemma}[\textit{Exchanging between $E_{Q^0}$ and $E_{P^0}$}]\label{lem: Q_TO_P} 
For an arbitrary function $f : \bar{\mathcal{Z}}_j\rightarrow\mathbb{R}$ with $E_{Q^0}[f^2(\bar{Z}_j)] < \infty $, $E_{Q^0}[f(\bar{Z}_j)] = E_{P^0}[ \lambda^{\dagger}_{j}(\bar{Z}_{j},S)f(\bar{Z}_j) \mid S \in \mathcal{S}_j]$ where $\lambda^{\dagger}_{j}(\bar{z}_j,s): = q^0(z_j\mid \bar{z}_{j-1})/p^0(z_j\mid \bar{z}_{j-1}, s) \lambda_{j-1}(\bar{z}_{j-1})$ and $\lambda_{j-1}$ denotes the Radon-Nikodym derivative of the marginal distribution of $\bar{Z}_{j-1}$ under sampling from $Q^0$ relative to the conditional distribution of $\bar{Z}_{j-1} \mid S \in \mathcal{S}_j$ under sampling from $P^0$.
\end{lemma}
\begin{proof}[Proof of Lemma S\ref{lem: Q_TO_P}:]
Using tower property and by the definition of $\lambda^{\dagger}_{j}$, we have
\begin{align*}
    E_{Q^0}[f(\bar{Z}_j)]
    & = E_{Q^0}[E_{Q^0}[f(\bar{Z}_j)\mid \bar{Z}_{j-1}]] \\
    & = E_{Q^0}\left[E_{P^0}\left[ \frac{q^0(Z_j \mid \bar{Z}_{j-1})}{p^0(Z_j \mid \bar{Z}_{j-1},  S\in \mathcal{S}_j)} f(\bar{Z}_j)\mid \bar{Z}_{j-1},S \in \mathcal{S}_j\right] \right]\\
    & = E_{P^0}\Bigg[\lambda_{j-1}(\bar{Z}_{j-1})E_{P^0}\bigg[ \frac{q^0(Z_j \mid \bar{Z}_{j-1})}{p^0(Z_j \mid \bar{Z}_{j-1}, S \in \mathcal{S}_j)} f(\bar{Z}_j)\mid \bar{Z}_{j-1},S \in \mathcal{S}_j\bigg] \mid  S \in \mathcal{S}_j \Bigg]\\
    & = E_{P^0}\Bigg[\lambda_{j-1}(\bar{Z}_{j-1})  E_{P^0}\bigg[ \frac{q^0(Z_j \mid \bar{Z}_{j-1})}{p^0(Z_j \mid \bar{Z}_{j-1}, S)} f(\bar{Z}_j)\mid \bar{Z}_{j-1},S\bigg] \mid  S \in \mathcal{S}_j \Bigg]\\
    & = E_{P^0}\left[ \lambda^{\dagger}_{j}(\bar{Z}_{j},S) f(\bar{Z}_j) \mid S \in \mathcal{S}_j\right]\\
    & = E_{P^0}\left[ \frac{\mathbbm{1}(S \in \mathcal{S}_j)}{P^0(S \in \mathcal{S}_j)}\lambda^{\dagger}_{j}(\bar{Z}_{j},S) f(\bar{Z}_j)\right]
\end{align*}
\end{proof}

\begin{proof}[Proof of $D_{P^0}$ in \eqref{eq:gradient_known}]
 Fix a function $h\in \mathcal{T}(P^0, \mathcal{P}_{\mathcal{Q},\beta^0})$ and submodel $\{P^{(\epsilon)} : \epsilon\in [0,\delta)\}$. Since $\mathcal{T}(P^0, \mathcal{P}_{\mathcal{Q},\beta^0})=\bigoplus_{j=0}^d \mathcal{T}(P^0, \mathcal{P}_{\mathcal{Q}_j,\beta^0_j})$, there exist $h_j\in \mathcal{T}(P^0, \mathcal{P}_{\mathcal{Q}_j,\beta^0_j})$, $j\in \{0\}\cup [d]$ , such that $h(z,s)=\sum_{j=0}^d h_j(\bar{z}_j,s)$. Based on previous results, for each $j\in \mathcal{J}$, there exists an $f_j\in \mathcal{T}(Q^0, \mathcal{Q}_j)$   such that $f_j(\bar{z}_j) = h_j(\bar{z}_j,s) + E_{P^0}[f_j(\bar{Z}_j) \mid \bar{z}_{j-1},s]$ for $(s,\bar{z}_{j-1}) \in \mathcal{S}_j \times \bar{\mathcal{Z}}^{\dagger}_{j-1}$. For each $\epsilon \in [0,\delta)$, let $Q^{(\epsilon)} \in \mathcal{Q}$ be such that $Q_j^{(\epsilon)}(\cdot \mid \bar{z}_{j-1}) =P^{(\epsilon)}_j(\cdot \mid \bar{z}_{j-1},S\in \mathcal{A}_j)$ for $Q^0$-almost all $\bar{z}_{j-1} \in \mathcal{Z}_{j-1}^\dagger$. By analogous arguemnts to those given in the proof of Lemma S\ref{lem: tangent_space}, $\{Q^{(\epsilon)}: \epsilon \in [0,\delta)\}$ has score $\sum_{j \in \mathcal{J}}f_j$ at $\epsilon=0$. As $\psi$ is pathwise differentiable at $Q^0$ relative to $\mathcal{Q}$, 
\begin{align}
    &\left.\frac{d}{d\epsilon} \phi(P^{(\epsilon)})\right|_{\epsilon=0} = \left.\frac{d}{d\epsilon} \psi\circ \theta(P^{(\epsilon)})\right|_{\epsilon=0} \nonumber \\
    &= \left.\frac{d}{d\epsilon} \psi(Q^{(\epsilon)})\right|_{\epsilon=0} \nonumber\\ 
    & =E_{Q^0}\left[D_{Q^0}(Z)\sum_{j\in\mathcal{J}} f_j(\bar{Z}_j)\right] \tag{by the pathwise differentiability of $\psi$}\\
    &=E_{Q^0}\left[\sum_{j\in\mathcal{J}} D_{Q^0,j}(\bar{Z}_j)f_j(\bar{Z}_j)\right]\\
    &= E_{P^0}\left[\sum_{j\in\mathcal{J}} \frac{\mathbbm{1}(S\in\mathcal{S}_j)}{P^0(S\in\mathcal{S}_j)} \lambda^{\dagger}_{j}(\bar{Z}_{j},S;\beta^0_j)D_{Q^0,j}(\bar{Z}_j)f_j(\bar{Z}_j)\right] \label{step_uselemS2_end}\\
\intertext{
Letting $b_j(\bar{z}_j,s):= \lambda^{\dagger}_{j}(\bar{z}_{j},s;\beta^0_j)D_{Q^0,j}(\bar{z}_j)$. The previous equation writes as, }
    &= E_{P^0}\left[\sum_{j\in\mathcal{J}} \frac{\mathbbm{1}(S\in\mathcal{S}_j)}{P^0(S\in\mathcal{S}_j)}b_j(\bar{Z}_j,S)f_j(\bar{Z}_j)\right] \nonumber\\
    &= E_{P^0}\bigg[\sum_{j\in\mathcal{J}} \frac{\mathbbm{1}(S\in\mathcal{S}_j)}{P^0(S\in\mathcal{S}_j)}b_j(\bar{Z}_j,S)\left\{h_j(\bar{Z}_j,S) + E_{P^0}\left[f_j(\bar{Z}_j)\mid \bar{Z}_{j-1},S\right]\right\}\bigg] \label{step_f_to_h}\\
    &= E_{P^0}\left[{\sum_{j\in\mathcal{J}} \frac{\mathbbm{1}(S\in\mathcal{S}_j)}{P^0(S\in\mathcal{S}_j)}b_j(\bar{Z}_j,S)}h_j(\bar{Z}_j,S) \right] \label{step_minus} 
\end{align}
where \eqref{step_f_to_h} is true by Lemma S\ref{lem: tangent_space} that for a submodel $Q^{(\epsilon)}$ with score $f_j(\bar{z}_j) \in \mathcal{T}(Q^0, \mathcal{Q}_j)$, there exists a submodel $P^{(\epsilon)}$ with score $h_j(\bar{z}_j,s)$ that takes the above form, and \eqref{step_minus} is true since 
\begin{align*}
    & E_{P^0}\left[\sum_{j\in\mathcal{J}} \frac{\mathbbm{1}(S\in\mathcal{S}_j)}{P^0(S\in\mathcal{S}_j)}b_j(\bar{Z}_j,S)E_{P^0}\left[f_j(\bar{Z}_j)\mid \bar{Z}_{j-1},S\right]\right] \\
    & =  E_{P^0}\Bigg[E_{P^0}\left[\sum_{j\in\mathcal{J}} \frac{\mathbbm{1}(S\in\mathcal{S}_j)}{P^0(S\in\mathcal{S}_j)}b_j(\bar{Z}_j,S)\mid \Bar{Z}_{j-1}, S\right]  E_{P^0}\left[f_j(\bar{Z}_j)\mid \bar{Z}_{j-1},S\right] \Bigg] \\
    & =  E_{P^0}\Bigg[\sum_{j\in\mathcal{J}} \frac{\mathbbm{1}(S\in\mathcal{S}_j)}{P^0(S\in\mathcal{S}_j)} \lambda_{j-1}(\bar{Z}_{j-1}) E_{P^0}\left[f_j(\bar{Z}_j)\mid \bar{Z}_{j-1},S\right]  \\
    & \hspace{5em} \cdot E_{P^0}\left[\frac{q^0(Z_j\mid \bar{Z}_{j-1})}{p^0(Z_j\mid \bar{Z}_{j-1},S)} D_{Q^0,j}(\bar{Z}_j)\mid \Bar{Z}_{j-1}, S\right]\Bigg] \\
    & = 0.
\end{align*}
It is straightforward to show that $E_{P^0}[\sum_{j \in \mathcal{S}_j} \frac{\mathbbm{1}(S \in \mathcal{S}_j)}{P^0(S \in \mathcal{S}_j)}b_j(\bar{Z}_j,S)] =0$. As $h\in \mathcal{T}(P^0,\mathcal{P}_{\mathcal{Q},\beta^0})$ was arbitrary, $\phi$ is pathwise differentiable at $P^0$ relative to $\mathcal{P}_{\mathcal{Q},\beta^0}$ with gradient $D_{P^0}$ in \eqref{eq:gradient_known}.

{Now we prove Lemma~\ref{lem:canonical_beta_known}. In what follows we use $\Pi_{P^0,\beta^0}\{\cdot \mid \mathcal{R}\}$ denote the $L_0^2(P^0)$-projection operator onto a subspace $\mathcal{R}$ of $L_0^2(P^0)$ under fixed $\beta^0$. The canonical gradient under $\mathcal{T}(P^0,\mathcal{P}_{\mathcal{Q},\beta^0})$ can be found by
\begin{align}
    \tilde{D}_{P^0}(z,s;\beta^0) &= \Pi_{P^0,\beta^0}\{ D_{P^0}\mid \mathcal{T}(P^0,\mathcal{P}_{\mathcal{Q},\beta^0})\}(z,s;\beta^0) \label{anyD}\\
    & = \Pi_{P^0,\beta^0}\{ D_{P^0}\mid \bigoplus_{j=0}^d \mathcal{T}(P^0,\mathcal{P}_{\mathcal{Q}_j,\beta^0_j})\}(\bar{z}_j,s;\beta^0_j) \nonumber \\
    & = \sum_{j\in\mathcal{J}}\Pi_{P^0,\beta^0}\{ D_{P^0}\mid \mathcal{T}(P^0,\mathcal{P}_{\mathcal{Q}_j,\beta^0_j})\}(\bar{z}_j,s;\beta^0_j)  \label{step_q_nonpar} \\
    & = \sum_{j\in\mathcal{J}}\Pi_{P^0,\beta^0}\{ D_{P^0,j}\mid \mathcal{T}(P^0,\mathcal{P}_{\mathcal{Q}_j,\beta^0_j})\}(\bar{z}_j,s;\beta^0_j),  \label{step_ortho}
\end{align}
where we choose to project $D_{P^0}$ onto $\mathcal{T}(P^0, \mathcal{P}_{\mathcal{Q},\beta^0})$ in \eqref{anyD}, where in fact, any valid gradient of $\phi$ can be used in \eqref{anyD} such as $D^\mathcal{A}_{P^0}$. Equation \eqref{step_q_nonpar} follows by the fact that $\mathcal{Q}$ is nonparametric and \eqref{step_ortho} is true since $D_{P^0,j}$ is orthogonal to subspaces $\mathcal{T}(P^0,\mathcal{P}_{\mathcal{Q}_m,\beta^0_m})$ for all $m \neq j$. We provide the proof below.
\begin{proof}[\textit{Proof of Equation~\ref{step_ortho}}.]
Now we show that it is always true that $D_{P^0,j}\perp g_{m} \textnormal{ for all } g_{m} \in \mathcal{T}(P^0,\mathcal{P}_{\mathcal{Q}_m,\beta^0_{m}})$ where $m\neq j$, $j\in\mathcal{J}$ and $m\in \mathcal{J}\cup\{0\}$. 
\begin{align*}
        &E_{P^0}\left[D_{P^0,j}(\bar{Z}_j,S;\beta^0_j) g_m(\bar{Z}_m,S)\right]\\
        & = 
\begin{cases}
    E_{P^0}\left[D_{P^0,j}(\bar{Z}_j,S;\beta^0_j)  E_{P^0}\left[g_m(\bar{Z}_m,S) \mid \bar{Z}_{m-1},S\right] \right]   =0, &\mbox{ if $j < m$}, \\
    E_{P^0}\left[E_{P^0}\left[D_{P^0,j}(\bar{Z}_j,S;\beta^0_j) \mid \bar{Z}_{j-1},S\right] g_m(\bar{Z}_m,S)  \right]   =0,&\mbox{if $j > m$.}
\end{cases}        
    \end{align*}
\end{proof}
\begin{sloppypar}
    Continuing in the line of Equation~\ref{step_ortho}, we claim that $\Pi_{P^0, {\beta^0}}\{D_{P^0,j} \mid \mathcal{T}(P^0, \mathcal{P}_{\mathcal{Q}_j,\beta^0_j})\}(\bar{z}_j,s';\beta^0_{j}) = \mathbbm{1}_{\mathcal{S}_j}(s')\mathbbm{P}_{j,s'}\tilde{d}_j(\bar{z}_j;\beta^0_{j})$, where $\mathbbm{P}_{j,s} a_j(\bar{z}_j;\beta^0_{j}):= a_j(\bar{z}_j;\beta^0_{j}) - E_{P^0}[a_j(\bar{Z}_j;\beta^0_{j})\mid \bar{z}_{j-1}, s]$ and $\tilde{d}_j(\bar{z}_j;\beta^0_j)$ takes the form in Lemma~\ref{lem:canonical_beta_known}. For ease of exposition, we focus on the case where $\mathcal{Z}_{j-1}^\dagger=\mathcal{Z}_{j-1}$ and use the fact that, as implied by Lemma S\ref{lem: tangent_space}, a generic element of $\mathcal{T}(P^0, \mathcal{P}_{\mathcal{Q}_j,\beta^0_j})$ takes the form $\mathbbm{1}_{\mathcal{S}_j}(S)(\mathbbm{P}_{j,S}a_j - g_j) + g_j$ for some $a_j \in \mathcal{T}(Q^0, \mathcal{Q}_j)$ and $g_j \in L_0^2(P^0_j)$. We first show that $\mathbbm{1}_{\mathcal{S}_j}(s')\mathbbm{P}_{j,s'}\tilde{d}_j \in \mathcal{T}(P^0,\mathcal{P}_{\mathcal{Q}_j,\beta^0_j})$, and then we show 
    \begin{align*}
        0 & = E_{P^0}\bigg[\left\{D_{P^0,j}(\bar{Z}_j,S;\beta^0_j) -  \mathbbm{1}_{\mathcal{S}_j}(S)\mathbbm{P}_{j,S}\tilde{d}_j(\bar{Z}_j;\beta^0_j) \right\}\\
        &\hspace{8em} \cdot \left\{\mathbbm{1}_{\mathcal{S}_j}(S)(\mathbbm{P}_{j,S}a_j(\bar{Z}_j;\beta^0_j) -g_j(\bar{Z}_j,S)) + g_j(\bar{Z}_j,S)\right\}\bigg] 
    \end{align*} for all $a_j \in \mathcal{T}(Q^0, \mathcal{Q}_j)$. 
\end{sloppypar}
\indent \textbf{Part I of showing that $\Pi_{P^0, {\beta^0}}\{D_{P^0,j} \mid \mathcal{T}(P^0, \mathcal{P}_{\mathcal{Q}_j,\beta^0_j})\}(\bar{z}_j,s';\beta^0_{j}) = \mathbbm{1}_{\mathcal{S}_j}(s')\mathbbm{P}_{j,s'}\tilde{d}_j(\bar{z}_j;\beta^0_{j})$: $\mathbbm{1}_{\mathcal{S}_j}(s') \mathbbm{P}_{j,s}\tilde{d}_j \in \mathcal{T}(P^0,\mathcal{P}_{\mathcal{Q}_j,\beta^0_{j,s}})$.} It is straightforward to show that $E_{Q^0}[\tilde{d}_j(\bar{Z}_{j};\beta^0_{j}) \mid \bar{z}_{j-1}] = 0$ everywhere. Since $Q^0$ is nonparametric, we have $\tilde{d}_j \in \mathcal{T}(Q^0,\mathcal{Q}_j)$. By Lemma S\ref{lem: tangent_space}, it follows that $\mathbbm{P}_{j,s}\tilde{d}_j \in \mathcal{T}(P^0,\mathcal{P}_{\mathcal{Q}_j,\beta^0_j})$ for any  $s \in \mathcal{S}_j$.\\
\indent \textbf{Part II of showing that $\Pi_{P^0, {\beta^0}}\{D_{P^0,j} \mid \mathcal{T}(P^0, \mathcal{P}_{\mathcal{Q}_j,\beta^0_j})\}(\bar{z}_j,s';\beta^0_{j}) = \mathbbm{1}_{\mathcal{S}_j}(s')\mathbbm{P}_{j,s'}\tilde{d}_j(\bar{z}_j;\beta^0_{j})$:} We wish to show that,
\begin{align*}
     0 & =E_{P^0}\bigg[\left\{D_{P^0,j}(\bar{Z}_j,S;\beta^0_j) -  \mathbbm{1}_{\mathcal{S}_j}(S)\mathbbm{P}_{j,S}\tilde{d}_j(\bar{Z}_j;\beta^0_j) \right\}\\
     & \hspace{6em}\cdot \left\{\mathbbm{1}_{\mathcal{S}_j}(S)(\mathbbm{P}_{j,S}a_j(\bar{Z}_j;\beta^0_j) -g_j(\bar{Z}_j,S)) + g_j(\bar{Z}_j,S)\right\}\bigg]\\
     & = E_{P^0}\left[\left\{D_{P^0,j}(\bar{Z}_j,S;\beta^0_j) -\mathbbm{P}_{j,S} \tilde{d}_{j}(\bar{Z}_j;\beta^0_j)\right\}\mathbbm{P}_{j,S}a_j(\bar{Z}_j;\beta^0_j)\mid S\in\mathcal{S}_j\right]\\
    & = E_{P^0}\bigg[\sum_{m\in\mathcal{S}_j} \mathbbm{1}(S=m )E_{P^0}\left[\left(D_{P^0,j} - \mathbbm{P}_{j,m}\tilde{d}_j(\bar{Z}_{j};\beta^0_j)\right)\mathbbm{P}_{j,m}a_j(\bar{Z}_j;\beta^0_j)\mid \bar{Z}_{j-1},S=m\right]\bigg].
\end{align*}
The conditional expectation in the expression above can be simplified to, 
\begin{align*}
    &E_{P^0}\left[\left(D_{P^0,j}(\bar{Z}_{j},S;\beta^0_j) - \mathbbm{P}_{j,m}\tilde{d}_j(\bar{Z}_{j};\beta^0_j)\right)\mathbbm{P}_{j,m}a_j(\bar{Z}_j;\beta^0_j)\mid \bar{Z}_{j-1},S=m\right]\\
    & = E_{P^0}\left[\left(\mathbbm{P}_{j,m}D_{P^0,j}(\bar{Z}_{j},S;\beta^0_j) - \mathbbm{P}_{j,m}\tilde{d}_j(\bar{Z}_{j};\beta^0_j)\right)\mathbbm{P}_{j,m}a_j(\bar{Z}_j;\beta^0_j)\mid \bar{Z}_{j-1},S=m\right] \tag{since $E_{P^0}[D_{P^0,j}(\bar{Z}_j,S;\beta^0_j) \mid \bar{z}_{j-1},S=m] =0$ for any $m\in\mathcal{S}_j$}\\
    & = E_{P^0}\left[\left(D_{P^0,j}(\bar{Z}_j,S;\beta^0_j)- \tilde{d}_j(\bar{Z}_j,S;\beta^0_j)\right) a_j(\bar{Z}_j;\beta^0_j)\mid \bar{Z}_{j-1},S=m\right] \\
    & \hspace{1em} - E_{P^0}\left[\left(D_{P^0,j}(\bar{Z}_j,S;\beta^0_j)- \tilde{d}_j(\bar{Z}_j,S;\beta^0_j)\right) \mid \bar{Z}_{j-1},S=m]\right] E_{P^0}\left[a_j(\bar{Z}_j;\beta^0_j)\mid \bar{Z}_{j-1},S=m\right].
\end{align*}
Continuing where we left off, we have
\begin{align*}
    0& = E_{P^0}\Bigg[\sum_{m\in\mathcal{S}_j} \mathbbm{1}(S=m )E_{P^0}\big[a_j(\bar{Z}_j;\beta^0_j) \Big(D_{P^0,j}(\bar{Z}_{j},S;\beta^0_j) - \tilde{d}_j(\bar{Z}_{j};\beta^0_j) \\
    & \hspace{6em} - E_{P^0}\left[D_{P^0,j}(\bar{Z}_{j},S;\beta^0_j) - \tilde{d}_j(\bar{Z}_{j};\beta^0_j) \mid \bar{Z}_{j-1},S=m\right]\Big) \mid \bar{Z}_{j-1},S=m\big]\Bigg]
\end{align*}
\begin{sloppypar}
\noindent Denoting $H(\bar{Z}_{j},m;\beta^0_j):= D_{P^0,j}(\bar{Z}_{j},m;\beta^0_j) - \tilde{d}_j(\bar{Z}_{j};\beta^0_j) - E_{P^0}\left[D_{P^0,j}(\bar{Z}_{j},S;\beta^0_j) - \tilde{d}_j(\bar{Z}_{j};\beta^0_j) \mid \bar{Z}_{j-1},S=m\right]$, we have 
\end{sloppypar}
\begin{align*}
   0 = \int \left\{\sum_{m\in\mathcal{S}_j} P^0(S=m)\frac{p^0(\bar{z}_{j-1}\mid S=m)}{q^0(\bar{z}_{j-1})}H(\bar{z}_{j},m;\beta^0_j)w^*_{j,m}(\bar{z}_j,m;\beta^0_{j,m})\right\}a_j(\bar{z}_j)dQ^0(\bar{z}_{j}),
\end{align*}
Given the fact that the equality remains valid for all $a_j \in \mathcal{T}(Q^0,\mathcal{Q}_j)$, it follows that the expression enclosed within the curly braces must vanish almost everywhere in $L_0^2(Q^0_j)$. Therefore solving the above is equivalent to solving a multidimensional Fredholm integral equation of the second kind:
\begin{align*}
    d_j(\bar{z}_j;\beta^0_j)
    & =\tilde{d}_j(\bar{z}_{j};\beta^0_j)  - \int \tilde{d}_j(z'_j,\bar{z}_{j-1};\beta^0_j) K(z_j, z'_j, \bar{z}_{j-1};\beta^0_j) dQ^0_j(z'_j\mid\bar{z}_{j-1}) ,
\end{align*}
\begin{sloppypar}
\noindent with kernel $K(z'_j, \bar{z}_{j}):= \sum_{m \in \mathcal{S}_j} r_{j,m}(\bar{z}_j;\beta^0_{j,m}) w^*_{j,m}(z'_j, \bar{z}_{j-1};\beta^0_{j,m})$ and  $d_j(\bar{z}_j;\beta^0_j)  = \sum_{m \in \mathcal{S}_j} r_{j,m}(\bar{z}_j;\beta^0_{j,m})D_{P^0,j}(\bar{z}_{j},m;\beta^0_{j,m})$. Since the kernel is Pincherle-Goursat (i.e. it factors in $z'_j$ and $z_j$ given values of $\bar{z}_{j-1}$), for any given values of $\bar{z}_{j-1}$, the Fredholm equation can be equivalently expressed as an algebraic system of ${|\mathcal{S}_j|}$ equations in ${|\mathcal{S}_j|}$ unknowns $A_jx = D_j$ where $x = (x_1, \ldots,x_{|\mathcal{S}_j|})^{\top}$ are unknowns, $A_j = I_{|\mathcal{S}_j|} - (a_{im})= I_{|\mathcal{S}_j|} - E_{Q^0}[\bar{r}_j(\bar{Z}_{j}){\bar{w}_j}^{*\top}(\bar{Z}_{j})\mid \bar{z}_{j-1}]$ and $D_j = [D_1^{\top}, \ldots, D_{|\mathcal{S}_j|}^{\top}]^{\top}$, with 
\end{sloppypar}
\begin{align*}
    a_{im}(\bar{z}_{j-1})
    & = \int r_{j,m}(\bar{z}_j)w^*_{j,i}(\bar{z}_j)dQ^0_j(z_j \mid \bar{z}_{j-1}) = \delta_{m}E_{Q^0_j}[r_j(\bar{Z}_j)w^*_{j,i}(\bar{Z}_j) w^*_{j,m}(\bar{Z}_j)\mid \bar{z}_{j-1}],\\
    D_m(\bar{z}_{j-1}) 
    & =  \int d_j(\bar{z}_j)w^*_{j,m}(\bar{z}_j)dQ^0_j(z_j\mid \bar{z}_{j-1}).
\end{align*}
We note that $A_j = M_j \Delta$ where $M_j$ is the matrix defined in the main text. Consequently, $A_j$ has a rank of ${|\mathcal{S}_j|}-1$. Thus, the solution to the aforementioned system of equations can be equivalently obtained by solving $x = \Delta^{-1} M_j^{-}D_j$, where $M_j^{-}$ denotes a generalized inverse of $M_j$.
\begin{align*}
    \tilde{d}_j(\bar{z}_j)
    &=  d_j(\bar{z}_j) + E_{Q^0_j}[d_j(\bar{Z}_j){\bar{w}^*_j}^\top(\bar{Z}_j)\mid \bar{z}_{j-1}]{(M_j^{-})}^{\top}(\bar{z}_{j-1}){\bar{w}_j}^{*\top}(\bar{z}_{j};\beta^0_j)r_j(\bar{z}_j;\beta^0_j)
\end{align*}
We $Q^0$-center the above to give the form of $\tilde{d}_j$ in the main text:
\begin{align*}
    &\tilde{d}_j(\bar{z}_j;\beta^0_j)\\
    &=\sum_{m \in \mathcal{S}_j} d_{j,m}(\bar{z}_j;\beta^0_{j,m} )- E_{P^0}\left[\sum_{m \in \mathcal{S}_j} d_{j,m}(\bar{Z}_j;\beta^0_{j,m})\mid  \bar{z}_{j-1}, S \in \mathcal{A}_j\right] \\
    & \quad  + E_{P^0}\big[\sum_{m \in \mathcal{S}_j} d_{j,m}(\bar{Z}_j;\beta^0_{j,m}) {\bar{w}_j}^{*\top}(\bar{Z}_{j};\beta^0_j)\mid  \bar{z}_{j-1}, S \in \mathcal{A}_j\big] \\
    & \hspace{1em}\cdot M^{-}_j(\bar{z}_{j-1};\beta^0_j)^{\top}  \bigg\{{\bar{w}_j}^{*\top}(\bar{z}_{j};\beta^0_j)r_j(\bar{z}_j;\beta^0_j)-E_{P^0}\left[{\bar{w}_j}^{*\top}(\bar{Z}_{j};\beta^0_j)r_j(\bar{Z}_{j};\beta^0_j)\mid  \bar{z}_{j-1}, S \in \mathcal{A}_j\right] \bigg\}.
\end{align*}
The above proof is adapted from the previous work \cite{gilbert1999maximum} and \cite{gilbert2000large}. Since $j$ is arbitrary, we have derived the canonical gradient of $\phi$ relative to $\mathcal{P}_{\mathcal{Q},\beta^0}$. }
\end{proof}

{Note in the proof above, we use $D_{P^0}$ as an initial gradient and proceed with the projection procedure. As an alternative, we can choose to project the gradient that only uses aligned sources $D^\mathcal{A}_{P^0}$ onto the tangent space $\mathcal{T}(P^0,\mathcal{P}_{\mathcal{Q},\beta^0})$.  }

\section{Proof of Lemma~\ref{lem: gradient_beta}}
\label{sec:app:lem_beta}
\begin{proof}[\textit{Proof of Lemma~\ref{lem: gradient_beta}}]
By the variation independence of $\beta^0$ and $Q^0$, Lemma 25.25 of \cite{van2000asymptotic} implies that the canonical gradient of $\beta^0$ under $\mathcal{T}(P^0, \mathcal{P}_{\mathcal{Q},\mathcal{B}})$ is $D_{\beta} = I^{-1}_{\beta^0} \dot{\ell}^*_{\beta}$, where $I_{\beta^0}:=  E_{P^0}[\dot{\ell}_{\beta}^*(Z,S;\beta^0){\dot{\ell}^{*\top}_{\beta}}(Z,S;\beta^0)]$ and $\dot{\ell}^*_{\beta}$ is the efficient score function of $\beta^0$. To find the efficient score $\dot{\ell}^*_{\beta}$, we first compute the score of $\beta^0$ when $Q^0$ is known and then project it onto the orthogonal complement of the nuisance tangent space $\mathcal{T}(P^0, \mathcal{P}_{\mathcal{Q},\beta^0})$ in a model where $Q^0$ is not known. To begin with, the loglikelihood of $(\beta^0,Q^0)$ at a given data point $(z,s)$ is
\begin{align*}
  \ell (\beta^0,Q^0 \mid z,s) &  = \sum_{j=1}^d \Big[\mathbbm{1}(s \in \mathcal{S}_j) \left\{\log w^*_{j,s}(\bar{z}_{j};\beta^0_{j,s}) + \log q_{j}^0(\bar{z}_{j})\right\}  \\
  &\hspace{3em}+ \mathbbm{1}(s \notin \mathcal{S}_j) \log p_j^0(\bar{z}_{j},s) + \log P^0(S=s)\Big].
\end{align*}
For fixed $s \in \mathcal{W}_j$ and $j \in \mathcal{J}$, the score for $\beta^0_{j,s}$, namely $\dot{\ell}_{\beta_{j,s}}(\bar{z}_{j},s';\beta^0_{j,s}):= \nabla_{\beta^0_{j,s}} \ell (\beta^0,Q^0 \mid z,s')$, takes the form

\begin{align*}
     & \dot{\ell}_{\beta_{j,s}}(\bar{z}_{j},s';\beta^0_{j,s}) =  \frac{\dot{w}_{j,s}(\bar{z}_{j},s';\beta^0_{j,s})}{w_{j,s}(\bar{z}_{j},s';\beta^0_{j,s})} -E_{P^0}\left[ \frac{\dot{w}_{j,s}(\bar{Z}_{j},S;\beta^0_{j,s})}{w_{j,s}(\bar{Z}_{j},S;\beta^0_{j,s})}\mid  \bar{z}_{j-1}, S=s'\right].
\end{align*}
For each $\beta^0_{j,s}$, the nuisance tangent space is $\mathcal{T}(P^0, \mathcal{P}_{\mathcal{Q},\beta^0}) = \bigoplus_{j=0}^d \mathcal{T}(P^0, \mathcal{P}_{\mathcal{Q}_j,\beta^0_j})$. The projection of score $\dot{\ell}_{\beta_{j,s}}$ onto $\mathcal{T}(P^0, \mathcal{P}_{\mathcal{Q},\beta^0})$ is, 
\begin{align}
    \Pi_{P^0, {\beta^0}}\{\dot{\ell}_{\beta_{j,s}} \mid \mathcal{T}(P^0, \mathcal{P}_{\mathcal{Q},\beta^0})\}(z,s;\beta^0) 
    & = \sum_{m=0}^d \Pi_{P^0, {\beta^0}}\{\dot{\ell}_{\beta_{j,s}} \mid \mathcal{T}(P^0, \mathcal{P}_{\mathcal{Q}_m,\beta^0_m})\}(z,s;\beta^0) \label{eq:orthogonal_all}\\
    & = \Pi_{P^0, {\beta^0}}\{\dot{\ell}_{\beta_{j,s}} \mid \mathcal{T}(P^0, \mathcal{P}_{\mathcal{Q}_j,\beta^0_j})\}(z,s;\beta^0). \label{eq:orthogonal_beta}
\end{align}
Similar to the arguments in the proof of Lemma~\ref{lem:canonical_beta_known}, Equation~\ref{eq:orthogonal_all} is true by the fact that $\mathcal{Q}$ is nonparametric, and so $\mathcal{T}(P^0, \mathcal{P}_{\mathcal{Q},\beta^0})$ is the orthogonal sum of $\mathcal{T}(P^0, \mathcal{P}_{\mathcal{Q}_m,\beta^0_m})$, $m=0,1,\ldots,d$. 
As we prove below, the latter equality holds since $\dot{\ell}_{\beta_{j,s}}$ is orthogonal to subspaces $\mathcal{T}(P^0, \mathcal{P}_{\mathcal{Q}_m,\beta^0_m})$ for all $j \neq m$, which makes it so that most of the projections are zero.
\begin{proof}[\textit{Proof of Equation~\ref{eq:orthogonal_beta}}.]
Now we show that it is always true that $\dot{\ell}_{\beta_{j}}(P^0) \perp g_{m} \textnormal{ for all } g_{m} \in \mathcal{T}(P^0,\mathcal{P}_{\mathcal{Q}_m,\beta^0_{m}})$ where $m\neq j$, $j\in\mathcal{J}$ and $m\in \mathcal{J}\cup\{0\}$. 
\begin{align*}
        &E_{P^0}\left[\dot{\ell}_{\beta_{j}}(\bar{Z}_j,S) g_m(\bar{Z}_m,S)\right]= 
\begin{cases}
    E_{P^0}\left[\dot{\ell}_{\beta_{j}}(\bar{Z}_j,S) E_{P^0}\left[g_m(\bar{Z}_m,S) \mid \bar{Z}_{m-1},S\right] \right]   =0, &\mbox{ if $j < m$}, \\
    E_{P^0}\left[E_{P^0}\left[\dot{\ell}_{\beta_{j}}(\bar{Z}_j,S)\mid \bar{Z}_{j-1},S\right] g_m(\bar{Z}_m,S)  \right]   =0,&\mbox{if $j > m$.}
\end{cases}        
    \end{align*}
Since $\beta^0_j$ is the concatenation of all $\beta^0_{j,s}$ for $s\in\mathcal{W}_j$, we have shown that for all $s\in\mathcal{W}_j$, $\Pi_{P^0, {\beta^0}}\{\dot{\ell}_{\beta_{j,s}} \mid \mathcal{T}(P^0, \mathcal{P}_{\mathcal{Q}_m,\beta^0_m})\}(z,s';\beta^0) =0$ when $m\neq j$.
\end{proof}
We claim that $\Pi_{P^0, {\beta^0}}\{\dot{\ell}_{\beta_{j,s}} \mid \mathcal{T}(P^0, \mathcal{P}_{\mathcal{Q}_j,\beta^0_j})\}(\bar{z}_j,s';\beta^0_{j}) =\mathbbm{1}_{\mathcal{S}_j}(s') \mathbbm{P}_{j,s'}a^*_j(\bar{z}_j;\beta^0_{j})$, where $a^*_j(\bar{z}_j)$ takes the form in Lemma~\ref{lem: gradient_beta}. The following proof shares great similarity with the one of Lemma~\ref{lem:canonical_beta_known}. . 
We first show that $\mathbbm{1}_{\mathcal{S}_j}(s')\mathbbm{P}_{j,s'}a^*_j \in \mathcal{T}(P^0,\mathcal{P}_{\mathcal{Q}_j,\beta^0_j})$, and then we show $$E_{P^0}[\{\dot{\ell}_{\beta_{j,s}}(\bar{Z}_j,S;\beta^0_{j,s}) - \mathbbm{P}_{j,S}a^*_j(\bar{Z}_j;\beta^0_{j}) \}\mathbbm{P}_{j,S}a_j(\bar{Z}_j;\beta^0_{j})\mid S\in\mathcal{S}_j] = 0$$ for all $a_j \in \mathcal{T}(Q^0, \mathcal{Q}_j)$.

\textbf{Part I of showing that $\Pi_{P^0, {\beta^0}}\{\dot{\ell}_{\beta_{j,s}} \mid \mathcal{T}(P^0, \mathcal{P}_{\mathcal{Q}_j,\beta^0_j})\}(\bar{z}_j,s';\beta^0_{j}) = \mathbbm{1}_{\mathcal{S}_j}(s')\mathbbm{P}_{j,s'}a^*_j(\bar{z}_j;\beta^0_{j})$: $\mathbbm{P}_{j,s}a^*_j \in \mathcal{T}(P^0,\mathcal{P}_{\mathcal{Q}_j,\beta^0_{j,s}})$.} It is straightfoward to show that $E_{Q^0}[a^*_j(\bar{Z}_{j};\beta^0_{j}) \mid \bar{z}_{j-1}] = 0$ everywhere. Since $Q^0$ is nonparametric, we have $a^*_j \in \mathcal{T}(Q^0,\mathcal{Q}_j)$. By Lemma S\ref{lem: tangent_space}, it follows that $\mathbbm{P}_{j,s}a^*_j \in \mathcal{T}(P^0,\mathcal{P}_{\mathcal{Q}_j,\beta^0_j})$ for any  $s \in \mathcal{W}_j$.

\textbf{Part II of showing that $\Pi_{P^0, {\beta^0}}\{\dot{\ell}_{\beta_{j,s}} \mid \mathcal{T}(P^0, \mathcal{P}_{\mathcal{Q}_j,\beta^0_j})\}(\bar{z}_j,s';\beta^0_{j}) = \mathbbm{1}_{\mathcal{S}_j}(s)\mathbbm{P}_{j,s'}a^*_j(\bar{z}_j;\beta^0_{j})$: $E_{P^0}[(\dot{\ell}_{\beta_{j,s}}(\bar{Z}_{j},S;\beta^0_{j,s}) - \mathbbm{P}_{j,S}a^*_j(\bar{Z}_{j};\beta^0_{j}) )\mathbbm{P}_{j,S}a_j(\bar{Z}_{j};\beta^0_{j})\mid S\in\mathcal{S}_j ] = 0$ for all $a_j \in \mathcal{T}(Q^0, \mathcal{Q}_j)$.} We wish to show that:
 \begin{align*}
   0  & = E_{P^0}\left[\left( \dot{\ell}_{\beta_{j,s}}(\bar{Z}_j,S) - \mathbbm{P}_{j,S}a^*_j(\bar{Z}_j) \right)\mathbbm{P}_{j,S}a_j(\bar{Z}_j)\mid S\in\mathcal{S}_j\right]\\
   & = E_{P^0}\bigg[ \sum_{m \in \mathcal{S}_j} \mathbbm{1}(S=m) E_{P^0}\left[\left( \dot{\ell}_{\beta_{j,s}}(\bar{Z}_j,S) - \mathbbm{P}_{j,m}a^*_j(\bar{Z}_j) \right)\mathbbm{P}_{j,m}a_j(\bar{Z}_j)\mid \bar{Z}_{j-1}, S=m\right]\bigg] \\
  &  = E_{P^0}\Bigg[\sum_{m \in\mathcal{S}_j}\mathbbm{1}(S=m) E_{P^0}\bigg[\bigg\{\bigg(\frac{\dot{w}_{j,s}(\bar{Z}_j,m;\beta^0_{j,s})}{w_{j,s}(\bar{Z}_j,m;\beta^0_{j,s})} - a^*_j(\bar{Z}_j) \bigg) \\
  & \hspace{13em}-  E_{P^0}\bigg[\frac{\dot{w}_{j,s}(\bar{Z}_j,m;\beta^0_{j,s})}{w_{j,s}(\bar{Z}_j,m;\beta^0_{j,s})} - a^*_j(\bar{Z}_j)  \mid \bar{Z}_{j-1} ,m \bigg]\bigg\} a_j(\bar{Z}_j) \mid \bar{Z}_{j-1},m \bigg] \Bigg]\\
  & =  \int \int \bigg\{\sum_{m \in\mathcal{S}_j} P^0(S=m)\frac{p^0(\bar{z}_{j-1}\mid S=m)}{q^0(\bar{z}_{j-1})}w^*_{j,m}(\bar{z}_j,m;\beta^0_{j,m})\\
  & \hspace{2em}\cdot \Bigg(\frac{\dot{w}_{j,s}(\bar{z}_j,m;\beta^0_{j,s})}{w_{j,s}(\bar{z}_j,m;\beta^0_{j,s})} - a^*_j(\bar{z}_j)  -E_{P^0}\left[\frac{\dot{w}_{j,s}(\bar{Z}_j,m;\beta^0_{j,s})}{w_{j,s}(\bar{Z}_j,m;\beta^0_{j,s})} - a^*_j(\bar{Z}_j)  \mid \bar{z}_{j-1} ,m \right] \Bigg)\bigg\} a_j(\bar{z}_j) dQ^0(\bar{z}_{j}).
\end{align*}
Again, the expression enclosed within the curly braces must vanish almost everywhere in $L_0^2(Q^0_j)$. Therefore solving the above is equivalent to solving a multidimensional Fredholm integral equation of the second kind:
\begin{align*}
    b_j(\bar{z}_j;\beta^0_j)
    & =a^*_j(\bar{z}_{j};\beta^0_j)  - \int a^*_j(z'_j,\bar{z}_{j-1};\beta^0_j) K(z_j, z'_j, \bar{z}_{j-1};\beta^0_j) dQ^0_j(z'_j\mid\bar{z}_{j-1}) ,
\end{align*}
\begin{sloppypar}
\noindent with kernel $K(z'_j, \bar{z}_{j}):= \sum_{m \in \mathcal{S}_j} r_{j,m}(\bar{z}_j;\beta^0_{j,m}) w^*_{j,m}(z'_j, \bar{z}_{j-1};\beta^0_{j,m})$ and $b_j(\bar{z}_j;\beta^0_j) = \sum_{m\in \mathcal{S}_j} r_{j,m}(\bar{z}_j;\beta^0_{j,m})\dot{\ell}_{\beta_{j,s}}(\bar{z}_{j},m)$. Following similar arguments in proof of Lemma~\ref{lem:canonical_beta_known}, the solution is 
\end{sloppypar}

\begin{align*}
    a^*(\bar{z}_j)
    &=  b_j(\bar{z}_j) + E_{Q^0_j}[b_j(\bar{Z}_j){\bar{w}^*_j}^\top(\bar{Z}_j)\mid \bar{z}_{j-1}]{(M_j^{-})}^{\top}(\bar{z}_{j-1}){\bar{w}_j}^{*\top}(\bar{z}_{j};\beta^0_j)r_j(\bar{z}_j;\beta^0_j)
\end{align*}
We $Q^0$-center the above to give the form of $a^*_j$ in the main text. To this end, we have proved that the projection of score of $\beta^0_{j,s}$ onto the nuisance tangent space is $\Pi_{P^0,\beta^0}\{\dot{\ell}_{\beta_{j,s}} \mid \mathcal{T}(P^0,\mathcal{P}_{\mathcal{Q},\beta^0})\}(\bar{z}_j,s';\beta^0_{j}) = \mathbbm{P}_{j,s'} a^*_j(\bar{z}_j;\beta^0_{j})$ and therefore,  the efficient score function takes the form $\dot{\ell}^*_{\beta_{j,s}}(\bar{z}_j,s';\beta^0_{j,s}) = \dot{\ell}_{\beta_{j,s}}(\bar{z}_j,s';\beta^0_{j,s}) - \mathbbm{P}_{j,s'} a^*_j(\bar{z}_j;\beta^0_{j})$ as specified in Lemma~\ref{lem: gradient_beta}.
Since $j$ and $s$ are arbitrary, we have derived the efficient score function of $\beta^0$ and let us denote it as $\dot{\ell}^*_{\beta}$. 

The canonical gradient of $\beta^0$ is $D^\beta_{P^0} = I^{-1}_{\beta^0} \dot{\ell}^*_{\beta}$.
\end{proof}

We conclude by noting that $I^{-1}_{\beta^0}$ is block-diagonal, which makes the canonical gradient of $\beta^0$ straightforward to compute. 
\begin{proof}[\textit{Proof of $I^{-1}_{\beta^0}$ being block-diagonal}]
    We start by noticing that $\tilde{I}_{\beta^0}:=E_{P^0}[\dot{\ell}_{\beta}(Z,S;\beta^0)\dot{\ell}^\top_{\beta}(Z,S;\beta^0)]$ is block-diagonal since it is straightforward to verify that $\frac{\partial }{\partial \beta_{j} \partial \beta_{m}}\ell(\beta^0,Q^0\mid z,s) =\vec{0}$ when $j\neq m$. Fix $j,m \in \mathcal{J}$ and $j \neq m$, the $\{j,m\}$-block of the Fisher's information matrix under unknown $Q^0$ is 
    \begin{align*}
        & I^{j,m}_{\beta^0}  = E_{P^0}\left[\dot{\ell}^*_{\beta_{j}}(Z,S;\beta^0_{j})\{\dot{\ell}^*_{\beta_{m}}(Z,S;\beta^0_{m})\}^\top\right]\\
        & = E_{P^0}\bigg[\left(\dot{\ell}_{\beta_{j}}(\bar{Z}_j,S;\beta^0_{j}) - \mathbbm{P}_{j,S} a^*_j(\bar{Z}_j;\beta^0_{j})\right) \left(\dot{\ell}_{\beta_{m}}(\bar{Z}_m,S;\beta^0_{m}) - \mathbbm{P}_{m,S} a^*_m(\bar{Z}_m;\beta^0_{m})\right)^\top\bigg]\\
        & = E_{P^0}\left[\dot{\ell}_{\beta_{j}} (\bar{Z}_j,S;\beta^0_{j})\dot{\ell}_{\beta_{m}}^\top(\bar{Z}_m,S;\beta^0_{m})\right] - E_{P^0}\left[\mathbbm{P}_{j,S}a^*_j(\bar{Z}_j;\beta^0_{j})\dot{\ell}^\top_{\beta_{m}}(\bar{Z}_m,S;\beta^0_{m})\right] \\
        & \quad -E_{P^0}\left[\dot{\ell}_{\beta_{j}} (\bar{Z}_j,S;\beta^0_{j})\mathbbm{P}^\top_{m,S}a^*_m(\bar{Z}_m;\beta^0_{m})\right] +E_{P^0}\left[\mathbbm{P}_{j,S}a^*_j (\bar{Z}_j;\beta^0_{j})\mathbbm{P^\top}_{m,S}a^*_m(\bar{Z}_m;\beta^0_{m})\right]\\
        & = 0^{c_j \times c_m}, 
    \end{align*}
    where each of the four terms above is a $c_j \times c_m$ matrix of zeros by tower property.
\end{proof}

\section{Proof of Theorem~\ref{thm: canonical}}
\label{sec:app:lem: canonical}
\begin{proof}[\textit{Proof of Theorem~\ref{thm: canonical}}]
We let { $\mathcal{T}_{\mathcal{B}}(P^0)$ denote the tangent space of $\mathcal{P}_{Q^0,\mathcal{B}}:= \{P_{Q^0,\beta} : \beta\in\mathcal{B}\}$ at $P^0$}. Throughout we'll suppose that the tangent space of $\mathcal{P}_{\mathcal{Q},\mathcal{B}}$ at $P^0$ writes as
\begin{align}
    \mathcal{T}(P^0,\mathcal{P}_{\mathcal{Q},\mathcal{B}}):= \left\{g + h : g\in \mathcal{T}(P^0,\mathcal{P}_{\mathcal{Q},\beta^0}),h\in \mathcal{T}_{\mathcal{B}}(P^0)\right\}. 
\end{align}
We denote the canonical gradient of $\phi$ under $\mathcal{P}_{\mathcal{Q},\mathcal{B}}$ as $D^\mathrm{eff}_{P^0}$. We begin by noting that, if $\{P_\epsilon:=P_{Q_\epsilon,\beta_\epsilon} : \epsilon\}\subset \mathcal{P}_{Q^0,\mathcal{B}}$ is a submodel with $P_{\epsilon=0}=P^0$, score $h\in \mathcal{T}_{\mathcal{B}}(P^0)$, and $Q_\epsilon=Q^0$ for all $\epsilon>0$, then $\psi\circ \theta(P_\epsilon)=\psi\circ \theta(P^0)$ for all $\epsilon$, and so
\begin{align}
     \left.\frac{d}{d\epsilon} \psi\circ \theta(P_\epsilon)\right|_{\epsilon=0}&= P^0 D_{P^0}^\mathrm{eff} h = 0.
     \label{eq:D0h}
\end{align}
We further note that, if $\{P_\epsilon:=P_{Q_\epsilon,\beta_\epsilon} : \epsilon\}\subset \mathcal{P}_{\mathcal{Q},\beta^0}$ is a submodel with $P_{\epsilon=0}=P^0$, score $g$, and $\beta_\epsilon=\beta^0$ for all $\epsilon>0$, then
\begin{align}
    \left.\frac{d}{d\epsilon} \psi\circ \theta(P_\epsilon)\right|_{\epsilon=0}&= P^0 D_{P^0} g = P^0 D_{P^0}^\mathrm{eff} g. \label{eq:D0g}
\end{align}
Combining the preceding two displays shows that the gradient $D_{P^0}^\mathrm{eff}$ must be a function $f\in L_0^2(P^0)$ that satisfies the following:
\begin{align}
    P^0 f g &= P^0 D_{P^0} g\ \textnormal{ for all }g\in \mathcal{T}(P^0,\mathcal{P}_{\mathcal{Q},\beta^0}), \label{eq:fcond1} \\
    P^0 f h&= 0\ \textnormal{ for all }h\in \mathcal{T}_{\mathcal{B}}(P^0). \label{eq:fcond2}
\end{align}
Furthermore, because Lemma S\ref{lem: tangent_space} shows that the submodels used to define \eqref{eq:D0h} and \eqref{eq:D0g} span the tangent space, any function $f\in L_0^2(P^0)$ that satisfies the above two conditions is a gradient. 
We now show that such a function does indeed exist provided the functional $\tilde{\gamma} : \mathcal{P}\rightarrow\mathbb{R}$ defined below is pathwise differentiable relative to $\mathcal{P}_{\mathcal{Q},\mathcal{B}}$:
\begin{align*}
    \tilde{\gamma}(P_{Q,\beta}):= \gamma(\beta).
\end{align*}
We note that $E_{P_{\underline{Q}^0, \beta}}[ \tilde{D}_{P^0}(Z,S;\beta^0)] = E_{P_{Q^0, \beta}}[ \tilde{D}_{P^0}(Z,S;\beta^0)]$ since $Q^0$ and $\underline{Q}^0$ agree on all the relevant conditional distributions, and so $\gamma(\beta)=E_{P_{Q^0, \beta}}[ \tilde{D}_{P^0}(Z,S;\beta^0)]$ for all $\beta$. Let $D^{\tilde{\gamma}}_{P^0}$ denote a gradient of this functional at $P^0$ relative to $\mathcal{P}_{\mathcal{Q},\mathcal{B}}$. Because the right-hand side of the above does not depend on $Q$, we have that, for any submodel $\{P_\epsilon:=P_{Q_\epsilon,\beta_\epsilon} : \epsilon\}\subset \mathcal{P}_{\mathcal{Q},\beta^0}$ with $P_{\epsilon=0}=P^0$, score $g$, and $\beta_\epsilon=\beta^0$ for all $\epsilon>0$, it holds that
\begin{align}
    \left.\frac{d}{d\epsilon} \tilde{\gamma}(P_\epsilon)\right|_{\epsilon=0}&= P^0 D^{\tilde{\gamma}}_{P^0} g = 0. \label{eq:Dtildeg}
\end{align}

We claim that the canonical gradient $D_{P^0}^\mathrm{eff}$ of $\psi\circ\phi$ relative to $\mathcal{P}_{\mathcal{Q},\mathcal{B}}$ is equal to $\tilde{D}_{P^0}-D^{\tilde{\gamma}}_{P^0}$. To show this, it suffices to establish that $\tilde{D}_{P^0}-D^{\tilde{\gamma}}_{P^0}$ belongs to $\mathcal{T}(P^0,\mathcal{P}_{\mathcal{Q},\mathcal{B}})$ and satisfies \eqref{eq:fcond1} and \eqref{eq:fcond2}. To see that $\tilde{D}_{P^0}-D^{\tilde{\gamma}}_{P^0}$ belongs to $\mathcal{T}(P^0,\mathcal{P}_{\mathcal{Q},\mathcal{B}})$, note that $D_{P^0}\in \mathcal{T}(P^0,\mathcal{P}_{\mathcal{Q},\beta^0})$ and $D^{\tilde{\gamma}}_{P^0}\in \mathcal{T}(P^0,\mathcal{P}_{\mathcal{Q},\mathcal{B}})$. Hence by Lemma S\ref{lem: tangent_space}, $\tilde{D}_{P^0}-D^{\tilde{\gamma}}_{P^0} \in \mathcal{T}(P^0,\mathcal{P}_{\mathcal{Q},\mathcal{B}})$.  To see that \eqref{eq:fcond1} is satisfied, note that \eqref{eq:Dtildeg} implies that, for all $g\in \mathcal{T}(P^0,\mathcal{P}_{\mathcal{Q},\beta^0})$, $P^0 [\tilde{D}_{P^0}-D^{\tilde{\gamma}}_{P^0}]g = P^0 \tilde{D}_{P^0} g$. We now show that \eqref{eq:fcond2} is satisfied. Fix a submodel $\{P_\epsilon:=P_{Q_\epsilon,\beta_\epsilon} : \epsilon\}\subset \mathcal{P}_{Q^0,\mathcal{B}}$ is a submodel with $P_{\epsilon=0}=P^0$, score $h\in \mathcal{T}_{\mathcal{B}}(P^0)$, and $Q_\epsilon=Q^0$ for all $\epsilon>0$. For any $\epsilon>0$,
\begin{align*} 
P^0 [\tilde{D}_{P^0}-D^{\tilde{\gamma}}_{P^0}]h&= P^0 \tilde{D}_{P^0} h- P^0D^{\tilde{\gamma}}_{P^0} h \\
    &= P^0 \tilde{D}_{P^0} h - \tilde{\gamma}(P_{Q^0,\beta_\epsilon}) + \tilde{\gamma}(P^0) + o(\epsilon) \\
    &= P^0 \tilde{D}_{P^0} h - P_{Q^0,\beta_\epsilon} \tilde{D} _{P^0} + P^0 \tilde{D}_{P^0} + o(\epsilon) \\
    &= P^0 \tilde{D}_{P^0} h - P_{Q^0,\beta_\epsilon} \tilde{D}_{P^0} + o(\epsilon) \\
    &= P^0 \tilde{D}_{P^0} h - P_\epsilon \tilde{D}_{P^0} + o(\epsilon) \\
    &= o(\epsilon).
\end{align*}
Taking $\epsilon\rightarrow 0$ on both sides shows that $P^0 [\tilde{D}_{P^0}-D^{\tilde{\gamma}}_{P^0}]=0$. As $h$ was arbitrary, \eqref{eq:fcond2} is satisfied. It is straightforward to verify that $D^{\tilde{\gamma}}_{P^0}$ takes the form in Lemma~\ref{lem: gradient_beta} by employing the delta method. 
\end{proof}

\section{{Proof of Corollary~\ref{corollary:rule}}}
\label{sec:app:proof_corollary}
{\begin{proof}[\textit{Proof of Corollary~\ref{corollary:rule}}]
    Note that $\mathcal{T}(P^0,\mathcal{P}_{\mathcal{Q},\mathcal{B}})$ equals the orthogonal sum of $\mathcal{T}(P^0,\mathcal{P}_{\mathcal{Q},\beta^0})$ 
 and $\mathcal{T}_{\mathcal{B}}^*(P^0)$. We have that
    \begin{align*}
        & \mathrm{var}(D^\mathrm{eff}_{P^0}(Z,S)) - \mathrm{var}(D^\mathcal{A}_{P^0}(Z,S))\\
        & = \mathrm{var}(\Pi_{P^0}\{   D^\mathcal{A}_{P^0} \mid \mathcal{T}(P^0,\mathcal{P}_{\mathcal{Q},\mathcal{B}})\}(Z,S))  - \mathrm{var}(D^\mathcal{A}_{P^0}(Z,S))\\
        & = \mathrm{var}(\Pi_{P^0}\{   D^\mathcal{A}_{P^0} \mid \mathcal{T}(P^0,\mathcal{P}_{\mathcal{Q},\beta^0}) \oplus \mathcal{T}^*_{\mathcal{B}}(P^0)\}(Z,S))  - \mathrm{var}(D^\mathcal{A}_{P^0}(Z,S))\\
        & = \mathrm{var}(\Pi_{P^0}\{   D^\mathcal{A}_{P^0} \mid \mathcal{T}(P^0,\mathcal{P}_{\mathcal{Q},\beta^0}) \}(Z,S))  +\mathrm{var}(\Pi_{P^0}\{   D^\mathcal{A}_{P^0} \mid \mathcal{T}^*_{\mathcal{B}}(P^0)\}(Z,S)) - \mathrm{var}(D^\mathcal{A}_{P^0}(Z,S))\\
        & = \mathrm{var}(\tilde{D}_{P^0}(Z,S)) + \mathrm{var}(\Pi_{P^0}\{   D^\mathcal{A}_{P^0} \mid {\mathcal{T}^*_{\mathcal{B}}(P^0)}\}(Z,S))  - \mathrm{var}(D^\mathcal{A}_{P^0}(Z,S))\\
        & = \mathrm{var}(\tilde{D}_{P^0}(Z,S)) + \mathrm{var}(\Pi_{P^0}\{   \tilde{D}_{P^0} +\mathcal{R}^\mathcal{A}_{P^0} \mid \mathcal{T}^*_{\mathcal{B}}(P^0)\}(Z,S))  - \mathrm{var}( \tilde{D}_{P^0}(Z,S)+ R^\mathcal{A}_{P^0}(Z,S)) \tag{by the definition of $\mathcal{R}^\mathcal{A}_{P^0}: = D^\mathcal{A}_{P^0} - \tilde{D}_{P^0}$}\\ 
        & = \mathrm{var}(\tilde{D}_{P^0}(Z,S)) + \mathrm{var}(\Pi_{P^0}\{   \mathcal{R}^\mathcal{A}_{P^0} \mid \mathcal{T}^*_{\mathcal{B}}(P^0)\}(Z,S))  - \mathrm{var}( \tilde{D}_{P^0}(Z,S)+ R^\mathcal{A}_{P^0}(Z,S)) \tag{since $\tilde{D}_{P^0}(Z,S)$ is orthogonal to $\mathcal{T}^*_{\mathcal{B}}(P^0)$}\\   
        & = \mathrm{var}(\tilde{D}_{P^0}(Z,S)) + \mathrm{var}(\Pi_{P^0}\{   \mathcal{R}^\mathcal{A}_{P^0} \mid \mathcal{T}^*_{\mathcal{B}}(P^0)\}(Z,S))  - \mathrm{var}( \tilde{D}_{P^0}(Z,S)) - \mathrm{var}( R^\mathcal{A}_{P^0}(Z,S)) \tag{since $\tilde{D}_{P^0}(Z,S)$ is orthogonal to $\mathcal{T}^*_{\mathcal{B}}(P^0)$}\\  
        & =  \mathrm{var}(\Pi_{P^0}\{   \mathcal{R}^\mathcal{A}_{P^0} \mid \mathcal{T}^*_{\mathcal{B}}(P^0)\}(Z,S))  - \mathrm{var}( R^\mathcal{A}_{P^0}(Z,S)) \tag{since $\tilde{D}_{P^0}(Z,S)$ is orthogonal to $\mathcal{R}^\mathcal{A}_{P^0}(Z,S)$}\\
        & < 0 \textnormal{ if $\mathcal{R}^\mathcal{A}_{P^0}  \notin \mathcal{T}^*_{\mathcal{B}}(P^0) $}.
        \end{align*}
\end{proof}} 
\section{{Examples illustrating additional efficiency gain from weakly aligned sources}}
\label{sec:app:examples of gains}
 {
    \begin{exmp} \textbf{Efficiency gains when estimating the $m^{th}$ central moment for univariate Gaussian distributions.}\\
    Suppose we observe draws of $X = (Z_1,S)$, where $S=1$ is the target population and $S=2$ is a source population. The goal is to estimate the $m^{th}$ central moment $E_{Q^0}[(Z_1 - E_{Q^0}[Z_1])^m]$. Suppose $Z_1$ is normally distributed and we have the prior knowledge that the source population is weakly aligned with the target population with a weight function of $w_{1,2}(z_1;\beta^0_{1,2}) = \exp (\beta^0_{1,2} z_1)$. In general, the incorporation of weakly aligned sources renders efficiency gain when $m>1$. 
    \begin{enumerate}[label=(\alph*)]
        \item When  $m=1$, suppose we are interested in estimating the mean $E_{Q^0}[Z_1]$. There is no efficiency gain from using weakly aligned sources. The proof is straightforward in the special case of $\beta^0 =0 $, which we will illustrate in the following. 
        \begin{proof}
            In this setting, it can be verified that, 
            \begin{align*}
                & D^\mathcal{A}_{P^0} - \Pi_{P^0,\beta^0}\{ D^\mathcal{A}_{P^0}\mid \mathcal{T}(P^0,\mathcal{P}_{\mathcal{Q},\beta^0})\} (z_1,s)\\
                & = \frac{\mathbbm{1}(s=1) - P^0(S=1)}{P^0(S=1)}(z_1 - E_{P^0}[Z_1])\\
                & = \frac{ - \mathbbm{1}(s=2) + P^0(S=2)}{P^0(S=1)}(z_1 - E_{P^0}[Z_1]),
                \intertext{which resides in the linear space spanned by the efficient score function}
                \dot{\ell}^*(z_1,s) & = \left(\mathbbm{1}(s=2) - P^0(S=2)\right)(z_1 - E_{P^0}[Z_1]).
            \end{align*}
        Hence, there is no efficiency gain from incorporating the source data. This can also be verified explicitly by showing that $\mathrm{var}_{P^0}(D^\mathcal{A}_{P^0})$ and $\mathrm{var}_{P^0}(D^\mathrm{eff}_{P^0})$ both equal $\mathrm{var}_{Q^0}(Z_1)/P^0(S=1)$. This example illustrates the case depicted by panel (b) in Figure~\ref{fig:tangent_space}, where the efficiency gain from incorporating weakly aligned sources and the cost of estimating $\beta^0$ cancel out. The same conclusion applies to the general case when $\beta^0\neq0$, since the direction of $\beta^0$ matters instead of the magnitude.
        \end{proof}
        \item When $m=2$ and we are interested in estimating the variance $\sigma^2:= E_{Q^0}[(Z_1 - E_{Q^0}[Z_1])^2]$, using weakly aligned sources renders efficiency gains in general. Indeed, it can be verified that 
        \begin{align*}
            & D^\mathcal{A}_{P^0} - \Pi_{P^0,\beta^0}\{ D^\mathcal{A}_{P^0}\mid \mathcal{T}(P^0,\mathcal{P}_{\mathcal{Q},\beta^0})\} (z_1,s) \\
            &  = \frac{ - \mathbbm{1}(s=2) + P^0(S=2)}{P^0(S=1)}\left[(z_1 - E_{Q^0}[Z_1])^2 -\sigma^2\right],
        \end{align*}
        which does not resides in $\mathcal{T}^*_{\mathcal{B}}(P^0)$. It can also be shown that $\mathrm{var}_{P^0}(D^\mathcal{A}_{P^0}) -\mathrm{var}_{P^0}(D^\mathrm{eff}_{P^0}) = P^0(S=2)/P^0(S=1)E_{Q^0}[(Z_1-E_{Q^0}[Z_1])^4] >0$. A similar argument shows the same conclusion applies to the general case for any $m > 1$.  
    \end{enumerate}
     \end{exmp}
     \begin{exmp}\textbf{Limited efficiency gains for complex weight functions.}\\
        Suppose the same setting with the weight function taking the form of $$w_{1,2}(z_1;\beta^0)= \exp(( z_1,z_1^2,\ldots, z_1^c)^\top\beta^0_{1,2}) $$ for some large $c\in\mathbbm{N}$. 
        We show that, for estimating a general estimand $\phi(P^0)$, the efficiency gain from incorporating weakly aligned data sources diminishes as $c \rightarrow \infty$. The proof is straightforward in the special case of $\beta^0 = 0$, which we will illustrate in the following.
        \begin{proof}
            It can verified that the efficient score function takes the following forms,
        \begin{align*}
            \dot{\ell}^*_{P^0}(z_1,s) & = 
            \begin{bmatrix}
                \left(\mathbbm{1}(s=2) - P^0(S=2)\right)(z_1 - E_{P^0}[Z_1])\\
           \left(\mathbbm{1}(s=2) - P^0(S=2)\right)\left(z_1^2 - E_{P^0}[Z_1^2]\right)\\
            \left(\mathbbm{1}(s=2) - P^0(S=2)\right)\left(z_1^3 - E_{P^0}[Z_1^3]\right)\\
            \vdots \\
            \left(\mathbbm{1}(s=2) - P^0(S=2)\right)\left(z_1^c - E_{P^0}[Z_1^c]\right)
            \end{bmatrix}.
        \end{align*}
        We emphasize on the $c$-dependency of the linear span of the scores of $\beta^0$ that $$\mathcal{T}^*_{\mathcal{B}_c}(P^0) = \{B \times \dot{\ell}^*_{P^0}(Z_1,S) : B \in \mathbbm{R}^c\}.$$ The score functions $\dot{\ell}^*_{P^0}$ can be transformed into Hermite polynomials via the Gram-Schmidt orthogonalization process with respect to the Gaussian weight $p^0$. Since the Hermite polynomials form an orthogonal basis for $L^2_0(P^0)$ that any function $h\in L^2_0(P^0)$ can be expressed as $h(z_1,s) = \sum_{i=1}^\infty e_i H_i(z,s)$ where $H_i$ denote the $i$-th degree Hermite polynomial basis, any functions in $L^2_0(P^0)$ can be approximated well by scores in $\mathcal{T}^*_{\mathcal{B}_c}(P^0)$ as  $c \rightarrow \infty$. As a result, 
         $$\lim_{c \rightarrow \infty}\Pi_{P^0}\{ \mathcal{R}^\mathcal{A}_{P^0}\mid \mathcal{T}^*_{\mathcal{B}_c}(P^0)\} = \Pi_{P^0}\{ \mathcal{R}^\mathcal{A}_{P^0}\mid L^2_0(P^0)\} =\mathcal{R}^\mathcal{A}_{P^0}, $$ 
         where the limit is taken in $L^2(P^0)$. Hence, as the density ratio model becomes arbitrarily complex, the projection residual $ \mathcal{R}^\mathcal{A}_{P^0}$ will essentially belong to $\mathcal{T}^*_{\mathcal{B}}(P^0)$, in the sense that $\mathcal{R}^\mathcal{A}_{P^0}$ is arbitrarily well approximated by its projection onto this space. 
        \end{proof} 
        This conclusion also applies in the general case when $\beta^0\neq 0$, where it can be shown that the score functions $\dot{\ell}^*_{P^0}$ can be transformed into scaled and shifted Hermite polynomials. 
     \end{exmp}}

\section{Form of gradients for estimands in Section~\ref{s:simulation}}
\label{sec:app:simulation_gradients}
Following \cite{kennedy2022semiparametric} and Equation~\eqref{eq:gradient_known}, a gradient of the average treatment effect $\phi(P^0)$ assuming $\beta^0$ is known takes the form of $D_{P^0}(z,s;\beta^0) = D^1_{P^0}(z,s;\beta^0) - D^0_{P^0}(z,s;\beta^0)$ where
\begin{align*}
    & D^a_{P^0}(z,s;\beta^0)\\
    &=  \frac{\mathbbm{1}(s \in \mathcal{S}_3)}{P^0(S \in \mathcal{S}_3)}\frac{\mathbbm{1}(z_2 = a) }{P^0(Z_2=a\mid z_1, S\in\mathcal{S}_3)} \frac{1}{{w^*}_{3,s}(z;\beta^0)}\cdot \frac{p^0(z_1\mid S \in \mathcal{A}_1 )}{p^0(z_1\mid S \in \mathcal{S}_3 )}\\
    & \quad \cdot \left(z_3 - E_{P^0}[Z_3 \mid Z_2 = a, z_1, S \in \mathcal{A}_3]\right) \\
    & \quad + \frac{\mathbbm{1}(s \in \mathcal{S}_1)}{P^0(S \in \mathcal{S}_1)} \frac{1}{{w^*}_{1,s}(z;\beta^0)}\left(E_{P^0}[Z_3 \mid Z_2 = a, z_1, S \in \mathcal{A}_3]- \phi_1(P^0)\right)
\end{align*}
Following Theorem~\ref{thm: canonical} gives the form of the canonical gradient.
\section{{Simulation setup}}
\label{sec:app:simulation_continued}
\begin{figure}[H]
    \centering
    \includegraphics[width=\linewidth]{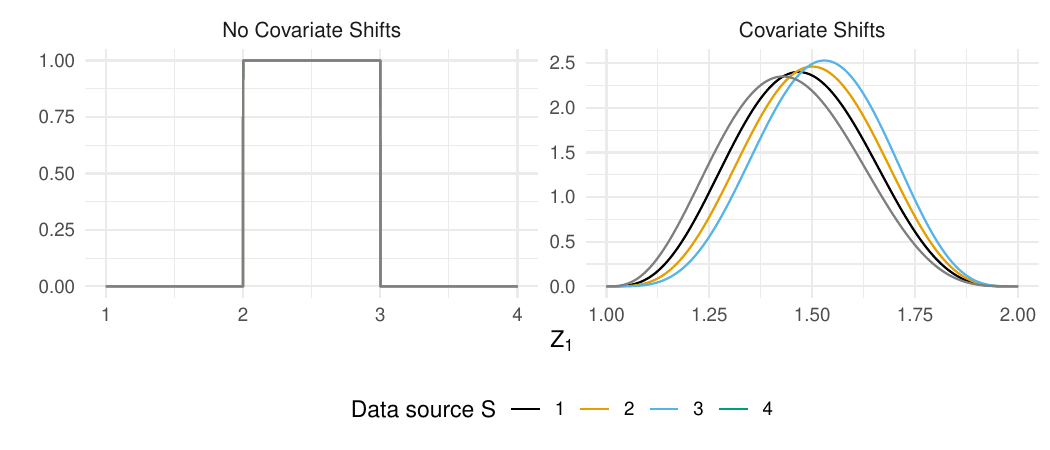}
    \caption{{The density plots of $Z_1$ under no shifts (left) and shifts (right).}}
    \label{fig:z_1 density}
\end{figure}

\begin{figure}[H]
    \centering
    \includegraphics[width=\linewidth]{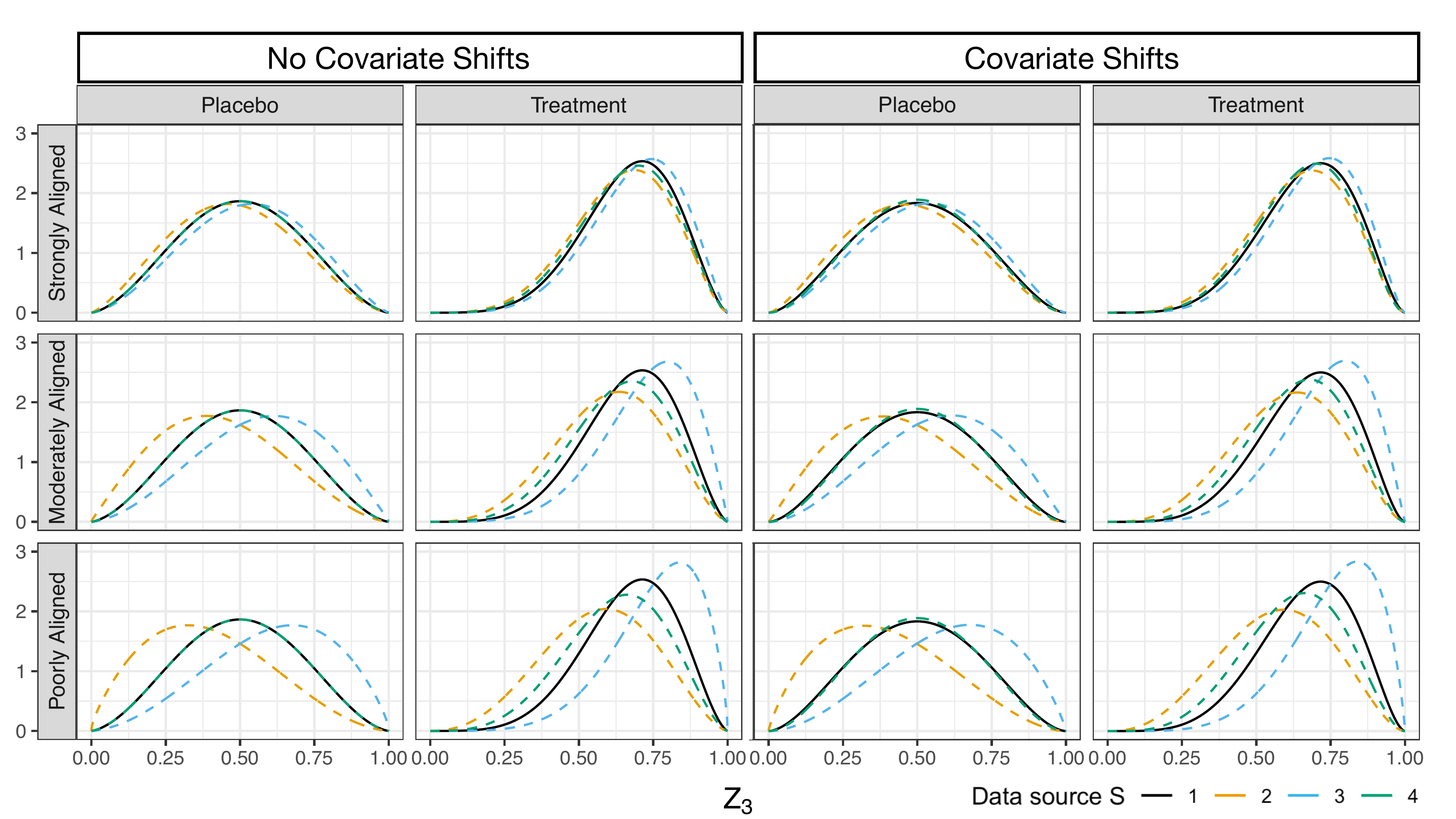}
    \caption{{The marginal density plots of $Z_3$ for each treatment group and data source under no covariates shifts (left), and covariate shifts (right). }}
    \label{fig:shifts_visual}
\end{figure}

{Table~\ref{tab:quantiles} reports various quantiles of the values of $\lambda^\dagger_3$ for each weakly aligned source $s$. In the absence of covariate shifts, $\lambda^\dagger_3$ is only upper bounded but not lower bounded. In fact, it tends to zero when $Z_3$ goes to zero. When running a Monte Carlo experiment with $n=10^8$, we observe the ranges of $\lambda_3^\dagger$ to be $[0.16, 1.42]$, $[0.14, 1.56]$ and $[0.38, 1.19]$ for data sources $S=2,3,4$ respectively. This violation in the lower bounds is further exacerbated by the increasing misalignment in $Z_3$, where we observe the ranges to be $[0.001, 4.08]$, $[0.002, 6.04]$ and $[0.039, 1.89]$ in the case of poor alignment. When there are shifts in $Z_1$, $\lambda^\dagger_3$ are unbounded. The bounds are violated more severely when $Z_3$ is poorly aligned (Figures~\ref{fig:z_1 density} and \ref{fig:shifts_visual}), in which we see the ranges of $\lambda^\dagger_3$ to be $[0.002, 855]$, $[0.001, 402]$ and $[0.04, 776]$ for data sources $S=2,3,4$ respectively.} 
\begin{table}[H]
\begin{center}
\caption{{Quantiles of $\lambda^\dagger_{3}(\bar{Z}_3,S)$ across different levels of alignment in $Z_3$ and $Z_1$.}} \label{tab:quantiles}
\begin{adjustbox}{width=\textwidth} 
{\begin{tabular}{lrrrrrrrrrrrr}
\toprule
       & \multicolumn{6}{c}{No shifts in $Z_1$} & \multicolumn{6}{c}{Shifts in $Z_1$}  \\
    \cmidrule(l){2-7} \cmidrule(l){8-13}
    &  $0.01^{\mathrm{th}}$ & $5^{\mathrm{th}}$ & $10^{\mathrm{th}}$&  $90^{\mathrm{th}}$ & $95^{\mathrm{th}}$ & $99.99^{\mathrm{th}}$&  $0.01^{\mathrm{th}}$ & $5^{\mathrm{th}}$ & $10^{\mathrm{th}}$&  $90^{\mathrm{th}}$ & $95^{\mathrm{th}}$ & $99.99^{\mathrm{th}}$\\
    \midrule
    \textbf{Strongly Aligned }\\
    \hspace{1em} $ S=2$ & 0.39 & 0.77 & 0.83 & 1.18 & 1.21 & 1.36 & 0.33 & 0.67 & 0.74 & 1.77 & 2.19 & 16.31\\
    \hspace{1em} $ S=3$ & 0.40 & 0.77 & 0.83 & 1.19 & 1.23 & 1.45 & 0.33 & 0.67 & 0.74 & 1.78 & 2.19 & 16.21\\
    \hspace{1em} $ S=4$ & 0.64 & 0.91 & 0.95 & 1.07 & 1.09 & 1.16 & 0.50 & 0.72 & 0.76 & 1.72 & 2.13 & 15.91\\
    \textbf{Moderately Aligned }\\
    \hspace{1em} $ S=2$ & 0.10 & 0.54 & 0.67 & 1.60 & 1.71 & 2.32 & 0.09 & 0.52 & 0.65 & 2.09 & 2.57 & 18.60\\
    \hspace{1em} $ S=3$ & 0.10 & 0.54 & 0.66 & 1.63 & 1.76 & 2.70 & 0.10 & 0.52 & 0.64 & 2.12 & 2.61 & 18.91\\
    \hspace{1em} $ S=4$ & 0.33 & 0.79 & 0.89 & 1.19 & 1.25 & 1.48 & 0.29 & 0.70 & 0.75 & 1.76 & 2.18 & 16.20\\
    \textbf{Poorly Aligned }\\
    \hspace{1em} $ S=2$ & 0.04 & 0.45 & 0.60 & 2.04 & 2.25 & 3.51 & 0.04 & 0.44 & 0.60 & 2.51 & 3.06 & 21.56\\
   \hspace{1em} $ S=3$ &  0.04 & 0.45 & 0.59 & 2.09 & 2.35 & 4.32 & 0.04 & 0.44 & 0.59 & 2.59 & 3.17 & 21.99\\
    \hspace{1em} $ S=4$ & 0.21 & 0.73 & 0.86 & 1.30 & 1.38 & 1.76 & 0.19 & 0.67 & 0.74 & 1.82 & 2.24 & 16.52\\
\bottomrule
\end{tabular}}
\end{adjustbox}
\end{center}
\end{table}


\section{{Misspecification of density ratio models}}
\label{sec:app:sens}
 \subsection{{Sensitivity analysis}}
       {
       We propose a sensitivity analysis procedure following the general approach outlined in \cite{diaz2013sensitivity} and \cite{luedtke2015statistics}. We define a bias term that quantifies the difference between the conditional average treatment effect for the target population and the one using weakly aligned sources with misspecified weight functions. Specifically, the bias function is
       \begin{align*}
           u(Z_1,S) := & E_{Q^0}[Z_3\mid Z_2=1,Z_1] - E_{Q^0}[Z_3\mid Z_2=0,Z_1] \\
           & -E_{P^0}\left[\frac{E_{Q^0}[\tilde{w}_{3,S}(\bar{Z}_3) \mid \bar{Z}_2]}{\tilde{w}_{3,S}(\bar{Z}_3)}Z_3\mid Z_2=1,Z_1,S=S\right] \\
           &  +  E_{P^0}\left[\frac{E_{Q^0}[\tilde{w}_{3,S}(\bar{Z}_3) \mid \bar{Z}_2]}{\tilde{w}_{3,S}(\bar{Z}_3)}Z_3\mid Z_2=0,Z_1,S=S\right],
       \end{align*} 
       where we use $\tilde{w}_{3,S} \neq w_{3,S}$ to denote the misspecified weight function for data source $S$. When Condition~\ref{weak_alignment} holds, $u(x,s) = 0$ for all $x$ and $s \in \mathcal{W}_3$. When Condition~\ref{weak_alignment} is violated, $u(X,S) \neq 0$ and is unknown. To perform sensitivity analyses, we suppose
       \begin{align}
           - \delta \leq u(X,S) \leq \delta \label{eq:conditionDelta}
       \end{align}
       with probability 1 over draws of $X$ and $S$ for some $\delta\ge 0$. Then it can be verified that $$- \delta \leq \psi(Q^0) - \Phi(P^0) \leq \delta,$$ where $\Phi(P^0)$ is defined using the misspecified weight functions $\tilde{w}_{3,S}$. 

       We provide the exact definition of $\Phi(P^0)$ in the next subsection, and show that $\Phi(P^0)$ is a KL-based projection estimand. Given that \eqref{eq:conditionDelta} holds, a valid 95\% confidence interval for $\psi(Q^0)$ is: $$[L_n - \delta, U_n + \delta],$$  where $L_n, U_n$ is the lower and upper 95\% confidence interval of $\Phi(P^0)$. Figure~\ref{fig:sens_delta} below shows how the 95\% confidence interval is impacted by different choices of $\delta$. While the confidence interval for the target-only estimator remains the same across different values of $\delta$, the ones for the data fusion estimator becomes wider as the magnitude of $\delta$ grows. For example, in the simulation setting from Section~\ref{s:simulation} with $\epsilon=0.2$, the fusion estimator has a wider CI compared to the target-only estimator when $\delta$ exceeds around 0.005. Hence, once the potential for heavily misspecified weight functions is taken into account, leveraging weakly aligned sources may widen confidence intervals.  
    \begin{figure}[H]
        \centering
        \includegraphics[scale = 0.2]{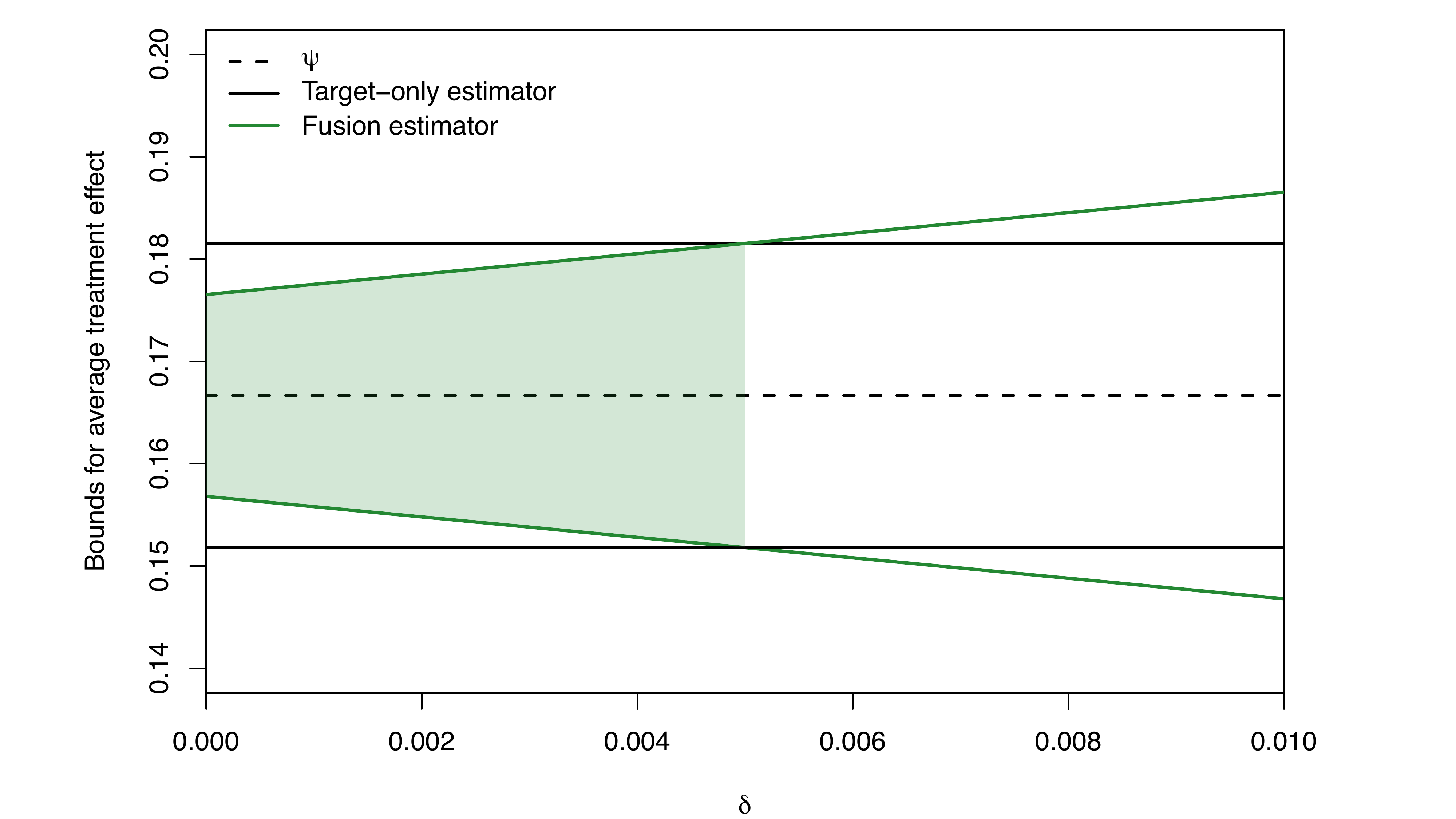}
        \caption{{95\% upper and lower confidence bounds for the average treatment effect across values of $\delta$ when data sources are strongly aligned ($\epsilon = 0.2$ in the simulation setting specified in Section~\ref{s:simulation}). The fusion estimator uses the fusion sets $\{0,1,2,3\}$. }}
        \label{fig:sens_delta}
    \end{figure}
    \subsection{{Estimand under misspecification}}
    \label{sec:app:sens_projection}
    Estimators under misspecified models often converge to projection-based estimands that minimize a population objective function \citep{muller2013risk,hansen2021inference}. In our case, when the weight functions are misspecified, our target parameter $\Phi(P^0)$ is a projection-based estimand. Without loss of generality, we assume that there is only one weakly aligned source and thus omit the dependence of weight function $w$ on data source $s$. For clarity, we focus on the case where only the weight function for the conditional distribution $Z_j$ is misspecified for a fixed $j \in [d]$, and denote it as $\tilde{w}_j(z;\gamma^0)$ for some function $\tilde{w}_j \neq w_j$ and $\gamma^0 \in \mathbbm{R}^c$. Then, our target estimand becomes  $\Phi: \mathcal{P}_{\mathcal{Q},\mathcal{B}} \rightarrow \mathbbm{R}$ defined as:
        $$\Phi := \phi \hspace{.1em}\circ \Pi,$$
        where the projection $\Pi$ maps a given distribution in $P \in \mathcal{P}_{\mathcal{Q},\mathcal{B}}$ to its best approximation $P_{Q,\gamma}: = \argmin_{\tilde{P} \in \mathcal{P}_{\mathcal{Q},\gamma}}\mathrm{KL}(P, \tilde{P})$  in 
        \begin{align*}
            \mathcal{P}_{\mathcal{Q},\gamma}:=\Big\{ \tilde{P} & \textnormal{ such that }\tilde{P}_i \in \mathcal{P}_{\mathcal{Q}_i,\beta_i} \textnormal{ for } i\neq j \textnormal{ and}, \\
            & \tilde{P}_j(\cdot \mid \bar{z}_{j-1}) = \frac{\tilde{w}_j(z;\gamma)}{E_{Q}[\tilde{w}_j(Z;\gamma) \mid \bar{z}_{j-1}]} Q_j(\cdot \mid \bar{z}_{j-1})\Big\}.
        \end{align*}
        The above is true since the true finite-dimensional parameter $\gamma^0$ maximizes the following log-likelihood, 
            \begin{align*}
                \gamma^0 
                & = \argmax_{\gamma \in \mathbbm{R}^c,\tilde{P}_\gamma \in \mathcal{P}_{\mathcal{Q},\gamma} }E_{P^0}\left[\log \frac{ d\tilde{P}_{\gamma}(Z)}{dP^0(Z)} \right]\\
                & = \argmin_{\gamma \in \mathbbm{R}^c,\tilde{P}_\gamma \in \mathcal{P}_{\mathcal{Q},\gamma} }E_{P^0}\left[\log \frac{dP^0(Z)}{d\tilde{P}_{\gamma}(Z)} \right]\\
                & = \argmin_{\gamma \in \mathbbm{R}^c,\tilde{P}_\gamma \in \mathcal{P}_{\mathcal{Q},\gamma} } \mathrm{KL}(P^0, \tilde{P}_\gamma)
            \end{align*}
        Therefore, under model misspecification, our estimand becomes $\Phi(P^0) = \phi(P_{Q^0,\gamma^0})$, where $P_{Q^0,\gamma^0}$ is the KL-projection of $P^0$ onto $\mathcal{P}_{\mathcal{Q},\gamma}$.\\~\\
        \subsection{{Simulation}}
        We examine the performances of the proposed estimators under misspecified weight functions. We focus on the setting where covariate $Z_1$ is identically distributed across all data sources, and the shifts in conditional distribution of the outcome vary. As a result, the KL-divergence between $P^0$ and $P_{Q^0,\gamma^0}$ is fully attributable to the conditional KL-divergence of the conditional outcome distributions. We misspecify the weight functions in two ways:  (i) interchanging weight functions between sources such that each source ends up with a wrong weight function, and (ii) assuming all weakly aligned sources $S=\{2,3,4\}$  are fully aligned. Specifically under setting (i),  we supply $w_{3,4}$ for data source $S=2$, $w_{3,2}$ for data source $S=3$, and $w_{3,3}$ for data source $S=4$. We create a sliding window by varying the extent of outcome shifts via $\epsilon$. For each scenario, 1000 Monte Carlo replications were conducted.\\~\\
        In scenario (i) where misspecified weight functions are supplied to each data sources,  
        treating weakly aligned sources as aligned sources leads to biased estimates and compromises coverage, especially when $\epsilon$ is large (Figure~\ref{fig:coverage_sens}). Among the four settings considered, the penalty associated with incorrectly treating data source $S=2$ as aligned is the least, since its source-specific average treatment effect is the closest to the target one. When supplying misspecified weight functions, coverage similarly decreases with the degree of misalignment. The only exception to this occurs when assuming weak alignment of data source $S=3$, where the degree of misalignment is mild enough to not harm coverage much.
         
        }
        \begin{figure}[H]
            \centering
            \includegraphics[scale = 0.7]{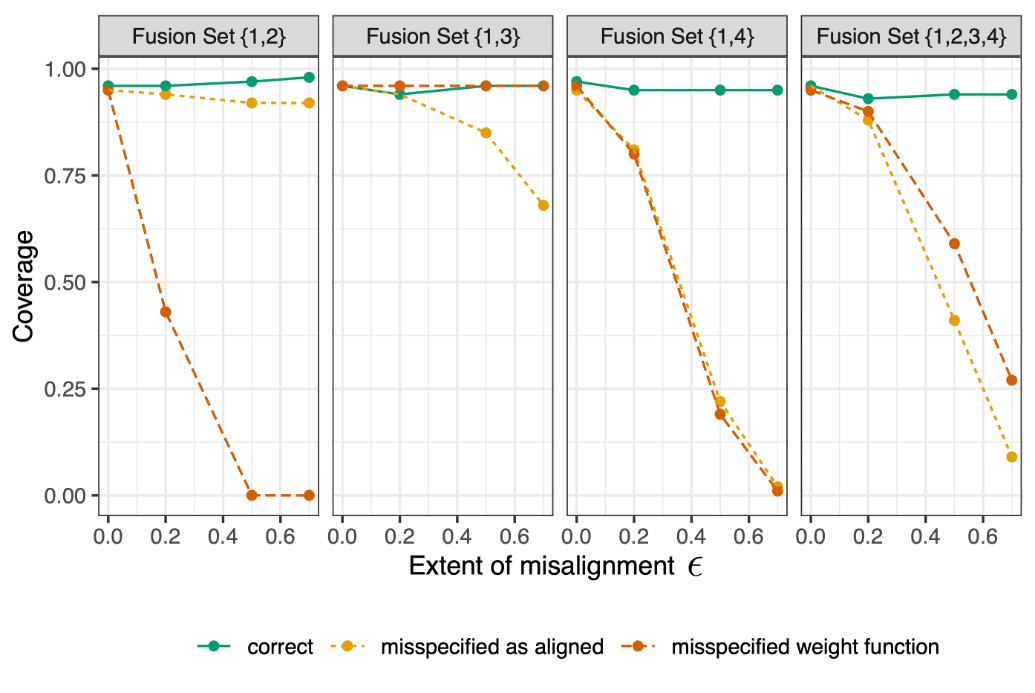}
            \caption{{Coverage of the proposed efficient fusion estimators under misspecified weight functions. Misalignment is greater when $\epsilon$ is larger.}}
            \label{fig:coverage_sens}
        \end{figure}

        {Under the setting of (ii) where all data sources are treated as aligned sources, the projection-based estimand becomes an average of source-specific average treatment effect over the fusion set. As detailed in Section~\ref{sec:app:sens_projection}, the KL-projection of $P^0$ onto $\mathcal{P}_{\mathcal{Q},\gamma}$ is simply  $P^0(Z_3\mid \bar{Z}_2, S\in\mathcal{S}_3)$  due to the equal size of all data sources. The proposed estimators give valid inference for this projection-based estimand as shown in Table~\ref{tab:results_projection}.}
        \begin{table}[H]
        \begin{center}
        {\caption{Bias, variance and coverage of fusion estimators for estimating the projection-based estimand of the target average treatment effect under setting (ii). Bias$^2$ and variance are scaled up by $10^5$ for clarity.}
        \label{tab:results_projection}
        \begin{adjustbox}{width=1\textwidth} 
        \begin{tabular}{lrrrrrrrrr}
        \toprule
               & \multicolumn{3}{c}{Strongly Aligned} & \multicolumn{3}{c}{Moderately Aligned} & \multicolumn{3}{c}{Poorly Aligned}  \\
            \cmidrule(l){2-4} \cmidrule(l){5-7} \cmidrule(l){8-10}
            Fusion set $\mathcal{S}_3$&  Bias$^2$ & Var & Coverage &  Bias$^2$ & Var & Coverage  &  Bias$^2$ & Var & Coverage\\
            \midrule
            $\{1,2\}$ &  0.00 & 3.23 & 0.95 & 0.00 & 3.72 & 0.95 & 0.00 & 4.44 & 0.93\\
            $\{1,3\}$ & 0.00 & 3.30 & 0.95 & 0.01 & 3.60 & 0.94 & 0.01 & 4.08 & 0.93\\
            $\{1,4\}$ &0.00 & 2.92 & 0.96 & 0.00 & 3.01 & 0.95 & 0.00 & 3.05 & 0.95\\
            $\{1,2,3,4\}$ & 0.00 & 1.65 & 0.95 & 0.00 & 1.73 & 0.95 & 0.00 & 1.88 & 0.95\\
         \bottomrule
        \end{tabular}
        \end{adjustbox}}
        \end{center}
        \end{table}
        

\section{Construction of estimators}
\label{sec:app:regularity}
{We propose the following procedure for constructing an one-step estimator using the canonical gradient derived in Theorem~\ref{thm: canonical}. To begin with, we estimate components of $P^0$ that are necessary for evaluating $D^\mathrm{eff}_{P^0}$, including density ratios and conditional expectations, as well as $\beta^0$. For the former two, we recommend using data-adaptive methods for flexible modeling to avoid misspecification. We detail the requirements on the convergence rates of the nuisance estimators in the next few paragraphs. 

For any weakly aligned data source $s\in\mathcal{W}_j$ and $j\in\mathcal{J}$, we can leverage the known form of weight function $w_{j,s}$, which relates $P^0(\cdot \mid z_{j-1},s)$ and $Q^0(\cdot \mid z_{j-1})$, to estimate the corresponding $\beta^0_{j,s}$. Specifically, we recommend using moment matching to obtain an initial estimate. This entails 
matching the moments of various basis functions associated with $\beta^0_{j,s}$ in the weight function $w_{j,s}$. When doing so, one can plug in an estimate for the corresponding normalizing function $W_{j,s}$. In order to ensure its consistency, we further construct an one-step estimator of $\beta^0$ using $D_{P^0}^\beta$ introduced in Lemma~\ref{lem: gradient_beta}. Through simulations, we found that using data adaptive estimates for $W_{j,s}$ often leads to a consistent initial estimate of $\beta^0$, leading to minimal differences between using the initial estimate alone or applying an additional one-step correction. }

{The one-step estimator will be asymptotically linear with the efficient influence function $D^\mathrm{eff}_{P^0}$ if
    \begin{enumerate}
    \item the remainder term $R(\widehat{P},P^0):= \phi(\widehat{P}) - \phi(P^0) + E_{P^0}\{D^\mathrm{eff}_{\widehat{P}}(X)\}$ is $o_p(n^{-1/2})$, and
    \item the empirical mean of $D^\mathrm{eff}_{\widehat{P}}(X)-D^\mathrm{eff}_{P^0}(X)$ is within $o_p(n^{-1/2})$ of the mean of this term when $X\sim P^0$.
    \end{enumerate}
 When the above two conditions hold, the one-step estimator $\hat{\phi}$ will be efficient among all regular estimators \citep{bickel1993efficient}. In order to satisfy the two conditions above,  we need to estimate the nuisance parameters sufficiently well and satisfy appropriate empirical process and consistency conditions. We refer reader's attention to conditions (2.2), (2.3) and (2.6) in \cite{klaassen2005efficient} for the necessary conditions needed. Lemmas (2.1), (2.2) and (2.3) in \cite{klaassen2005efficient} are also helpful. We provide such conditions when the target estimand is the average treatment effect and follow the notations introduced in Section~\ref{s:simulation}. In addition, we denote $\pi_{z_2}(z_1): = P^0(Z_2 =z_2\mid Z_1, S\in\mathcal{S}_3)$, $\mu(z_2,Z_1): = E_{Q^0}[Y\mid Z_2 = z_2, Z_1]$ and let $\hat{\pi}_{z_2}(z_1)$ and $\mu_{\hat{P}}(z_2,Z_1)$ denote the corresponding estimates using observed data. We use $M^-_{a,b}$ to denote the entry of $M^-$ that corresponds to data source $a$ and $b$. 
 \setcounter{cond}{2}
 \begin{cond}  We denote the $L^2_0(P^0)$ norm as $\Vert \cdot \Vert$. Under the following conditions, the remainder term satisfies $R(\hat{P}, P^0) = o_p(n^{- 1/2})$.
     \begin{enumerate}[label=\textbf{3\alph*},ref=3\alph*,leftmargin=*] 
        \item \label{cond:rem1} ${\left\Vert \mu(z_2,Z_1) - \mu_{\hat{P}}(z_2,Z_1)\right\Vert} \left\{{\left\Vert \hat{\lambda}_{1}(Z_1)-\lambda_{1}(Z_1)\right\Vert} +{\left\Vert \hat{\pi}_{z_2}(Z_1)-\pi_{z_2}(Z_1)\right\Vert}\right\} = o_p(n^{-1/2})$ for $z_2 = \{0,1\}$.
        \item \label{cond:rem2}  ${\left\Vert \mu(z_2,Z_1) - \mu_{\hat{P}}(z_2,Z_1)\right\Vert}{\left\Vert \hat{w}^*_{j,s}(\bar{Z}_j;\hat{\beta}_{j,s}) - w^*_{j,s}(\bar{Z}_j;\beta^0_{j,s}) \right\Vert} = o_p(n^{-1/2})$ for $z_2 = \{0,1\}$, $s\in\mathcal{S}_j$ and $j\in\{1,3\}$.   
        \item \label{cond:rem3} $\Vert \sum_{s\in\mathcal{S}_j} E_{\hat{P}_s}\left[\hat{r}_j(\bar{Z}_j;\hat{\beta})(Z_j - \mu_{\hat{P}_j})\mid \bar{Z}_{j-1}\right] \cdot$\\
         $\left(\Delta^{-1} - E_{\hat{P}}[\hat{r}_j(\bar{Z}_j;\hat{\beta}_j)\bar{\hat{w}}^*_j(\bar{Z}_j;\hat{\beta}_j){\bar{\hat{w}}^*_j(\bar{Z}_j;\hat{\beta}_j)}^\top \mid \bar{Z}_{j-1}, S\in\mathcal{A}_j]\right)^-_{sm} \Vert $\\ $ {\left\Vert  \frac{\hat{p}(\bar{Z}_j \mid S=m)}{\hat{p}(\bar{Z}_j \mid S \in \mathcal{S}_j)} - \frac{p^0(\bar{Z}_j \mid S=m)}{p^0(\bar{Z}_j \mid S \in \mathcal{S}_j)}\right\Vert} = o_p(n^{-1/2})$ for  $m\in\mathcal{S}_j$ and $j \in \{1,3\}$.
        \item \label{cond:rem4} ${\left\Vert r_{j,m}(\bar{Z}_j) - \hat{r}_{j,m}(\bar{Z}_j)\right\Vert}{\left\Vert E_{P^0}\left[\frac{\dot{w}_{j,m}(\bar{Z}_j;\hat{\beta}_{j,m})}{w_{j,m}(\bar{Z}_j;\hat{\beta}_{j,m})}\mid \bar{Z}_{j-1}, S\right]  - E_{P^0}\left[\frac{\dot{w}_{j,m}(\bar{Z}_j;\beta^0_{j,m})}{w_{j,m}(\bar{Z}_j;\beta^0_{j,m})} \mid \bar{Z}_{j-1}, S\right]\right\Vert} = o_p(n^{-1/2})$ for  $m\in\mathcal{S}_j$ and $j \in\{1,3\}$.      
        \item \label{cond:rem5} ${\left\Vert r_{j,m}(\bar{Z}_j) - \hat{r}_{j,m}(\bar{Z}_j)\right\Vert}{\left\Vert E_{\hat{P}}\left[\frac{\dot{w}_{j,m}(\bar{Z}_j;\hat{\beta}_{j,m})}{w_{j,m}(\bar{Z}_j;\hat{\beta}_{j,m})} \mid \bar{Z}_{j-1}, S\right] - E_{P^0}\left[\frac{\dot{w}_{j,m}(\bar{Z}_j;\beta^0_{j,m})}{w_{j,m}(\bar{Z}_j;\beta^0_{j,m})} \mid \bar{Z}_{j-1}, S\right]\right\Vert} = o_p(n^{-1/2})$ for  $m\in\mathcal{S}_j$ and $j \in\{1,3\}$.   
        \item \label{cond:rem6} $\Vert \sum_{s\in\mathcal{S}_j} E_{\hat{P}_s}\left[\hat{a}_j (\bar{Z}_j;\hat{\beta}_{j})\mid \bar{Z}_{j-1}\right] \cdot  $\\
         $\left(\Delta^{-1} - E_{\hat{P}}[\hat{r}_j(\bar{Z}_j;\hat{\beta}_j)\bar{\hat{w}}^*_j(\bar{Z}_j;\hat{\beta}_j){\bar{\hat{w}}^*_j(\bar{Z}_j;\hat{\beta}_j)}^\top \mid \bar{Z}_{j-1}, S\in\mathcal{A}_j]\right)^-_{sm} \Vert $ \\$\cdot {\left\Vert \frac{\hat{p}(\bar{Z}_j \mid S=m)}{\hat{p}(\bar{Z}_j \mid S \in \mathcal{S}_j)} - \frac{p^0(\bar{Z}_j \mid S=m)}{p^0(\bar{Z}_j \mid S \in \mathcal{S}_j)}\right\Vert} = o_p(n^{-1/2})$ for  $m\in\mathcal{S}_j$ and $j \in \{1,3\}$. 
        \end{enumerate}
 \end{cond}
 The remainder term of the one-step estimator constructed via $\tilde{D}_{\hat{P}}$ is $o_p(n^{-1/2})$ under Conditions S\ref{cond:rem1}, S\ref{cond:rem2} and S\ref{cond:rem3}. Specifically, if the estimated conditional outcome regressions, propensity scores and density ratio of $Z_1$ achieves the rate of $o_p(n^{-1/4})$, then Condition S\ref{cond:rem1} is satisfied. It can be also verified that if the normalizing functions $W_{j,m}$ and conditional regressions in Condition~S\ref{cond:rem2} and S\ref{cond:rem3} can be both estimated at $n^{-1/4}$ rate, then Conditions~S\ref{cond:rem2} and S\ref{cond:rem3} will be satisfied. In the meantime, Conditions~S\ref{cond:rem4} to S\ref{cond:rem6} ensures the remainder term associated with estimating $\beta^0$ is $o_p(n^{-1/2})$. We recommend using flexible data-adaptive models for estimating these nuisance parameters to achieve the required rates. 
 \begin{proof}
    The remainder term writes as 
    \begin{align*}
        & R(\hat{P},P^0)\\
        & = \hat{\phi} - \phi + E_{P^0}\left[\tilde{D}_{\hat{P}}(Z,S;\hat{\beta}) - E_{\hat{P}}[\tilde{D}_{\hat{P}}(Z,S;\hat{\beta}) \dot{\ell}_{\hat{P}}(Z,S;\hat{\beta})] D^\beta_{\hat{P}}(Z,S;\hat{\beta})\right].
    \end{align*}
We will study the remainder term for the average treatment effect $\phi:=\phi_1 -\phi_0$ where $\phi_a: = E_{P^0}[[Z_3\mid Z_2 = a, Z_1, S\in\mathcal{A}_3]\mid S\in\mathcal{A}_1]$. For clarity, we focus on estimating $\phi_1$, as the same results apply to the estimation of $\phi_0$. We first study $R_1$, which corresponds to the remainder term associated with estimating $\phi_1$ under a fixed $\hat{\beta}$:
   \begin{align*}
       &R_1(\hat{P},P^0)\\
       & =\hat{\phi}_1 - \phi_1 + E_{P^0}\left[\tilde{D}_{\hat{P}}(Z,S;\hat{\beta})\right]\\ 
       & = \hat{\phi}_1 - \phi_1 + E_{P^0}\left[\sum_{j\in\mathcal{J}} \mathbbm{1}(S \in \mathcal{S}_j)\{\tilde{d}_j(\bar{Z}_j;\hat{\beta}_j) -E_{\hat{P}}[\tilde{d}_j(\bar{Z}_j;\hat{\beta}_j)\mid \bar{Z}_{j-1},S]\}\right]\\ 
       & = \hat{\phi}_1 - \phi_1 \\
       & \quad + E_{P^0}\left[\sum_{j\in\mathcal{J}}\{\tilde{d}_j(\bar{Z}_j;\hat{\beta}_j) -E_{\hat{P}}[\tilde{d}_j(\bar{Z}_j;\hat{\beta}_j)\mid \bar{Z}_{j-1},S \in \mathcal{S}_j]\} \mid S\in\mathcal{S}_j\right]P^0(S\in\mathcal{S}_j).
\end{align*}
For each $j\in\mathcal{J}$, we can write $ \tilde{d}_j(\bar{Z}_j;\hat{\beta}_j) -E_{\hat{P}}[\tilde{d}_j(\bar{Z}_j;\hat{\beta}_j)\mid \bar{Z}_{j-1},S \in \mathcal{S}_j] = (I)_j + (II)_j$ where 
\begin{align*}
    (I)_j&= \hat{r}_j(\bar{Z}_j;\hat{\beta}_j)D_{\hat{Q},j}(\bar{Z}_j) - E_{\hat{P}}[\hat{r}_jD_{\hat{Q},j}(\bar{Z}_j) \mid \bar{Z}_{j-1},S\in\mathcal{S}_j]  \\
    & = \hat{r}_j(\bar{Z}_j;\hat{\beta}_j)D_{\hat{Q},j}(\bar{Z}_j) \\
    (II)_j&= E_{\hat{P}}\left[ \hat{r}_j(\bar{Z}_j;\hat{\beta}_j)D_{\hat{Q},j}(\bar{Z}_j) {\bar{\hat{w}}^*_j}^\top(\bar{Z};\hat{\beta})\mid \bar{Z}_{j-1}, S\in\mathcal{A}_j\right]\hat{M}^-_j(\bar{Z}_{j-1};\hat{\beta}_j)\\
    & \quad \cdot \left\{ {\bar{\hat{w}}^*_j}^\top(\bar{Z};\hat{\beta})\hat{r}_j(\bar{Z}_j;\hat{\beta}_j) - {\bar{w}^*_j}^\top(\bar{Z};\beta^0)r_j(\bar{Z}_j;\beta^0_j)\right\}
\end{align*}
Then we have
\begin{align*}
    R_1(\hat{P},P^0)& = \hat{\phi}_1 - \phi_1 + E_{P^0}\left[\sum_{j\in\mathcal{J}}\{ (I)_j-(II)_j\} \mid S\in\mathcal{S}_j\right]P^0(S\in\mathcal{S}_j)\\
    & = \underbrace{\hat{\phi}_1 - \phi_1 + E_{P^0}\left[\sum_{j\in\mathcal{J}} (I)_j \mid S\in\mathcal{S}_j\right]P^0(S\in\mathcal{S}_j)}_{(A)} \\
    & \quad - \underbrace{E_{P^0}\left[\sum_{j\in\mathcal{J}} (II)_j \mid S\in\mathcal{S}_j\right]P^0(S\in\mathcal{S}_j)}_{(B)}.
\end{align*}

We denote $\hat{f}(\bar{Z}_2):=\frac{\hat{p}(Z_1 \mid S\in\mathcal{A}_1)}{\hat{p}(Z_1 \mid S\in\mathcal{S}_3)}\frac{\mathbbm{1}(Z_2 = 1)}{\hat{P}(Z_2 = 1 \mid Z_1, S\in\mathcal{S}_3)}$. Then
\begin{align*}
    (A)& = E_{P^0}\bigg[\mathbbm{1}(S\in\mathcal{S}_3)\hat{f}(\bar{Z}_2) E_{P^0}\left[ \hat{\lambda}_3(\bar{Z}_3;\hat{\beta}_3)(Z_3 - \mu_{\hat{P}}(1,Z_1))\mid Z_2 = 1, Z_1, S\in\mathcal{S}_3\right]  \bigg] \\
    & \quad  + E_{P^0}\left[\mathbbm{1}(S\in\mathcal{S}_1)\hat{\lambda}_1(Z_1;\hat{\beta}_1)(\mu_{\hat{P}}(1,Z_1) - \hat{\phi}_1) \right] + \hat{\phi_1} - \phi_1\\
    & = E_{P^0}\left[\mathbbm{1}(S\in\mathcal{S}_3) \left( \hat{f}(\bar{Z}_2) -f(\bar{Z}_2)\right)\left(\mu(1,Z_1) - \mu_{\hat{P}}(1,Z_1)\right)\right]\\
    & \quad + E_{P^0}\bigg[\mathbbm{1}(S\in\mathcal{S}_3)\hat{f}(\bar{Z}_2) \cdot \\
    & \hspace{4em}E_{P^0}\left[\left(\hat{\lambda}_3(\bar{Z}_3;\hat{\beta}_3) - \lambda_3(\bar{Z}_3;\beta^0_3)\right)(Z_3 - \mu(1,Z_1))\mid Z_2=1, Z_1, S\in\mathcal{S}_3\right]\bigg]\\
    & \quad + E_{P^0}\left[\mathbbm{1}(S \in\mathcal{S}_1)\left( \hat{\lambda}_1(Z_1;\hat{\beta}_1) - \lambda_1(Z_1;\beta_1^0)\right)(\mu(1,Z_1) - \mu_{\hat{P}}(1,Z_1))\right].
\end{align*}
By Cauchy-Schwarz inequality, the above is bounded by (up to a multiplicative factor),
\begin{align*}
    (A)& \leq \left\{\left\Vert \ \hat{\lambda}_1(Z_1)  - \lambda_1(Z_1) \right\Vert + \left\Vert \ \hat{\pi}_1(Z_1) -\pi_1(Z_1) \right\Vert \right\}\left\Vert \mu(1,Z_1) - \mu_{\hat{P}}(1,Z_1)\right\Vert\\
    & \quad  + \left\Vert \mu(1,Z_1) - \mu_{\hat{P}}(1,Z_1)\right\Vert \left\Vert \hat{\lambda}_3(\bar{Z}_3;\hat{\beta}_3) - \lambda_3(\bar{Z}_3;\beta^0_3) \right\Vert  \\
    & \quad + \left\Vert \mu(1,Z_1) - \mu_{\hat{P}}(1,Z_1)\right\Vert \left\Vert \hat{\lambda}_1(Z_1;\hat{\beta}_1) - \lambda_1(Z_1;\beta^0_1) \right\Vert.
\end{align*}
Under Condition S\ref{cond:rem1} and S\ref{cond:rem2}, $(A) = o_p(n^{-1/2})$. For the other term $(B)$, we note 
\begin{align*}
    (B) & =  \sum_{j\in\mathcal{J}}E_{P^0}\left[ \mathbbm{1}(S\in\mathcal{S}_j) (II)_j\right]\\
    & =  \sum_{j\in\mathcal{J}}E_{P^0}\bigg[ \mathbbm{1}(S\in\mathcal{S}_j) E_{\hat{P}}\left[ \hat{r}_j(\bar{Z}_j;\hat{\beta}_j)D_{\hat{Q},j}(\bar{Z}_j) {\bar{\hat{w}}^*_j}^\top(\bar{Z};\hat{\beta})\mid \bar{Z}_{j-1}, S\in\mathcal{A}_j\right]\\
    & \quad \cdot \hat{M}^-_j(\bar{Z}_{j-1};\hat{\beta}_j)\left\{ {\bar{\hat{w}}^*_j}^\top(\bar{Z};\hat{\beta})\hat{r}_j(\bar{Z}_j;\hat{\beta}_j)  - {\bar{w}^*_j}^\top(\bar{Z};\beta^0)r_j(\bar{Z}_j;\beta^0_j)\right\}\bigg]\\
    & =  \sum_{j\in\mathcal{J}}E_{P^0}\bigg[ \mathbbm{1}(S\in\mathcal{S}_j) \sum_{m\in\mathcal{S}_j}\sum_{s\in\mathcal{S}_j}\big\{  E_{\hat{P}_s}\left[\hat{r}_j(\bar{Z}_j;\hat{\beta})(Z_j - \mu_{\hat{P}_j})\mid \bar{Z}_{j-1}\right] \\
    & \quad \cdot \left( \Delta^{-1} - E_{\hat{P}}\left[ \hat{r}_j \bar{w}_j^* {\bar{w}_j^* }^\top \mid\bar{Z}_{j-1}, S\in\mathcal{A}_j\right]\right)^-_{sm} (\hat{w}^*_{j,m}(\bar{Z};\hat{\beta}_{j,m})\hat{r}_j(\bar{Z}_j;\hat{\beta}_j) - 1 )\big\} \bigg]\\
    & \leq  \sum_{j\in\mathcal{J}} \sum_{m\in\mathcal{S}_j}
    {\left\Vert  (\hat{w}^*_{j,m}(\bar{Z};\hat{\beta}_{j,m})\hat{r}_j(\bar{Z}_j;\hat{\beta}_j) - {w}^*_{j,m}(\bar{Z};{\beta}^0_{j,m}){r}_j(\bar{Z}_j;{\beta}^0_j))\right\Vert}\cdot\\
    & \quad  {\left\Vert \sum_{s\in\mathcal{S}_j} E_{\hat{P}_s}\left[\hat{r}_j(\bar{Z}_j;\hat{\beta})(Z_j - \mu_{\hat{P}_j})\mid \bar{Z}_{j-1}\right] \left( \Delta^{-1} - E_{\hat{P}}\left[ \hat{r}_j \bar{w}_j^* {\bar{w}_j^* }^\top \mid\bar{Z}_{j-1}, S\in\mathcal{A}_j\right]\right)^-_{sm} \right\Vert}\\
    & =  \sum_{j\in\mathcal{J}} \sum_{m\in\mathcal{S}_j}
    {\left\Vert  \frac{\hat{p}(\bar{Z}_j \mid S=m)}{\hat{p}(\bar{Z}_j \mid S \in \mathcal{S}_j)} - \frac{p^0(\bar{Z}_j \mid S=m)}{p^0(\bar{Z}_j \mid S \in \mathcal{S}_j)}\right\Vert}\cdot\\
    & \quad  {\left\Vert \sum_{s\in\mathcal{S}_j} E_{\hat{P}_s}\left[\hat{r}_j(\bar{Z}_j;\hat{\beta})(Z_j - \mu_{\hat{P}_j})\mid \bar{Z}_{j-1}\right] \left( \Delta^{-1} - E_{\hat{P}}\left[ \hat{r}_j \bar{w}_j^* {\bar{w}_j^* }^\top \mid\bar{Z}_{j-1}, S\in\mathcal{A}_j\right]\right)^-_{sm} \right\Vert}\\
    & = o_p(n^{-1/2}) \textnormal{ under Condition S\ref{cond:rem3}}.
\end{align*}
Next, we study the term $R_2$, which corresponds to the remainder term associated with estimating $\beta^0$:
\begin{align*}
    R_2(\hat{P},P^0) & = E_{P^0}\left[ E_{\hat{P}}[\tilde{D}_{\hat{P}}(Z) \dot{\ell}_{\hat{P}}(Z,S;\hat{\beta})] D^\beta_{\hat{P}}(Z,S;\hat{\beta})\right].
\end{align*}
Provided that $E_{\hat{P}}[\tilde{D}_{\hat{P}}(Z) \dot{\ell}_{\hat{P}}(Z,S;\hat{\beta})]$ is bounded, and $\hat{I}$ is invertible, it suffices to study the term $E_{P^0}[\dot{\ell}^*_{\hat{P}}(Z,S;\hat{\beta})]$. For clarity, we study the the efficient score function for a specific $\beta^0_{j,m}$, which is the corresponding parameter for measuring the shift in $Z_j\mid \bar{Z}_{j-1}$ between source data source $m$ and the target population. It can be verified that $E_{P^0}\left[\hat{\dot{\ell}}_{\beta_{j,m}}^*(\bar{Z}_j,S;\hat{\beta}_{j,m})\right] = (C) - (D)$ where
\begin{align*}
 (C)& =  E_{P^0}\bigg[ \mathbbm{1}(S\in\mathcal{S}_j)\left(\mathbbm{1}(S=m) - \hat{r}_{j,m}(\bar{Z}_j;\hat{\beta}_{j})\right)\\
 & \hspace{5em} \cdot \left(\frac{\dot{w}_{j,m}(\bar{Z}_j;\hat{\beta}_{j,m})}{w_{j,m}(\bar{Z}_j;\hat{\beta}_{j,m})} - E_{\hat{P}}\left[\frac{\dot{w}_{j,m}(\bar{Z}_j;\hat{\beta}_{j,m})}{w_{j,m}(\bar{Z}_j;\hat{\beta}_{j,m})} \mid \bar{Z}_{j-1}, S\right]\right)\bigg] \\
  (D)& =  E_{P^0}\bigg[ \mathbbm{1}(S\in\mathcal{S}_j) E_{\hat{P}}\left[\hat{a}_j(\bar{Z}_j;\hat{\beta}_j) {\bar{\hat{w}}^*_j}^\top(\bar{Z}_j;\hat{\beta})\mid \bar{Z}_{j-1}, S\in\mathcal{A}_j\right]\hat{M}^-_j(\bar{Z}_{j-1};\hat{\beta}_j)\\
  & \hspace{5em} \cdot \left\{ {\bar{\hat{w}}^*_j}^\top(\bar{Z}_j;\hat{\beta})\hat{r}_j(\bar{Z}_j;\hat{\beta}_j) - {\bar{w}^*_j}^\top(\bar{Z};\beta^0)r_j(\bar{Z}_j;\beta^0_j)\right\}\bigg],
\end{align*}
where by Cauchy-Schwarz inequality, $(D) = o_p(n^{-1/2})$ under Condition S\ref{cond:rem6}. We denote $\nu_{j,m}(\bar{z}_{j-1},s;\beta^0_{j,m}):=E_{P^0}\left[\frac{\dot{w}_{j,m}(\bar{Z}_j;\beta^0_{j,m})}{w_{j,m}(\bar{Z}_j;\beta^0_{j,m})} \mid \bar{z}_{j-1}, s\right]$, then
\begin{align*}
    (C)   & = E_{P^0}\Bigg[\mathbbm{1}(S\in\mathcal{S}_j)\left(\mathbbm{1}(S=m) - \hat{r}_{j,m}(\bar{Z}_j;\hat{\beta}_{j})\right)  \left(\frac{\dot{w}_{j,m}(\bar{Z}_j;\hat{\beta}_{j,m})}{w_{j,m}(\bar{Z}_j;\hat{\beta}_{j,m})} -  \nu_{j,m}(\bar{Z}_{j-1},S;\beta^0_{j,m})\right)\Bigg] \\
    & \quad - E_{P^0}\Bigg[\mathbbm{1}(S\in\mathcal{S}_j)\left(\mathbbm{1}(S=m) - \hat{r}_{j,m}(\bar{Z}_j;\hat{\beta}_{j})\right) \\
    & \hspace{5em}\cdot \left(E_{\hat{P}}\left[\frac{\dot{w}_{j,m}(\bar{Z}_j;\hat{\beta}_{j,m})}{w_{j,m}(\bar{Z}_j;\hat{\beta}_{j,m})} \mid \bar{Z}_{j-1}, S\right] -  \nu_{j,m}(\bar{Z}_{j-1},S;\beta^0_{j,m})\right)\Bigg]\\
    \intertext{By Cauchy-Schwarz inequality, $(C)$ is bounded up to a multiplicative factor by }
    &  {\left\Vert r_{j,m}(\bar{Z}_j) - \hat{r}_{j,m}(\bar{Z}_j)\right\Vert}{\left\Vert E_{P^0}\left[\frac{\dot{w}_{j,m}(\bar{Z}_j;\hat{\beta}_{j,m})}{w_{j,m}(\bar{Z}_j;\hat{\beta}_{j,m})}\mid \bar{Z}_{j-1}, S\right]  -  \nu_{j,m}(\bar{Z}_{j-1},S;\beta^0_{j,m})\right\Vert}  \\
    & \quad + {\left\Vert r_{j,m}(\bar{Z}_j) - \hat{r}_{j,m}(\bar{Z}_j)\right\Vert}{\left\Vert E_{\hat{P}}\left[\frac{\dot{w}_{j,m}(\bar{Z}_j;\hat{\beta}_{j,m})}{w_{j,m}(\bar{Z}_j;\hat{\beta}_{j,m})} \mid \bar{Z}_{j-1}, S\right] - \nu_{j,m}(\bar{Z}_{j-1},S;\beta^0_{j,m})\right\Vert}   \\
    & = o_p(n^{-1/2}) \textnormal{ under Conditions S\ref{cond:rem4} and S\ref{cond:rem5}}. 
\end{align*}
 \end{proof}}
\section{Data analysis: estimand of interest, gradients and continued results}
\label{sec:app:analysis_table_continued}
For each selected amino acid features as well as the genetic distance feature, we studied the associations between PT80 and these covariates in least squares projections onto univariate working linear regression models. We take the genetic distance $X$ as an example. We are interested in estimating 
\begin{align*}
    \theta^0 : = \argmin \theta E_{Q^0}[l(X,Y;\theta)],
\end{align*}
where $l(X,Y;\theta)$ denotes the loss function that is given by
\begin{align*}
        l(x,y;\theta)& := \left[ y  - \mu(x;\theta)\right]^2,
\end{align*}
where $\mu(x;\theta) = \theta_0 + \theta_1 x$ denotes the mean function. We adopt the Z-estimation framework and our problem is equivalent to solving 
\begin{align*}
     0 & = E_{P^0}[E_{P^0}[m_{\theta}(X,Y)\mid X, S \in \mathcal{A}_3] \mid S \in \mathcal{A}_1],
\end{align*}
where $m_{\theta'}(x,y) = \frac{\partial l(x,y;\theta)}{\partial \theta}\mid_{\theta = \theta'}$. A gradient of $\theta^0$ assuming $\beta^0$ is known is thus given by
\begin{align*}
    & D_{P^0}(x,y,s;\theta^0,\beta^0)\\
    &  = - \left( \frac{\partial  E_{P^0}[E_{P^0}[m_{\theta}(X,Y)\mid X, S \in \mathcal{A}_3] \mid S \in \mathcal{A}_1]}{\partial \theta} \mid_{\theta = \theta^0}\right)^{-1} F_{P^0}(x,y,s;\theta^0,\beta^0),
\end{align*}
where
\begin{align*}
& F_{P^0}(x,y,s;\theta,\beta)\\ 
    &  = \frac{\mathbbm{1}(s \in \mathcal{S}_3)}{P^0(S \in \mathcal{S}_3)} \frac{1}{w^*_{3,s}(x,y;\beta)}\frac{dP^0(x \mid S\in \mathcal{A}_1)}{dP^0(x \mid S\in \mathcal{S}_3)}\left(m_{\theta} - E_{P^0}[m_{\theta}\mid x, S \in \mathcal{A}_3]\right) \\
    & \quad +\frac{\mathbbm{1}(s \in \mathcal{S}_1)}{P^0(S \in \mathcal{S}_1)} \frac{1}{w^*_{1,s}(x;\beta)} \left( E_{P^0}[m_{\theta}\mid x, S \in \mathcal{A}_3] - E_{P^0}[E_{P^0}[m_{\theta}\mid x, S \in \mathcal{A}_3]\mid S \in \mathcal{A}_1]\right).
\end{align*}

The canonical gradient of $\theta^0$ can be constructed using the result in Theorem~\ref{thm: canonical}.

{\begin{longtable}{llrrrrr} 
\caption{Estimated coefficient using the HVTN 703 and HVTN 704 data. Estimation results are presented as estimates (standard errors).}\label{beta_PT80_appendix} \\
\toprule
& & \multicolumn{2}{c}{Augmenting HVTN 703} & \multicolumn{2}{c}{Augmenting HVTN 704} \\
\cmidrule(l){3-4} \cmidrule(l){5-6}
& & 703 only & Efficient & 704 only & Efficient\\
Residue & Site & (N=43) & (N=98) & (N=55) & (N=98) \\
\midrule
\endfirsthead
\toprule
& & \multicolumn{2}{c}{Augmenting HVTN 703} & \multicolumn{2}{c}{Augmenting HVTN 704} \\
\cmidrule(l){3-4} \cmidrule(l){5-6}
& & 703 only & Efficient & 704 only & Efficient\\
Residue & Site & (N=43) & (N=98) & (N=55) & (N=98) \\
\midrule
\endhead
\bottomrule
\endfoot
I & 371 &  &  & 0.10 (0.33) & 0.44 (0.20)\\
I & 467 & 0.59 (0.32) & 0.21 (0.23) & -0.51 (0.43) & -0.20 (0.22)\\
K & 432 &  &  & 0.48 (0.32) & 0.48 (0.18)\\
K & 476 & -0.27 (0.31) & -0.05 (0.22) & -0.02 (0.31) & -0.07 (0.24)\\
K & 97 & 0.49 (0.36) & 0.47 (0.20) &  & \\
L & 369 &  &  & -0.62 (0.38) & -0.05 (0.33)\\
L & 426 &  &  & 0.09 (0.45) & 0.21 (0.21)\\
M & 426 &  &  & 0.43 (0.41) & 0.25 (0.22)\\
N & 279 & 0.37 (0.36) & 0.47 (0.23) & 0.44 (0.31) & 0.41 (0.17)\\
N & 461 &  &  & -0.28 (0.43) & -0.15 (0.27)\\
N & 463 & 0.28 (0.40) & 0.31 (0.28) & 0.33 (0.34) & 0.46 (0.22)\\
N & 465 &  &  & -0.43 (0.32) & -0.56 (0.17)\\
N & 474 & 0.22 (0.32) & 0.03 (0.22) & -0.32 (0.30) & -0.18 (0.18)\\
P & 369 &  &  & 0.58 (0.34) & 0.23 (0.28)\\
R & 432 &  &  & -0.25 (0.30) & -0.38 (0.18)\\
R & 476 & 0.27 (0.31) & 0.43 (0.29) & 0.13 (0.31) & 0.18 (0.17)\\
S & 198 & -0.26 (0.32) & -0.19 (0.21) &  & \\
S & 278 &  &  & -0.14 (0.31) & -0.08 (0.15)\\
S & 365 & 0.43 (0.32) & 0.12 (0.26) & -0.21 (0.35) & -0.02 (0.25)\\
T & 198 & 0.16 (0.31) & 0.36 (0.21) &  & \\
T & 278 &  &  & 0.28 (0.31) & 0.25 (0.20)\\
T & 281 & -0.70 (0.38) & -0.65 (0.24) & -0.33 (0.37) & -0.43 (0.25)\\
T & 283 &  &  & 0.25 (0.33) & 0.05 (0.18)\\
T & 455 & -0.04 (0.32) & -0.10 (0.18) &  & \\
T & 461 &  &  & 0.61 (0.28) & 0.48 (0.21)\\
T & 465 & 0.31 (0.37) & 0.25 (0.24) & 0.41 (0.32) & 0.55 (0.21)\\
T & 467 & -0.55 (0.31) & -0.13 (0.23) & 0.38 (0.36) & 0.02 (0.22)\\
V & 371 &  &  & -0.10 (0.33) & -0.43 (0.20)\\
V & 455 & 0.33 (0.37) & 0.30 (0.22) &  & \\
\bottomrule
\end{longtable}}

\end{appendices}

\bibliography{reference}
\end{document}